\documentclass[article]{JHEP3} 

\usepackage{epsfig}
\usepackage{amssymb}
\usepackage{cancel}
\usepackage{cite}

\def\beq{\begin{equation}}
\def\beqn{\begin{eqnarray}}
\def\eeq{\end{equation}}
\def\eeqn{\end{eqnarray}}
\def\abs#1{\left|#1\right|}

\newcommand\sss{\scriptscriptstyle}

\newcommand\as{\alpha_{\sss S}}

\newcommand\IMC{I_{\sss MC}}
\newcommand\GenMCk{{\cal F}_{\mbox{\tiny MC}}^{(k)}}
\newcommand\clH{{\mathbb H}}
\newcommand\clS{{\mathbb S}}
\newcommand\kt{k_{\sss T}}
\newcommand\mt{m_{\sss T}}
\newcommand\pt{p_{\sss T}}
\newcommand\vpt{\vec{p}_{\sss T}}
\newcommand\ptBorn{p_{\sss T}^{Born}}
\newcommand\ptD{p_{\sss T}^{(D)}}
\newcommand\ptcut{p_{\sss T}^{{\rm cut}}}
\newcommand\stepf{\Theta}
\newcommand\Kern{{\cal K}}
\newcommand\KernMC{{\cal K}_{\sss\rm MC}}
\newcommand\kinHi{\Xi_{\clH,i}}
\newcommand\kinSi{\Xi_{\clS,i}}
\newcommand\kinHz{\Xi_{\clH,0}}
\newcommand\kinSz{\Xi_{\clS,0}}
\newcommand\kinHo{\Xi_{\clH,1}}
\newcommand\kinSo{\Xi_{\clS,1}}

\newcommand\kinHN{\Xi_{\clH,N}}
\newcommand\kinSN{\Xi_{\clS,N}}
\newcommand\tb{\bar{t}}
\newcommand\muME{\mu_{{\rm ME}}}
\newcommand\muimn{\mu_{i,{\min}}}
\newcommand\muimx{\mu_{i,{\max}}}
\newcommand\muSimn{\mu_{i,{\min}}^{(\clS)}}
\newcommand\muSimx{\mu_{i,{\max}}^{(\clS)}}
\newcommand\muHimn{\mu_{i,{\min}}^{(\clH)}}
\newcommand\muHimx{\mu_{i,{\max}}^{(\clH)}}
\newcommand\muSNmn{\mu_{N,{\min}}^{(\clS)}}

\newcommand\muHNmn{\mu_{N,{\min}}^{(\clH)}}

\newcommand\ea{{\em a}}
\newcommand\eb{{\em b}}
\newcommand\ec{{\em c}}
\newcommand\ed{{\em d}}
\newcommand\ee{{\em e}}

\newcommand{\HWs}{{\sc HERWIG6}}
\newcommand{\HWpp}{{\sc Herwig++}}
\newcommand{\PY}{{\sc Pythia}}
\newcommand{\PYs}{{\sc Pythia6}}
\newcommand{\PYe}{{\sc Pythia8}}
\newcommand{\madgraph}{{\sc MadGraph}}
\newcommand{\madloop}{{\sc MadLoop}}

\newcommand{\amcatnlo}{a{\sc MC@NLO}}

\newcommand{\alpgen}{{\sc Alpgen}}
\newcommand{\fastjet}{{\sc FastJet}}

\preprint{
 CERN-PH-TH/2012-247 \\
 ZU-TH 21/12
 }
\title{Merging meets matching in MC@NLO}

\author{Rikkert Frederix\\
Institut f\"ur Theoretische Physik, Universit\"at Z\"urich,
Winterthurerstrasse 190,\\ CH-8057 Z\"urich, Switzerland
}
\author{Stefano Frixione%
  \thanks{On leave of absence from INFN, Sezione di Genova, Italy.}\\
  PH Department, TH Unit, CERN, CH-1211 Geneva 23, Switzerland\\
  ITPP, EPFL, CH-1015 Lausanne, Switzerland
}

\abstract{
The next-to-leading order accuracy for MC@NLO results exclusive 
in $J$ light jets is achieved if the computation is based 
on matrix elements that feature $J$ and $J\!+\!1$ QCD partons.
The simultaneous prediction of observables which are exclusive in 
different light-jet multiplicities cannot simply be obtained by summing 
the above results over the relevant range in $J$; rather, a suitable 
merging procedure must be defined. We address the problem of such a merging,
and propose a solution that can be easily incorporated into existing
MC@NLO implementations. We use the automated \amcatnlo\ framework to
illustrate how the method works in practice, by considering the 
production at the 8~TeV LHC of a Standard Model Higgs in association 
with up to $J=2$ jets, and of an $e^+\nu_e$ pair or a $t\tb$ pair 
in association with up to $J=1$ jet.
}


\begin{document}


\section{Introduction and definitions\label{sec:intro}}
Let us consider the process
\beq
P_1+P_2\;\longrightarrow\;S+M~{\rm jets}\,,
\label{hadproc}
\eeq
where $P_1$ and $P_2$ can be either hadrons or leptons, and in order to 
simplify the discussion we assume that the final state is defined at the
parton level (that is, before the hadronization phase in an event
generator). $S$ is a set of particles which does not contain any QCD 
massless partons, and the $M$ jets are light, i.e., obtained by clustering
light quarks and gluons. If the definition of a given observable $O$
explicitly involves $J$ jets, with $0\le J\le M$ (with or without a 
further dependence on the four-momenta of the particles in $S$), 
we refer to such an observable as exclusive in $J$ jets, and 
inclusive in the remaining $M-J$ jets; when $J=0$, the observable
is typically called fully inclusive.

In Monte Carlos (MCs) based on the leading-order (LO) approximation, the
description of the process in eq.~(\ref{hadproc}) stems from the 
underlying tree-level matrix elements relevant to the process
\beq
a_1+a_2\;\longrightarrow\;S+i~{\rm partons}\,,
\label{partproc}
\eeq
with $a_1$ and $a_2$ partons or leptons (according to the identities
of $P_k$), and $i$ a {\em fixed} number with $0\le i\le M$; the shower
must generate at least $M-i$ partons, in order to obtain a final
state that can contribute to eq.~(\ref{hadproc}). The most 
straightforward approach is that of using matrix elements with
$i=0$ (which, for later use, we suppose to be of ${\cal O}(\as^b)$), 
and thus of letting the shower generate all partons. The apparent
simplicity of this procedure has the drawback that cross sections
are severely underestimated in the hard regions (i.e., for energetic
and well-separated jets). This problem, which obviously worsens when 
increasing the c.m.~energy, stems from the fact that showers give
only a leading-logarithmic (LL) approximation to the full matrix
elements that would be the correct description in those phase-space
regions. As a result, one may say that with $i=0$ one obtains
a LL accuracy for observables exclusive in $J$ jets, with $J\ge 1$;
for fully-inclusive observables, the accuracy is LO+LL.
In order to partly remedy to the lack of hard emissions in Monte Carlos,
the shower scale (which is directly related to the largest hardness
attainable) can be increased in the latter to very large values. While 
this is formally correct, it entails the stretching of Monte Carlo 
simulations outside their range of validity, and thus a loss of
predictive power.

If one is interested in $J$-jet exclusive observables, for $J$ given
and fixed, it is best to use matrix elements with $i=J$, which promotes
such observables to LO+LL accuracy. This has the disadvantage that
the event samples generated in this way cannot be used for observables
exclusive in less than $J$ jets, since in doing so one would be forced
to integrate matrix elements over soft and collinear regions, where
they diverge. There are technical implications as well. Firstly, the
matrix-element divergences must be avoided by introducing (unphysical)
cutoffs, chosen so as not to bias physical predictions. Secondly, 
one requires that the largest jet hardness generated in the shower phase 
be less than that at the matrix-element level, which can be typically
done with a suitable choice of the shower scale. This requirement is 
not necessary in order to avoid double counting (which cannot occur in this
context: since each shower emission is associated with one power of $\as$, 
after a single emission one has already a contribution of relative
${\cal O}(\as^{i+1})$), but rather in order to maintain the expected 
LO+LL accuracy of the $J$-jet exclusive observables.
The two issues discussed here are specific manifestations of the
general problem of the {\em matching} between matrix-element calculations
and parton shower simulations.

A first improvement on the situation presented in the preceding paragraph
aims at the LO+LL-accurate description of fully inclusive {\em and} $1$-jet
exclusive observables, in the context of a single simulation; the 
corresponding techniques are generally known as matrix element corrections 
(MECs), although this terminology has been lately adopted to identify
other, and more sophisticated, approaches. MECs rely on the computations
of $i=0$ and $i=1$ matrix elements, which are combined by either 
``switching'' between the two descriptions they
underpin~\cite{Seymour:1994we,Seymour:1994df}, or by directly
including the information on $i=1$ in the showers, which is still
initiated starting from an $i=0$ configuration~\cite{Bengtsson:1986hr,
Gustafson:1987rq,Miu:1998ju}. These consistent combinations of matrix
elements characterized by different multiplicities are the first
examples of {\em merging} procedures\footnote{We point out that
in ref.~\cite{Miu:1998ju} the approach of refs.~\cite{Seymour:1994we,
Seymour:1994df} is called matching, and that of refs.~~\cite{Bengtsson:1986hr,
Gustafson:1987rq,Miu:1998ju} merging. This naming convention becomes a source
of confusion when one considers the problem of NLO corrections, and hence
is not adopted here.}.

Extensions of MECs to arbitrarily-large multiplicities have attracted 
considerable interest, in view of their relevance to Tevatron and LHC 
phenomenology, where multi-jet processes are ubiquitous. There are now 
well-established procedures, such as CKKW~\cite{Catani:2001cc,Krauss:2002up},
\mbox{CKKW-L}~\cite{Lonnblad:2001iq,Lavesson:2005xu} (and their later
improvements, see refs.~\cite{Hoeche:2009rj,Hamilton:2009ne,Lonnblad:2011xx}),
and MLM~\cite{Alwall:2007fs}. These merging approaches are
based on the computations of all tree-level matrix elements 
(which is why they can be referred to as attaining an LO 
merging) with $0\le i\le N$, where $N$ is as large as a 
computer can handle; for $J\le N$ the $J$-jet exclusive observables
are LO+LL accurate. Since LO-merging techniques use tree-level
results, cutoffs are necessary that prevent infrared (IR) divergences 
from occurring. The avoidance of biases in physical results is much more 
difficult to achieve than in the case of the two parton multiplicities
relevant to MECs, and the interplay between the merging
and matching conditions is in fact non trivial. The matching can
be expressed in terms of the IR cutoffs, which can be collectively
(and slightly improperly) called matching scale\footnote{It is
in fact a matching {\em and} merging scale. We stick to the 
standard notation.}. The dependence 
of observables on the matching scale never completely
vanishes, but in practice it is sufficiently small numerically
(consistently with the proof~\cite{Catani:2001cc} that for CKKW
in $e^+e^-$ collisions it is suppressed by two logarithmic powers
w.r.t.~the nominal accuracy of the shower). We point out that
such a dependence is the unavoidable consequence of the fact that
a cutoff-free cross section at relative ${\cal O}(\as^N)$
can only be achieved by a full N$^N$LO computation (i.e., which
includes loop corrections). Clever matching conditions reduce the
cutoff dependence, but cannot eliminate it.

In parallel to the development of LO-merging procedures, the problem
was considered of matching next-to-leading order (NLO) QCD computations
with parton showers. This spurred quite a lot of theoretical 
activity~\cite{Dobbs:2001gb,Frixione:2002ik,Chen:2001nf,
Kurihara:2002ne,Nason:2004rx,Nagy:2005aa,Bauer:2006mk,Nagy:2007ty,
Giele:2007di,Bauer:2008qh,Lavesson:2008ah,Hoche:2010pf,Hoeche:2011fd},
but currently the only proposals that are systematically applied 
to hadroproduction processes are MC@NLO~\cite{Frixione:2002ik}
and POWHEG~\cite{Nason:2004rx} (possibly in their SHERPA
implementations, refs.~\cite{Hoeche:2011fd} and~\cite{Hoche:2010pf}
respectively). The problem is the analogue
of the LO matching for $J$-jet exclusive observables, with $J$ fixed,
discussed above. The novelty is that the underlying matrix elements
are of ${\cal O}(\as^{b+J})$ {\em and} of  ${\cal O}(\as^{b+J+1})$.
The former are the same as those relevant to the LO matching, but play
the role of Born contributions in this context; the latter include the real 
corrections, their IR subtraction terms, and the one-loop contributions.
Although real matrix elements have parton multiplicities one unit larger
than the other matrix elements involved, which is also what happens in 
MECs, the problem at hand is a matching, and not a merging, one. In fact,
its aim is not that of enlarging the number of jet multiplicities which
can be predicted to LO+LL accuracy, but rather that of promoting $J$-jet
exclusive observables from LO+LL to NLO+LL accuracy; the fact that
$(J+1)$-jet exclusive observables are indeed predicted at LO+LL is beside
the point, since this fact is a spinoff of the method, rather than its
primary motivation. Technically, what happens is that one set of IR 
divergences (those arising from one parton becoming soft, or two 
partons becoming collinear) are cancelled without the need of introducing
a cutoff. Apart from the case of zero jet at the Born level ($J=0$),
cutoffs are still required in order to prevent the cross sections from 
diverging because of multi-parton IR configurations (e.g., two partons 
becoming soft, or three partons becoming collinear); however, their
impact on fully-showered events is much reduced w.r.t.~the analogous 
situation at the LO  (loosely speaking, by a factor of $\as$ -- see 
e.g.~ref.~\cite{Frederix:2011ig} for a discussion on this point).

The scope of LO-merging procedures is larger than that of NLO matching,
since in the latter the relevance of matrix-element information is
limited to a couple of jet multiplicities (of which only one is
truly NLO, as mentioned above). Still, NLO-matched approaches must be
the method of choice whenever possible, because they are capable of
giving predictions affected by theoretical uncertainties (such as
cutoff biases and scale dependences) significantly smaller than at
the LO. It is therefore clear that the situation can be improved
by combining the two strategies, in order to merge consistently event 
samples which are individually matched with parton showers to NLO accuracy 
(this, we call for brevity an NLO-merging approach). Merging at NLO
requires one to tackle two issues. Firstly, there is the purely theoretical
problem of devising an acceptable solution, which has stimulated much
work lately~\cite{Lavesson:2008ah,Hamilton:2010wh,Hoche:2010kg,
Giele:2011cb,Alioli:2011nr,Hoeche:2012yf,Gehrmann:2012yg}\footnote{We
expect, in particular, that refs.~\cite{Hoeche:2012yf,Gehrmann:2012yg}
may have features in common with the present paper, since use is made
there of the MC@NLO formalism, as it is also advocated here. A detailed
comparison between the two approaches is beyond the scope of this work, 
not least because they differ by higher orders already at the un-merged 
level~\cite{Frixione:2002ik,Hoeche:2011fd}.}. Secondly,
in order for the above solution not to remain an academic achievement,
the computations must be feasible of all matrix elements involved in the 
relevant NLO cross sections (i.e., up to relatively large multiplicities).
This is indeed the case thanks to the extremely high level of automation
achieved in the past few years for the one-loop~\cite{Denner:2005nn,
Berger:2008sj,Ellis:2008ir,Hirschi:2011pa,Bevilacqua:2011xh,Becker:2011vg,
Cullen:2011ac,Cascioli:2011va,Agrawal:2011tm,Bern:2011ep,Badger:2012pg}
and the subtracted real~\cite{Gleisberg:2007md,Frederix:2008hu,
Czakon:2009ss,Frederix:2009yq,Hasegawa:2009tx,Frederix:2010cj} 
contributions, both of which have greatly benefitted from the 
thorough understanding of tree-level-amplitude
calculations~\cite{Stelzer:1994ta,Caravaglios:1995cd,Yuasa:1999rg,
Kanaki:2000ey,Moretti:2001zz,Krauss:2001iv,Mangano:2002ea,Fujimoto:2002sj,
Cafarella:2007pc,Boos:2009un,Alwall:2011uj}.

The goal of this paper is that of proposing a prescription for an
NLO merging built upon the matching achieved according to the
MC@NLO formalism~\cite{Frixione:2002ik}, and which requires only very
minimal modifications to the latter. This allows us to easily implement 
such a merging scheme into the existing \amcatnlo\ framework (see 
e.g.~refs.~\cite{Frederix:2011ss,Frederix:2011ig} for recent applications),
and to test it by considering, at the 8~TeV LHC, the production of
$S=H$ (Standard Model Higgs), $S=e^+\nu_e$, and $S=t\tb$,
for observables exclusive in up to one or two jets at the NLO.
The paper is organized as follows: in sect.~\ref{sec:merg}
we describe our approach; in sect.~\ref{sec:res} we present
sample predictions; finally, in sect.~\ref{sec:conc} we
draw our conclusions.

\section{Merging at the NLO\label{sec:merg}}

\subsection{Outline of the procedure\label{sec:descr}}
Let us consider the process in eq.~(\ref{partproc}), and interpret it
as the ${\cal O}(\as^{b+i})$ Born contribution to a $(J=i)$-jet exclusive 
cross section -- hence, all final-state partons are hard and well separated. 
The total transverse momentum vanishes because of momentum conservation:
\beq
\ptBorn=\abs{\vpt(S)+\sum_{j=1}^i \vpt(j)}=0\,.
\eeq
The ${\cal O}(\as^{b+i+1})$ real corrections feature $(S+i+1)$-body
final states, which define the $\clH$-event configurations in MC@NLO.
For such final states, $\ptBorn$ is different from zero\footnote{$\ptBorn$
is not an observable, but can be defined operatively in both fixed-order
computations and MC simulations. However, we shall not need to do that
here.}; this property will be exploited in what follows as an intuitive 
way to measure the ``extra'' radiation w.r.t.~a given (Born) kinematic 
configuration\footnote{Other variables can be devised so as to distinguish
Born from real-emission configurations; they are all equivalent for
the sake of the present discussion.}.
The $(i+1)^{th}$ parton can be arbitrarily soft or collinear to any 
other parton (which implies $\ptBorn\simeq$~small), but also hard 
and well separated (where $\ptBorn\simeq$~large). All the 
other ${\cal O}(\as^{b+i+1})$ contributions to the cross section have 
an $(S+i)$-body kinematics, identical to that of the Born; these are 
the $\clS$-event configurations in MC@NLO. 

After processing hard events with parton showers, one will obtain 
configurations quite different from those of the $\clS$ and $\clH$ 
events; in particular, final-state multiplicities will have greatly 
increased. However, these differences may be irrelevant to physics
observables, which may be almost identical, in shape and normalization,
to those resulting from an NLO parton-level computation\footnote{In some 
cases, hadronization effects can blur this picture, and is therefore 
convenient to consider them switched off in MCs for the time being.}. For 
this not to be the case, two conditions must be fulfilled. Firstly, the 
observable
must be IR-sensitive (i.e., large logarithms can appear in the coefficients
of its perturbative expansion). Secondly, one must be in an IR phase-space 
region, where partons are soft and/or collinear (which causes those 
logarithms to grow large); this corresponds to having $\ptBorn\simeq$~small.
When this happens, the shape of the observable is determined by the MC
(large logarithms are resummed), while its normalization is still 
dictated by the underlying NLO matrix elements (thanks to the unitarity 
property of the shower). This implies, in particular, that the value of
$\ptBorn$ of the configuration emerging from the shower can be
markedly different from that relevant to the $\clH$ event from where
the shower started (which is trivially true for $\clS$ events, since
they have $\ptBorn=0$). On average, one can say that in the IR regions
$\clS$ and $\clH$ events provide the normalization, while the kinematics
is controlled by the MC. 

Let us now consider the hard regions, where
$\ptBorn\simeq$~large. $\clS$ events do not contribute there, since in
order to do so the shower would have to provide all the extra radiation 
leading to $\ptBorn$ (which is still possible, but at the price of choosing 
unjustifiably large shower scales). On the other hand, $\clH$ events do 
contribute; more specifically, the values of $\ptBorn$ before and after the 
shower do not differ significantly. Thus, on average, in the hard regions
$\clH$ events provide one with both the normalization and the kinematic
configurations. Finally, it should be stressed that the characteristics
of the $\clS$ and $\clH$ events discussed here are quite directly
related to the fact that MC@NLO is designed to perturb in a minimal
manner both the MC and the matrix-element results (in particular,
there are no contributions of relative ${\cal O}(\as^2)$ which are
not of MC origin).

The above observations underpin the proposal for the NLO-merging 
strategy that we sketch here.
\begin{enumerate}
\item For any given Born multiplicity, except for the largest one 
 considered, there must not be contributions to the hard regions. This
 implies, in particular, that in such regions real emissions must not
 occur, and the corresponding matrix elements must rather be viewed 
 as defining the Born process for the next (i.e., one unit larger) 
 multiplicity.
\item Suitable choices of veto scales in showers must be made for
 consistency with item 1.
\item Further conditions can be imposed which are similar to those
 used in LO-merging procedures; for a given multiplicity, the combination 
 of $\clH$- and $\clS$-event contributions plays essentially the same role 
 as a tree-level matrix element in LO mergings.
\end{enumerate}
A few comments are in order here. The definition of a ``hard'' region
is realised by introducing a (generalised) matching scale,
analogously to what is done in LO mergings. Item~1. is achieved 
by cuts at the matrix-element level, defined by means of the 
matching scale. Since MC@NLO and NLO parton-level predictions are quite 
similar in the hard regions, by getting rid of the latter, one also 
eliminates the former without the need of any extra conditions at
the MC level. However, the hard region defined here may still be 
populated by the Monte Carlo (whose choice of shower scales
is made {\em a priori}, and independently of any merging procedure)
when showering $\clS$ events, hence item 2. Note that the relevant scale 
choices there are easily worked out by taking into account what is done
in item 1., owing to the interplay between $\clH$ and $\clS$ events.

As far as item 3.~is concerned, let us suppose that, for a given
$(S+i)$-body Born kinematics, MC@NLO were used to obtain only
$i$-jet exclusive observables. This is effectively as if 
the $(i+1)^{th}$ parton were never resolved, and always integrated
over. It thus suggests to formally treat the combination of $\clS$ 
and $\clH$ events as equivalent to $(S+i)$-body tree-level matrix
elements in an LO merging which implies, among other things, the 
reweighting of these events by suitable combinations of Sudakov factors
(or equivalent suppressions, as in the MLM procedure). By construction, 
such factors must then be obtained using $(S+i)$-body configurations 
(possibly effective), also in the case of $\clH$ events. One readily observes 
that Sudakov reweightings are used in an LO merging (together with conditions 
on showers) in order to prevent different multiplicities from double 
counting which, in the present context, is supposed to be guaranteed by
items 1.~and 2.. This is correct; the idea is indeed that of employing
reweighting factors which contribute to a relative ${\cal O}(\as^2)$
(hence, beyond NLO), so as the perturbative expansions of the cross
sections that define $\clS$ and $\clH$ events are identical to the
original ones up to ${\cal O}(\as^{b+i+1})$. This is 
a condition whose application guarantees that the 
accuracy of the MC@NLO calculation is not spoiled, and which
results in the insertion into the MC@NLO short distance cross 
sections of an extra relative ${\cal O}(\as)$ term, that we shall call
$d\sigma_i^{(\Delta)}$. It also implies that Sudakov reweightings 
are totally general, and do not constrain the type of observables 
that one predicts starting from a given multiplicity -- it should be clear
that the assumption made at the beginning of this paragraph has
the sole role of simplifying the picture. Still, it will remain
true (essentially because of item 1.) that $i$-jet exclusive observables
will receive the dominant contributions from MC@NLO samples associated 
with an underlying $(S+i)$-body Born kinematics. This is directly
related to the fact that, when Sudakov reweightings are applied, 
the term $d\sigma_i^{(\Delta)}$ is numerically small; in fact,
as we shall explicitly show later, the procedures in items 1.~and 2.~are 
sufficient to obtain smooth results for most observables.

\subsection{Technicalities\label{sec:tech}}
The contributions to the process in eq.~(\ref{partproc}) 
will be denoted by:
\beq
T_i\,,\;\;\;\;\;V_i\,,
\label{matri}
\eeq
where $T_i$ are the ${\cal O}(\as^{b+i})$ tree-level matrix elements,
and $V_i$ is the ${\cal O}(\as^{b+i+1})$ sum of the finite part of the 
one-loop amplitude times that of the Born, plus the finite remainders of the 
soft and collinear subtractions. The MC@NLO cross section for what  we shall 
call the \underline{\em $i$-parton sample} can thus be written 
schematically as follows:
\beqn
d\sigma_{i}&=&d\sigma_{\clS,i}+d\sigma_{\clH,i}\,,
\label{xsecalli}
\\
d\sigma_{\clS,i}&=&T_i+V_i-T_i\Kern+T_i\KernMC\,,
\label{xsecSi}
\\
d\sigma_{\clH,i}&=&T_{i+1}-T_i\KernMC\,,
\label{xsecHi}
\eeqn
where $\Kern$ and $\KernMC$ indicate symbolically the kernels relevant
to the NLO and MC subtractions respectively. In order to simplify the
notation, in eqs.~(\ref{xsecSi}) and~(\ref{xsecHi}) we have understood 
the interface-to-MC's $\IMC$ of ref.~\cite{Frixione:2002ik} (in turn 
equivalent to the generating functionals $\GenMCk$ of 
ref.~\cite{Frixione:2003ei}), since they will not play any role in
what follows. In fact, all manipulations of the short-distance cross 
sections that will be carried out below are relevant to hard events, 
i.e.~at the parton level and before the shower phase (with one exception, 
discussed in sect.~\ref{sec:Sud}).

For our merging scheme we introduce a function:
\beqn
D(\mu)=\left\{
\begin{array}{ll}
1               &\phantom{aaaa} \mu\le\mu_1\,,\\
{\rm monotonic} &\phantom{aaaa} \mu_1<\mu\le\mu_2\,,\\
0               &\phantom{aaaa} \mu > \mu_2\,,\\
\end{array}
\right.
\label{Ddef}
\eeqn
with $\mu_1\le\mu_2$ two arbitrary mass scales, whose role
can be roughly summarized as follows:
\beqn
\mu\le\mu_1&&\phantom{aaaaaa}{\rm soft~(MC-dominated)},
\nonumber\\
\mu_1 < \mu\le\mu_2&&\phantom{aaaaaa}{\rm intermediate},
\nonumber\\
\mu > \mu_2&&\phantom{aaaaaa}{\rm hard~(ME-dominated)}.
\nonumber
\eeqn
Although one may want to choose a smooth function $D$ for numerical
reasons, it is more transparent from the physics viewpoint to
adopt a sharp version:
\beq
D(\mu)=\stepf\left(\mu_Q-\mu\right)\,,\;\;\;\;\;\;\;\;\mu_Q=\mu_1=\mu_2\,,
\label{Dsharp}
\eeq
which is a particular case of eq.~(\ref{Ddef}). In eq.~(\ref{Dsharp}),
the identification of $\mu_Q$ as the matching scale is obvious.
We shall also denote by
\beq
d_j
\eeq
the scale (with canonical dimension equal to one, i.e.~mass) 
at which a given $S$+partons configuration passes from being reconstructed 
as an $j$-jet one to being reconstructed as an $(j-1)$-jet one, according 
to a $\kt$ jet-finding algorithm~\cite{Catani:1993hr} (in other words, 
there are $j$ jets of hardness $d_j-\varepsilon$, and $(j-1)$ jets of 
hardness $d_j+\varepsilon$, with $\varepsilon$ arbitrarily small). 
For example:
\beq
a_1+a_2\;\longrightarrow\;S+a_3\phantom{aaaa}\Longrightarrow\phantom{aaaa}
d_1=\pt(a_3)\equiv\pt(S)\,.
\label{d1expl}
\eeq
In general, for $n$ final-state partons one will have
\beq
d_n\le d_{n-1}\le\ldots\le d_2\le d_1\,.
\label{dord}
\eeq
It will also turn out to be convenient to define
\beq
d_j=\sqrt{s}\,,\;\;\;\;\;\;\;\;j\le 0\,,
\label{dmax}
\eeq
with $\sqrt{s}$ the parton c.m.~energy, i.e.~the largest energy
scale available event-by-event.

Equations~(\ref{xsecalli})--(\ref{xsecHi}) imply that the $i$-parton
MC@NLO sample gets contributions from both the $i$- and the $(i+1)$-parton 
tree-level matrix elements. This is what usually happens in MC@NLO, when 
eq.~(\ref{xsecalli}) is used to compute the $i$-jet exclusive cross section
with NLO+LL accuracy. However, in the context where MC@NLO samples
with different multiplicities must be consistently merged, this
implies that a given tree-level matrix element $T_i$ will contribute to 
both the $i$-parton sample (as Born contribution) and to the $(i-1)$-parton 
sample (as real correction). This fact is peculiar of the merging at
the NLO (since at the LO one imposes that $T_i$ contributes solely
to the $i$-jet exclusive cross section), and can lead to problems 
of double-counting nature even before considering the matching
to showers. In order to avoid these problems, we introduce
the following rule:
\begin{itemize}
\item[{\em R.1}] For any given $(S+i)$-body kinematic configuration 
(with $i\ge 1$) at the matrix-element level, the sum of the contributions 
due to $T_i$ to the $i$- and $(i-1)$-parton samples must be equal to $T_i$ 
(possibly times a factor smaller than one if this helps prevent the 
reconstruction of a number of hard jets smaller than $(i-1)$).
\end{itemize}
We now proceed to incorporate rule {\em R.1} into the MC@NLO
short-distance cross sections. In order to be definite, let
us consider the merging of the $i$-parton samples with
\beq
0\le i\le N\,,
\eeq
that is, the largest final-state multiplicity will be $N+1$ 
partons, relevant to the real corrections to the $N$-parton 
sample. We then formally define modified MC@NLO formulae in the following 
way:
\beqn
d\bar{\sigma}_{i}&=&d\sigma_{i}\,
D(d_{i+1})\,\left(1-D(d_i)\right)\,\stepf\left(d_{i-1}-\mu_2\right)
\phantom{aaaa}i\le N-1\,,\phantom{aa}
\label{modi}
\\
d\bar{\sigma}_{N}&=&d\sigma_{N}\,\left(1-D(d_N)\right)\,
\stepf\left(d_{N-1}-\mu_2\right)\,,
\label{modN}
\eeqn
with $d\sigma_{i}$ given in eq.~(\ref{xsecalli}). As it will be discussed
in what follows, these expressions are incorrect; however, their intuitive
meaning is easy to grasp. Thus, we discuss here their physics contents,
and refine them later. 
The $D$ and $\stepf$ functions in eq.~(\ref{modi}) imply that,
out of the $i+1$ jets in the $i$-parton sample ($i<N$), there 
are at least:
\begin{itemize}
\item[A.] $i-1$ jets harder than $\mu_2$ (owing to 
$\stepf\left(d_{i-1}-\mu_2\right)$);
\item[B.] one jet harder than $\mu_1$ (owing to $\left(1-D(d_i)\right)$);
\item[C.] one jet softer than $\mu_2$ (owing to $D(d_{i+1})$).
\end{itemize}
Note that one jet is degenerate (i.e., has zero four momentum)
in the case of $\clS$ events, and that condition C. does not apply 
to $\clH$ events when $i=N$.
We stress again that no shower is involved yet, and the jets are thus 
defined at the matrix-element level. Furthermore, according to the definition
of $d_i$, it would be more appropriate to talk about a hardness scale
at which one resolves $i$ or $(i-1)$ jets; in practice, a less precise
language is acceptable, since no confusion is possible here.

It is easier to start analysing the implications of the previous formulae by 
considering first a sharp $D$ function, eq.~(\ref{Dsharp}).
In such a case, items A.--C. imply that there are $i$ jets harder than 
$\mu_Q$, and one jet softer than $\mu_Q$. The idea is that the matrix
element description of the former $i$ jets is adequate (and of NLO accuracy), 
while the latter one jet will be heavily affected by MC showers. Consistently 
with this picture, one will not want the MC to generate emissions harder 
than $\mu_Q$. When a generic and smooth $D$ function is considered instead
(eq.~(\ref{Ddef})), the interpretation is basically the same, only 
slightly more involved. In particular, the description of the $(i-1)$ 
jets harder than $\mu_2$ is still a fully matrix-element one. In the 
intermediate-hardness region $(\mu_1,\mu_2)$, item B. implies the presence 
of an extra jet, with ``probability'' given by $1-D$. This is more
correctly interpreted as our confidence in the correctness of a matrix-element 
description for such a jet, which is maximal (i.e., equal to one) for a 
hardness equal to $\mu_2$, and minimal (i.e., equal to zero) for a hardness
equal to $\mu_1$. This damping factor $1-D$ is arbitrary, and must be
compensated. This will happen thanks the contribution to the $i$-jet
cross section due to the $(i-1)$-parton sample, the extra jet being 
generated in the intermediate region $(\mu_1,\mu_2)$ by means of MC radiation.
This compensation is consistent with the idea that in the 
intermediate-hardness region both the matrix element and the MC 
description are on equal footing. Finally, the case of item C. is 
analogous to that of item B., but specular. In particular, the 
matrix-element description of the extra jet relevant here is turned off 
with probability $D$, so that an MC description is generally dominant.

The latter point is not only motivated by physical arguments, but is 
actually necessary in the context of a well-behaved NLO computation.
In fact, while the last two factors on the r.h.s.~of eq.~(\ref{modi}) 
limit the hardness of $i$ jets from below, the factor $D(d_{i+1})$ limits
that of the $(i+1)^{th}$ jet from above. It is clear that for such
a jet (which generally corresponds to the softest parton in the event) 
there must not be a lower bound on hardness, because of infrared safety.
On the other hand, the bound from above prevents one from having
an $(i+1)$-jet configuration generated by the $i$-parton sample,
since this would effectively be an LO (rather than an NLO) prediction.
The exception is of course that of the largest parton multiplicity
available to the calculation, simply because one cannot do better
than LO there; this is the reason for the special case considered
in eq.~(\ref{modN}).

We can now check the consistency of eqs.~(\ref{modi}) 
and~(\ref{modN}) with rule {\em R.1}. As it can be seen from 
eqs.~(\ref{xsecSi})and~(\ref{xsecHi}), $T_i$ enters the $\clS$ events 
of the $i$-parton sample, and the $\clH$ events of the $(i-1)$-parton 
sample. Explicitly, one obtains:
\beqn
d\bar{\sigma}_{\clS,i}&=&\ldots+T_i\,
\left(1-D(d_i)\right)\,\stepf\left(d_{i-1}-\mu_2\right)\,,
\label{eq:temp1}
\\
d\bar{\sigma}_{\clH,i-1}&=&\ldots+T_i\,
D(d_{i})\,\left(1-D(d_{i-1})\right)\,\stepf\left(d_{i-2}-\mu_2\right)\,,
\label{eq:temp2}
\eeqn
where in eq.~(\ref{eq:temp1}) we have exploited the fact that, for
$\clS$ events in the $i$-parton sample (with $i<N$), $d_{i+1}=0$ and hence
$D(d_{i+1})=1$ (when $i=N$, this $D$ factor simply does not appear
in eq.~(\ref{modN})). By using
\beq
1=\stepf\left(d_{i-1}-\mu_2\right)+\stepf\left(\mu_2-d_{i-1}\right)
\eeq
and the properties of the $D$ function, eq.~(\ref{eq:temp2}) can be
rewritten as follows:
\beqn
d\bar{\sigma}_{\clH,i-1}&=&\ldots+T_i\,
D(d_{i})\,\stepf\left(d_{i-1}-\mu_2\right)
\nonumber\\*&+&T_i
D(d_{i})\,\left(1-D(d_{i-1})\right)\,\stepf\left(\mu_2-d_{i-1}\right)\,
\stepf\left(d_{i-2}-\mu_2\right)\,.
\label{eq:temp3}
\eeqn
Therefore:
\beqn
d\bar{\sigma}_{\clH,i-1}+d\bar{\sigma}_{\clS,i}&=&\ldots+T_i\,
\stepf\left(d_{i-1}-\mu_2\right)
\label{eq:temp4}
\\*&+&T_i
D(d_{i})\,\left(1-D(d_{i-1})\right)\,\stepf\left(\mu_2-d_{i-1}\right)\,
\stepf\left(d_{i-2}-\mu_2\right)\,.
\nonumber
\eeqn
This result indeed obeys rule {\em R.1}. In fact, the first term
on the r.h.s.~of eq.~(\ref{eq:temp4}) is a contribution to the hard 
region for $(i-1)$ jets; the hardness of the remaining one jet can
be either small (thus playing the role of an NLO correction to
an $(i-1)$-jet cross section), or large (thus being a Born contribution
to an $i$-jet cross section). In both cases, the use of a matrix
element description is fully justified, and $T_i$ appears with its 
proper weight (which implies that there is no double counting of 
matrix-element origin in the combination of $d\bar{\sigma}_{\clH,i-1}$ 
and $d\bar{\sigma}_{\clS,i}$).
On the other hand, in the second term on the r.h.s.~of eq.~(\ref{eq:temp4}) 
$T_i$ is multiplied by a non-trivial weight factor. However, that term
corresponds to having only $(i-2)$ jets in the hard region, while one
of them is forced to be in the intermediate region (owing to
\mbox{$(1-D(d_{i-1}))\stepf(\mu_2-d_{i-1})$}), and another one to be either
in the intermediate or in the soft region (owing to $D(d_{i})$).
As was discussed before, in this situation one should not expect the 
matrix elements to give the only correct description, and it therefore 
appears desirable that $T_i$ be multiplied by a number smaller than one.

Ultimately, effects such as that giving rise to the second term on the 
r.h.s.~of eq.~(\ref{eq:temp4}) can be ascribed to the systematics
of the merging scheme. This can be checked e.g.~by changing the values
of $\mu_1$ and $\mu_2$, and the functional form of $D$ in the intermediate
region. For example, one can see immediately that the sharp $D$ form of
eq.~(\ref{Dsharp}) simply sets the term above identically equal to zero
(as it must happen, since with eq.~(\ref{Dsharp}) there is no intermediate
region). It has to be stressed that, in the case of a smooth $D$, 
eq.~(\ref{eq:temp4}) will only give an upper bound to the merging
systematics. In fact, we expect a compensating effect, mainly due
to the $\clS$ events of the $(i-1)$-parton sample, giving rise through
showers to at least two extra jets, one in the intermediate region,
and one in the soft region. Finally, the particular source of systematics
we are discussing here is essentially due to the fact that two jets can
simultaneously be present in the intermediate region. In order to
avoid this, one can think to variants of the prescription given
in eq.~(\ref{modi}) such as:
\beqn
D(d_{i+1})\,\left(1-D(d_i)\right)&\longrightarrow&
\stepf\left(\mu_1-d_{i+1}\right)\,\left(1-D(d_i)\right)
\nonumber\\*&+&
D(d_{i+1})\,\stepf\left(d_{i}-\mu_2\right)\,.
\eeqn
This option and its possible variants will not be 
considered in this paper.

What was done so far brings out the physics contents of eqs.~(\ref{modi}) 
and~(\ref{modN}). As they stand, however, those equations are ambiguous,
since the two components of $d\sigma_{i}$, namely $d\sigma_{\clS,i}$
and $d\sigma_{\clH,i}$, are associated with different kinematics
configurations ($2\to S+i$ and $2\to S+i+1$ respectively), and
one needs to specify which of these is used in the computation of 
the $d_i$'s. Let us denote by
\beq
\kinSi\,,\;\;\;\;\;\;\;\;\kinHi
\eeq
these kinematic configurations (the notation reminds one that they
are associated with the $\clS$ and $\clH$ events of the $i$-parton sample).
It is fairly obvious that $\kinSi$ must be used in the $d\sigma_{\clS,i}$
contribution to eq.~(\ref{modi}), while $\kinHi$ must be used in the 
$d\sigma_{\clH,i}$ bit. In fact, by doing so the manipulations carried 
out in eqs.~(\ref{eq:temp1})--(\ref{eq:temp4}) are still correct,
since one can always identify $\Xi_{\clH,i-1}$ with $\kinSi$.
As an explicit example of the cross sections that one obtains with
these kinematic assignments, we write here the results relevant to
the simplest case of the merging of the two lowest parton multiplicities
($N=1$):
\beqn
d\bar{\sigma}_{\clS,0}&=&T_0+V_0-T_0\Kern+T_0\KernMC\,,
\label{H0S}
\\
d\bar{\sigma}_{\clH,0}&=&\Big[T_1-T_0\KernMC\Big] D(d_1(\kinHz))\,,
\label{H0H}
\\
d\bar{\sigma}_{\clS,1}&=&\Big[T_1+V_1-T_1\Kern+T_1\KernMC\Big]
\left(1-D(d_1(\kinSo))\right)\,,
\label{H1S}
\\
d\bar{\sigma}_{\clH,1}&=&\Big[T_2-T_1\KernMC\Big]
\left(1-D(d_1(\kinHo))\right)\,.
\label{H1H}
\eeqn
Here, we have used eq.~(\ref{dmax}), $d_1(\kinSz)=0$, and the properties 
of the $D$ function in eq.~(\ref{Ddef}) (obviously, $\mu_2<d_0$).

While the above prescription removes the ambiguity in eqs.~(\ref{modi}) 
and~(\ref{modN}), and gives those equations an operative meaning, it leads
to cross sections affected by double counting {\em within} the $i$-parton
sample (i.e., even before merging different multiplicities), as one
can readily prove by using eqs.~(\ref{H0S})--(\ref{H1H}), and by 
proceeding e.g.~as is done in appendix B of ref.~\cite{Frixione:2002ik}. 
However, the correct formulae can easily be
obtained by means of a few simple modifications. We shall
illustrate them in the following, starting from the case $N=1$
in order to simplify the discussion, and moving next to the
fully general case.

\subsubsection{Merging $0$- and $1$-parton samples\label{sec:01merg}}
The correct, non-double-counting versions of eqs.~(\ref{H0S})--(\ref{H1H})
read as follows:
\beqn
d\bar{\sigma}_{\clS,0}&=&T_0+V_0-T_0\Kern+T_0\KernMC D(d_1(\kinHz))\,,
\label{H0Sc}
\\
d\bar{\sigma}_{\clH,0}&=&\Big[T_1-T_0\KernMC\Big] D(d_1(\kinHz))\,,
\label{H0Hc}
\\
d\bar{\sigma}_{\clS,1}&=&\Big[T_1+V_1-T_1\Kern+T_1\KernMC\Big]
\left(1-D(d_1(\kinSo))\right)\,,
\label{H1Sc}
\\
d\bar{\sigma}_{\clH,1}&=&T_2\left(1-D(d_1(\kinHo))\right)
-T_1\KernMC\left(1-D(d_1(\kinSo))\right)\,,
\label{H1Hc}
\eeqn
where changes have occurred in the definitions of $d\bar{\sigma}_{\clS,0}$
and of $d\bar{\sigma}_{\clH,1}$. The factor $D(d_1(\kinHz))$ in
eq.~(\ref{H0Hc}) limits the hardness of the final-state parton as prescribed
by the function $D$ (with a sharp $D$, eq.~(\ref{Dsharp}), the parton 
relative transverse momentum will obey $\pt<\mu_Q$, see eq.~(\ref{d1expl})).
While this condition is imposed at the matrix-element level, one should
keep in mind that the MC subtraction term, \mbox{$T_0\KernMC$}, appears
in eq.~(\ref{H0Hc}) in order to prevent double counting at
the NLO. Hence, consistency demands that its modification due to the
$D$-dependent prefactor be accompanied by a prescription for the
shower scale that limits emissions within the same hardness range.
Given the NLO accuracy of the MC subtraction terms, this can be
conveniently done by means of the LH-interface~\cite{Boos:2001cv}
parameter {\sc SCALUP}, which will be chosen event-by-event in a
random manner (so as to avoid biases) according to the inverse of
the function $D$ (for example, with a sharp $D$ function and {\sc SCALUP}
having the meaning of relative $\pt$, such a scale will be always
set equal to $\mu_Q$). The modifications of the shower scale and
of the MC subtraction term in $\clH$ events imply that the MC 
subtraction term must be modified in $\clS$ events as well; this
is the reason for the factor $D(d_1(\kinHz))$ in eq.~(\ref{H0Sc}).
As far as the $1$-parton sample is concerned (eqs.~(\ref{H1Sc})
and~(\ref{H1Hc})), the factors \mbox{$1-D$} limit from below
what is essentially the relative $\pt$ of the Born-level parton -- 
in the case of a sharp $D$ function, this is therefore equivalent 
to imposing hard Born-level cuts. Thus, it should be intuitively clear,
and could be formally proven using again the techniques of appendix B of 
ref.~\cite{Frixione:2002ik}, that the proper \mbox{$1-D$} prefactor
for the $\clH$-event MC subtraction term is that in eq.~(\ref{H1Hc}),
and not that in eq.~(\ref{H1H}).

\subsubsection{The general case\label{sec:allmerg}}
What is done in sect.~\ref{sec:01merg} is sufficient to sketch 
the procedure one has to follow in order to convert the naive prescriptions
of eqs.~(\ref{modi}) and~(\ref{modN}) into correct expressions for
MC@NLO short-distance cross sections. We obtain:
\beqn
d\bar{\sigma}_{\clS,0}&=&T_0+V_0-T_0\Kern+T_0\KernMC D(d_1(\kinHz))\,,
\label{H0Sfin}
\\
d\bar{\sigma}_{\clH,0}&=&\Big[T_1-T_0\KernMC\Big]\,D(d_1(\kinHz))\,,
\label{H0Hfin}
\\
d\bar{\sigma}_{\clS,i}&=&\Big[T_i+V_i-T_i\Kern+T_i\KernMC 
D(d_{i+1}(\kinHi))\Big]
\label{HiSfin}
\\*&&\times
\left(1-D(d_i(\kinSi))\right)\,
\stepf\left(d_{i-1}(\kinSi)-\mu_2\right)\,,
\nonumber
\\
d\bar{\sigma}_{\clH,i}&=&\Big[T_{i+1}\left(1-D(d_i(\kinHi))\right)
\stepf\left(d_{i-1}(\kinHi)-\mu_2\right)
\label{HiHfin}
\\*&&-
T_i\KernMC \left(1-D(d_i(\kinSi))\right)
\stepf\left(d_{i-1}(\kinSi)-\mu_2\right)\Big]\,D(d_{i+1}(\kinHi))\,,
\nonumber
\\
d\bar{\sigma}_{\clS,N}&=&\Big[T_N+V_N-T_N\Kern+T_N\KernMC\Big]
\label{HNSfin}
\\*&&\times
\left(1-D(d_N(\kinSN))\right)\stepf\left(d_{N-1}(\kinSN)-\mu_2\right)\,,
\nonumber
\\
d\bar{\sigma}_{\clH,N}&=&T_{N+1}\left(1-D(d_N(\kinHN))\right)
\stepf\left(d_{N-1}(\kinHN)-\mu_2\right)
\label{HNHfin}
\\*&-&
T_N\KernMC\left(1-D(d_N(\kinSN))\right)
\stepf\left(d_{N-1}(\kinSN)-\mu_2\right)\,.
\nonumber
\eeqn
We stress that eqs.~(\ref{H0Sfin}) and~(\ref{H0Hfin}) are redundant,
since they are just eqs.~(\ref{HiSfin}) and~(\ref{HiHfin}) respectively,
with $i=0$; we report them explicitly only for the sake of clarity.
Furthermore, eqs.~(\ref{HNSfin}) and~(\ref{HNHfin}) are identical
to eqs.~(\ref{HiSfin}) and~(\ref{HiHfin}) respectively, with $i=N$,
except for the fact that the hardness of the $(N+1)^{th}$ parton
is not bounded from above. This is correct, since there is no higher 
multiplicity whose Born-level kinematics could compensate for the lack 
of hard emissions in the $N$-parton sample.

As was already mentioned in sect.~\ref{sec:01merg}, the equations
above entail specific choices of shower scales. We have chosen:
\beqn
&&\min(d_{i}(\kinSi),\ptD)\,,
\label{shscS}
\\
&&\max(d_i(\kinHi)-d_{i+1}(\kinHi)\,,
       d_i(\kinHi)-\ptD),
\label{shscH}
\eeqn
for $\clS$ and $\clH$ $i$-sample events respectively\footnote{The settings
relevant to the $Q^2$-ordered \PYs~\cite{Sjostrand:2006za} may be 
different. We have postponed the study of this issue to a forthcoming work.}. 
We have introduced the quantity
\beq
\ptD=D^{-1}(r)\,,
\eeq
with $r$ a random number, to be generated event by event; one has
$\ptD=\mu_Q$ with a sharp $D$ function. In the case of
the $N$-parton sample, the $\ptD$-dependent part in eqs.~(\ref{shscS})
and~(\ref{shscH}) must be dropped. The $\ptD$ dependence in eq.~(\ref{shscS})
has already been discussed in sect.~\ref{sec:01merg}. That on 
$d_{i}(\kinSi)$ is intuitively clear: since such a quantity is directly
related to the hardness of the softest Born-level jet, one does not want the 
shower to generate jets harder than that. This argument can be substantiated 
analytically in the context of the toy model of ref.~\cite{Frixione:2002ik}, 
where one can actually show that any monotonically-growing function of
$d_{i}$, subject to the conditions $1/2d_{i}\le f(d_{i})\le d_{i}$,
will do. For un-merged MC@NLO samples, the toy model does not give
any prescription for the shower-scale choice of $\clH$ events; when
merging, however, one obtains the constraint that the scale be either 
a monotonically-decreasing function of the hardness of the real emission,
or a constant, which motivates eq.~(\ref{shscH}) (but does not 
determine it uniquely). Incidentally, such a choice is 
just the generalisation of what was done in 
ref.~\cite{Frixione:2002ik}, where it was motivated simply by the
argument that the shower scale has to be related to the hardness ``left''
in the system after the first emission.

\subsubsection{Sudakov reweighting\label{sec:Sud}}
The formulae presented in sect.~\ref{sec:allmerg} achieve
the strategy described in items 1. and 2. of sect.~\ref{sec:descr}.
As far as item 3. there is concerned, the basic idea has already
been discussed, which is essentially that of following the
CKKW prescription~\cite{Catani:2001cc} with a reweighting of
the short-distance cross sections by a combination of Sudakov
factors. We implement this by defining modified
MC@NLO cross sections as follows:
\beqn
d\hat{\sigma}_{\clS,i}&=&\Big[d\bar{\sigma}_{\clS,i}+
d\sigma_i^{(\Delta)}\Big]\,\Delta_i\!\left(\muSimn,\muSimx\right)\,,
\label{SudiS}
\\
d\hat{\sigma}_{\clH,i}&=&d\bar{\sigma}_{\clH,i}\,\,
\Delta_i\!\left(\muHimn,\muHimx\right)\,,
\label{SudiH}
\eeqn
with $d\bar{\sigma}_{\clS,i}$ and $d\bar{\sigma}_{\clH,i}$ given 
in eqs.~(\ref{HiSfin}) and~(\ref{HiHfin}) respectively (or
eqs.~(\ref{HNSfin}) and~(\ref{HNHfin}) when $i=N$). The
compensating factor alluded to in sect.~\ref{sec:descr},
necessary in order to avoid double counting in the presence 
of the Sudakovs, reads as follows:
\beq
d\sigma_i^{(\Delta)}=-T_i\left(1-D(d_i(\kinSi))\right)\,
\stepf\left(d_{i-1}(\kinSi)-\mu_2\right)
\Delta_i^{(1)}\!\left(\muSimn,\muSimx\right)\,,
\label{Delsig}
\eeq
where by $\Delta_i^{(1)}$ we have denoted the ${\cal O}(\as)$
term in the perturbative expansion of $\Delta_i$. We point 
out that the quantities $\Delta_i$ are products of ordinary
Sudakov factors, which we define following the CKKW prescription
of reconstructing the most probable shower history, according to the
jet-finding algorithm that determines the $d_i$. When doing that,
we use the $(S+i)$-body kinematic configurations when dealing with
$\clS$ events (eq.~(\ref{SudiS})), and the $(S+i+1)$-body kinematic 
configurations for $\clH$ events (eq.~(\ref{SudiH})), as one would
naively expect. However, in the latter case the softest of the $d$'s
is discarded. This is in keeping with what has been discussed at
the end of sect.~\ref{sec:descr}, that $\clH$ events have to be
treated on the same footing as $\clS$ ones as far as the multiplicity
relevant to the definition of Sudakovs is concerned. The scales 
entering the Sudakov factors in eqs.~(\ref{SudiS})--(\ref{Delsig})
are defined as follows:
\beqn
\muSimx&=&\max\Big\{\muME,d_1(\kinSi)\Big\}\,,
\label{Sudsc1}
\\
\muHimx&=&\max\Big\{\muME,d_1(\kinHi)\Big\}\,,
\label{Sudsc2}
\\
\muSimn&=&\min(d_{i}(\kinSi),\ptD)
\phantom{aaaaaaaaaa}i<N\,,
\label{Sudsc3}
\\
\muHimn&=&\min(d_{i}(\kinHi),\ptD)
\phantom{aaaaaaaaaa}i<N\,,
\label{Sudsc3b}
\\
\muSNmn&=&d_N(\kinSN)\,,
\label{Sudsc4}
\\
\muHNmn&=&d_N(\kinHN)\,.
\label{Sudsc5}
\eeqn
Here, $\muME$ is a hard scale, which can be generically associated 
with matrix-elements computations (e.g.~an NLO parton-level result
which corresponds to a given MC@NLO simulation); explicit examples 
will be given below. Note that, in the case of a sharp $D$ function,
the r.h.s.~of eqs.~(\ref{Sudsc3}) and~(\ref{Sudsc3b}) are equal 
to $\ptD=\mu_Q$.

In the CKKW procedure, a reweighting by $\as$ factors is also
performed. Although this would technically be possible in the 
context of MC@NLO, the complications it entails (owing to the
more involved dependence on the renormalization scale of NLO cross 
sections w.r.t.~that of LO ones) do not seem justified, in view of
the fact that in MC@NLO there is already an ${\cal O}(\as)$ cancellation
between matrix elements and Monte Carlo effects. On the other hand,
it is probably best to choose a renormalization scale whose definition 
exploits the CKKW-like considerations which lead to eqs.~(\ref{SudiS})
and~(\ref{SudiH}). We adopt therefore what is used in the
MINLO procedure~\cite{Hamilton:2012np}, which is tailored for
NLO computations, and in view of its connections with CKKW:
\beq
\bar{\mu}=\left(\muME^b\,\prod_{j=1}^i d_j\right)^{1/(i+b)}\,.
\label{mubar}
\eeq
We remind the reader that $b$ is the power of $\as$ that appears
in the Born contribution to the $0$-parton sample; furthermore,
in eq.~(\ref{mubar}) both $\muME$ and the $d_j$'s are meant to
be computed with the kinematics proper of either $\clS$ or $\clH$ events.
For consistency with ref.~\cite{Hamilton:2012np}, we also set the
factorization scale equal to $d_i$. We remark that both renormalization 
and factorization scales are set equal to $\muME$ for NLO mergings
that do not include the Sudakov reweighting discussed here, and
for un-merged MC@NLO predictions.

Reference~\cite{Hamilton:2012np} also suggests a simplification
in the implementations of eqs.~(\ref{SudiS}) and~(\ref{SudiH}),
which is useful because of a known issue with out-of-the-box
CKKW (see e.g.~ref.~\cite{Mrenna:2003if,Lavesson:2005xu}).
Namely, Sudakov reweighting leads to better results from the
numerical viewpoint if the Sudakov factors used in the short-distance
cross sections are equal to those that enter the Monte Carlo which is
matched to the matrix elements. However, analytical NLL Sudakov,
such as those considered in ref.~\cite{Catani:2001cc}, are appealing
precisely because, being MC-independent, they give one the possibility
of an error-free, easy, and universal implementation. The differences
induced by analytical or actual-MC Sudakovs will grow with the
distance between the largest and smallest scales entering them.
We can therefore envisage the following possibility: in
eqs.~(\ref{SudiS}) and~(\ref{SudiH}), when $i<N$ we use the identity:
\beq
\Delta_i\!\left(\muimn,\muimx\right)=
\left[\Delta_i\!\left(\ptD,d_{i}\right)\stepf\!\left(d_{i}-\ptD\right)
+\stepf\!\left(\ptD-d_{i}\right)\right]
\Delta_i\!\left(d_{i},\muimx\right)\,,
\label{CKKWMLM}
\eeq
having taken eqs.~(\ref{Sudsc3}) and~(\ref{Sudsc3b}) into account. 
The Sudakov factors outside the square brackets on the r.h.s.~of 
eq.~(\ref{CKKWMLM}) are then computed with the NLL analytical forms
(which, given the scales used, is what is 
done in ref.~\cite{Hamilton:2012np}). On
the other hand, the Sudakov factors inside the square brackets
are effectively computed using an MLM-type rejection procedure,
which is supposed to be a good approximation of the use of
actual-MC Sudakovs (in that it exploits information obtained
from the Monte Carlo). When such a procedure is applied at the LO, 
one matches partons with jets\footnote{We point out that the verb 
``to match'' used in context of the MLM procedure does not have the 
same meaning as in the rest of this paper.}. In order to extend
this idea to the NLO, we must make use of jets, reconstructed at the
level of the hard subprocess (which corresponds to the matrix-element
level), instead of partons; this preserves IR safety. We then require
these hard-subprocess jets to match parton-level jets after shower,
in essentially the same way as in the original MLM 
procedure~\cite{Alwall:2007fs}. The only difference is that, while
MLM uses a cone jet-finding algorithm, we adopt
a $\kt$ one~\cite{Catani:1993hr}, as is also done in the MLM 
LO-implementation in \madgraph~\cite{Alwall:2008qv}, and
consistently with the construction of the $d_i$'s. More specifically,
if jets are defined by the $\kt$ algorithm with a given radius $R_0$,
we tag a jet after shower as matched with a hard-subprocess-level
one if the two are less than \mbox{$1.5 R_0$} apart in the $\eta-\phi$ 
plane. All the results we shall present in sect.~\ref{sec:res}
have been obtained with $R_0=1$ (large radii have to be preferred,
but there does not appear to be a strict constraint on what to choose.
For example, we have verified that with $R_0=0.8$ our results are
unchanged). Finally, after having obtained a set 
of shower-level matched jets, we impose that in the $i$-parton
sample there be exactly $i$ jets with hardness larger than $\ptD$.
In practice, as a further measure to ensure that there be no
biases after applying the MLM rejection, we generate hard events
by relaxing the conditions enforced by the $1-D$ factors in
eqs.~(\ref{HiSfin})--(\ref{HNHfin}). For $i<N$, this implies
a loss of efficiency, while for $i=N$ it requires that an MLM condition 
be imposed as well, with hardness $d_N$. In particular, one demands 
that, if there are $N$ or $N+1$ jets at the hard-subprocess level,
then these should match the $N$ or $N+1$ hardest ones after shower.
We finally point out that the MLM conditions above are identical
in the cases of $\clS$ and $\clH$ events, again consistently with 
the treatment of these two event classes in the CKKW-type procedure
set up here.

\section{Results\label{sec:res}}
In this section, we present results obtained with the NLO-merging
procedure described in sect.~\ref{sec:merg} which are relevant to the
production of a Standard Model Higgs (denoted by $H$ henceforth),
of an $e^+\nu_e$ pair, and of a $t\tb$ pair, at the 8~TeV LHC.
We concentrate in particular on $H$ production, which is an ideal 
test case since it features a very large amount of radiation, in this 
way helping expose any problems in the matching and merging techniques. 
Furthermore, the matrix elements relevant to this process are relatively 
simple, and thus fast to evaluate. We consider the cases of sharp
and smooth $D$ functions, and of Sudakov reweighting; we study
the merging with $N=1$ and with $N=2$. As far as $e^+\nu_e$ and
$t\tb$ pair production are concerned, we limit ourselves to presenting
a few key observables with $N=1$ and Sudakov reweighting, which is
sufficient to show the generality and flexibility of the procedure.
The latter is also guaranteed by its implementation in the automated
\amcatnlo\ framework, which has been used to obtain all the results 
shown below.

The relevant information are the following:
\begin{enumerate}
\item[\ea.] The merged result.
\item[\eb.] The $i$-parton-sample results, $0\le i\le N$.
\item[\ec.] The un-merged (``standalone'') MC@NLO results for the 
 corresponding multiplicities.
\item[\ed.] The merged result with different matching scales
and/or merging conditions.
\item[\ee.] The merged result with different $N$.
\end{enumerate}
For each observable, we present all the results above in one
plot, with the following layout. The main frame displays \ea,
superimposed to \eb, thus allowing one to check the origin of the features
of the merged results, and to see the interplay among the various
$i$-parton-sample contributions. An upper inset shows the ratios $(c/a)$;
in this way, it is easy to assess how much the merged results differ from 
the standalone MC@NLO ones, in both shape and normalization. 
It should be kept in mind that the latter, if obtained with
an underlying $(S+J)$-parton description, are physical only
for observables that feature at least $J$ hard jets (be them 
obtained by explicit cuts, or by effective ones enforced by other 
constraints, such as reconstructing the $\pt$ of recoiling objects).
A lower inset shows the ratios $(d/a)$ and/or $(e/a)$ -- the idea
here is that of assessing the merging systematics, and of studying 
the effect of increasing the largest multiplicity that enters the 
merging procedure. Finally, in the cases of $H$ and $t\tb$ 
production we have also compared our results with those of 
\alpgen~\cite{Mangano:2002ea} (including parton showers 
and MLM merging). The latter have been renormalized by a 
process-dependent overall factor, at the sole purpose of 
rendering them more visible in the figures, where they typically appear
as ratios in the lower insets.

All the Monte Carlo simulations have been performed with
\HWs~\cite{Corcella:2000bw}; however, the merging procedure has been
implemented so as no or minor changes are foreseen in the cases of
\HWpp~\cite{Bahr:2008pv} and \PYe~\cite{Sjostrand:2007gs}
(while the case of $Q^2$-ordered \PYs~\cite{Sjostrand:2006za}
may require further consideration), which are currently available 
and under testing respectively in \amcatnlo. Underlying events
have not been generated. Jets are reconstructed (by clustering
all final-state stable hadrons) with 
the $\kt$~\cite{Catani:1993hr}, anti-$k_T$~\cite{Cacciari:2008gp},
and Cambridge/Aachen~\cite{Wobisch:1998wt,Dokshitzer:1997in}
algorithms as implemented in \fastjet~\cite{Cacciari:2005hq}, 
for several different jet radii -- our default choice, to be shown in 
the plots, is the $\kt$ algorithm with $R=0.6$. We have used the PDF set
MSTWnlo200868cl~\cite{Martin:2009iq}.

\subsection{Standard Model Higgs production\label{sec:resH}}
Our runs have been performed with $m_H=125$~GeV; we have set 
$\muME$ equal to the Higgs transverse mass. We have also considered
$\muME=H_{\sss T}/2$, and found the same patterns as with our default
choice; hence, the corresponding results will not be presented here.
The one-loop matrix elements relevant to the $1$- and $2$-parton
samples have been taken from the MCFM code~\cite{Campbell:2010cz}
as implemented in ref.~\cite{Campbell:2012am}.
We have studied the following four merging scenarios:
$N\!=\!1$, sharp-$D$, non-Sudakov-reweighted; 
$N\!=\!1$, smooth-$D$, non-Sudakov-reweighted; 
$N\!=\!1$, sharp-$D$, Sudakov-reweighted; and
$N\!=\!2$, sharp-$D$, Sudakov-reweighted.

\begin{figure}[htb!]
  \begin{center}
        \epsfig{file=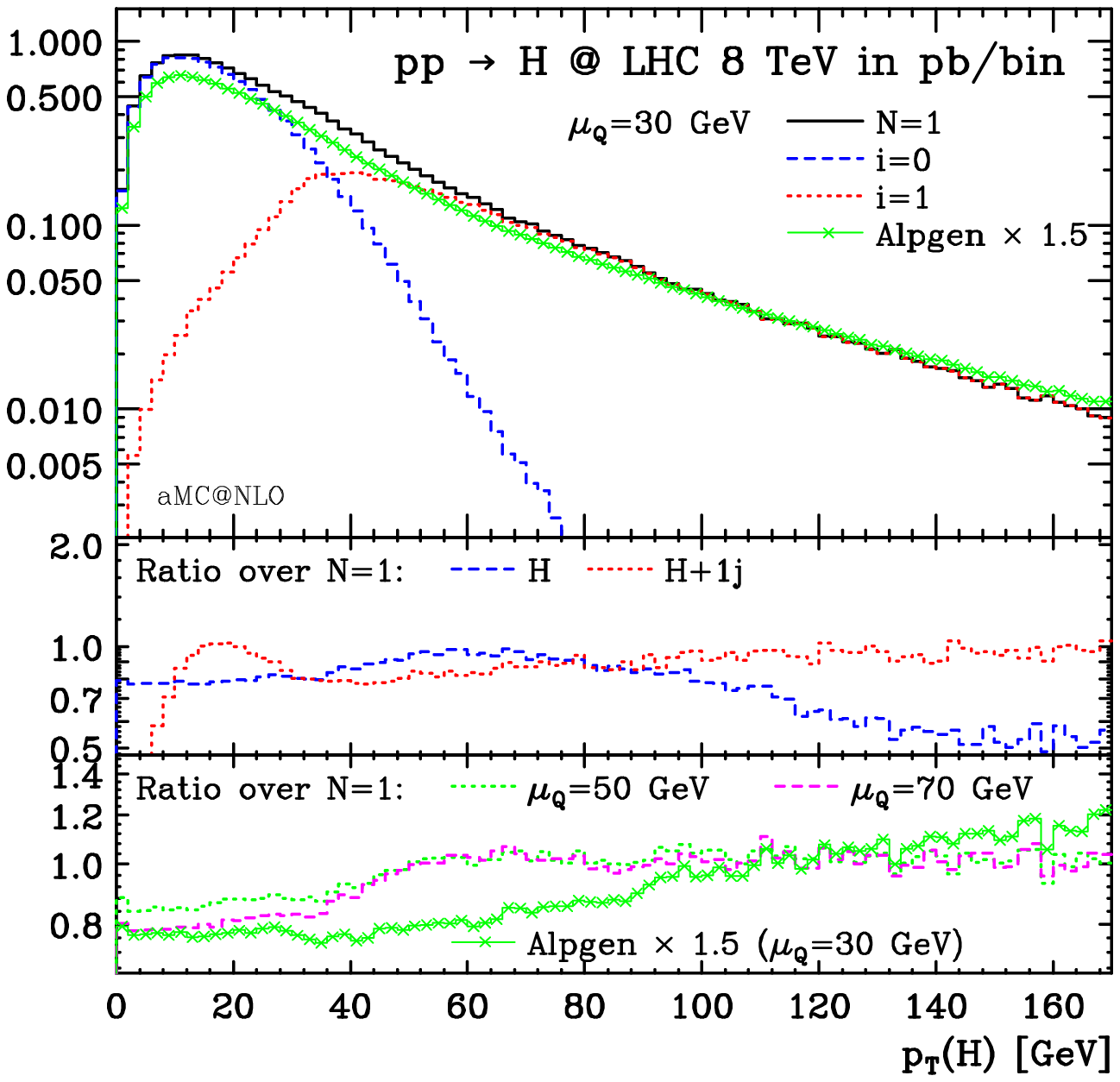, width=0.49\textwidth}
        \epsfig{file=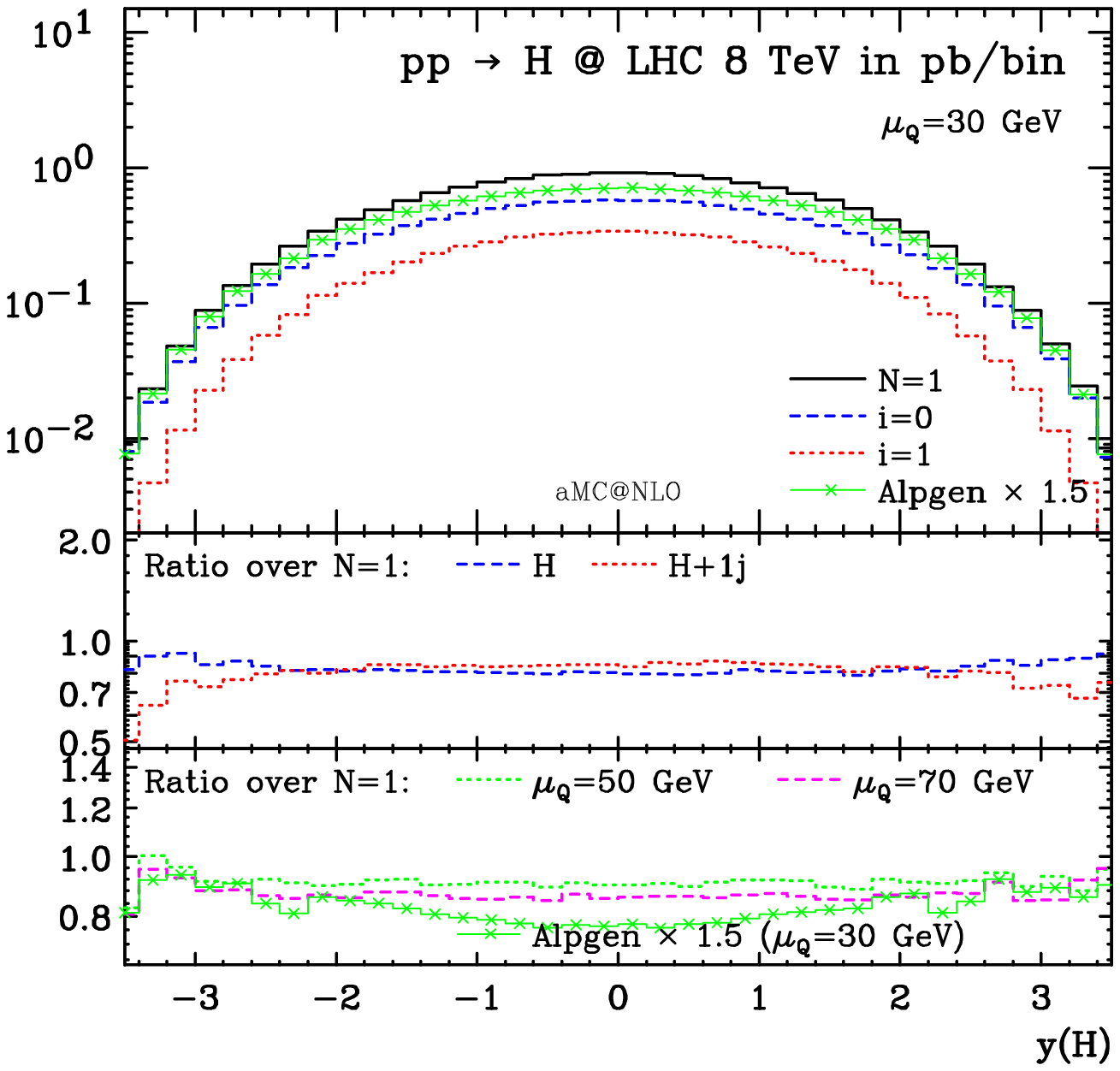, width=0.49\textwidth}
  \end{center}
  \begin{center}
        \epsfig{file=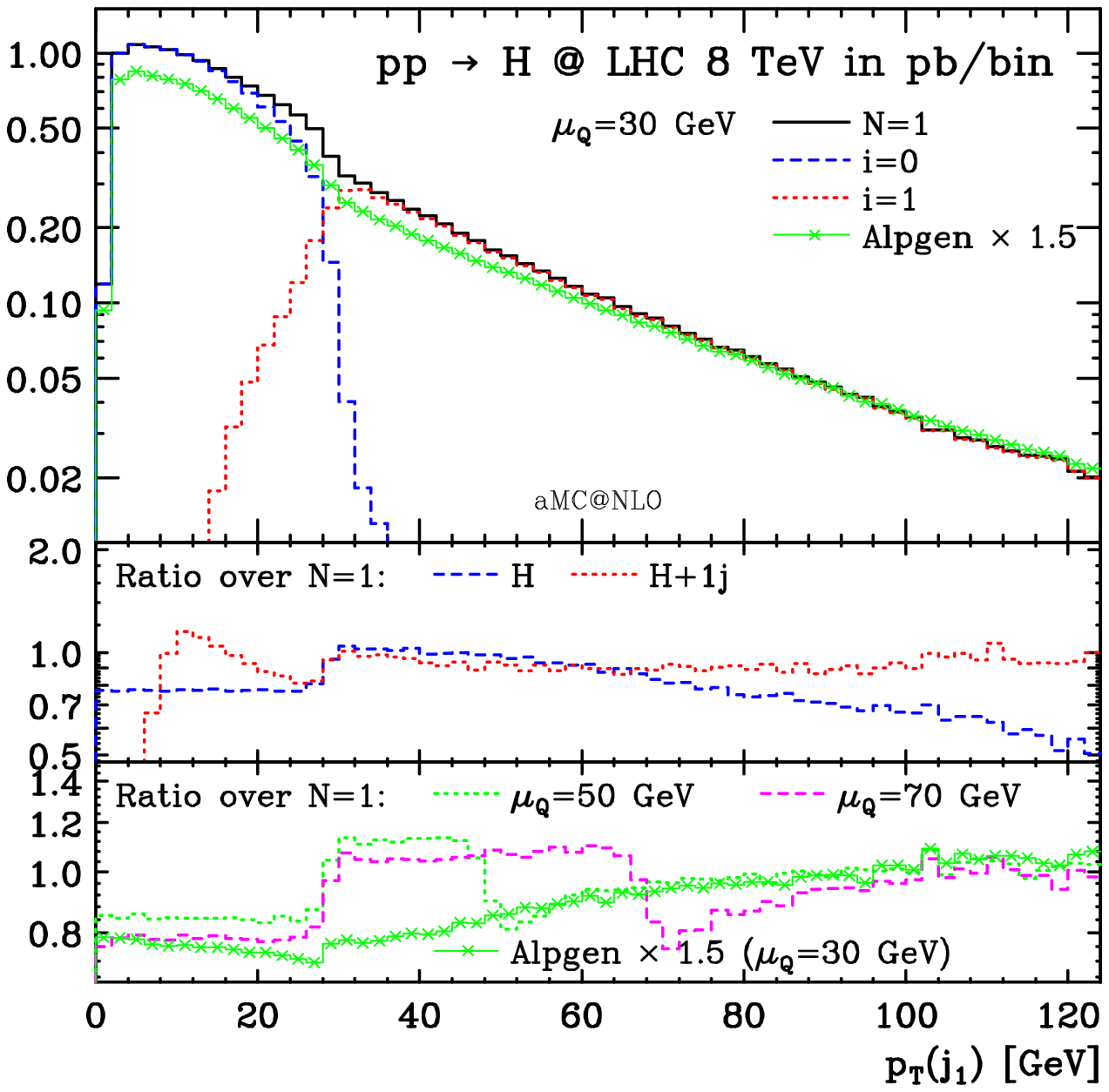, width=0.49\textwidth}
        \epsfig{file=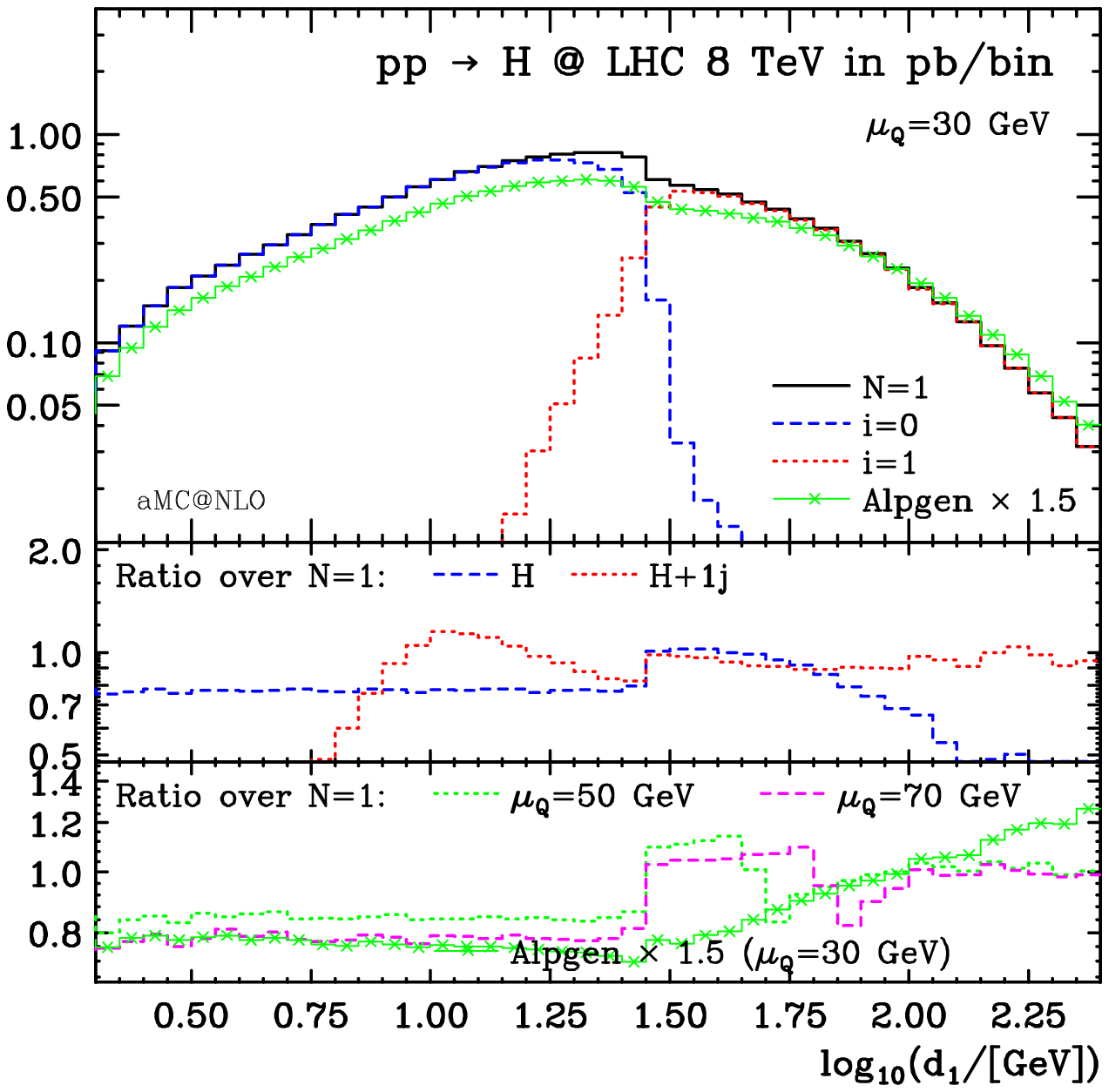, width=0.49\textwidth}
  \end{center}
  \vspace{-20pt}
  \caption{Higgs production, with $N=1$, sharp $D$ function, and without
    Sudakov reweighting. Higgs $\pt$ (upper left), Higgs rapidity
    (upper right), hardest-jet $\pt$ (lower left), and $d_1$ (lower right) are
    shown.}
\label{fig:1}
\end{figure}

We start by discussing the former case, for the three choices 
$\mu_Q=30$, $50$, and $70$~GeV.
Sample observables are shown in fig.~\ref{fig:1}; the merged
results in the main frame correspond to $\mu_Q=30$~GeV, in order
to facilitate the direct comparison with \alpgen, where the matching
scale has also been set equal to 30~GeV, which is fairly typical for
this process. The two observables displayed on the top panels of
fig.~\ref{fig:1} are representative of the behaviour of all observables
which are not directly related to jet transverse momenta: namely, the $0$- 
and $1$-parton samples merge smoothly. For the Higgs transverse momentum,
this smoothness, and the agreement with the standalone $H+1j$ prediction
at large $\pt(H)$ (which results naturally from the merging procedure), 
gives an effective constraint on the total rate, which can only act as
a normalization effect at small $\pt(H)$, where differences can be
seen w.r.t~the standalone $H+0j$ prediction (while the shapes are
identical, as they should be by construction). We note that normalization
effects have to be expected in this case, where the fact that the 
shower may cause ``leaks'' into larger exclusive multiplicities
w.r.t.~those of the underlying parton cross sections is not compensated
by any suppression (e.g.~by Sudakov reweighting). In spite of this,
and of the fairly severe conditions posed by a sharp $D$ function,
it is reassuring that these effects are smaller than 20\%, as shown 
by the two lower insets of the upper-left panel. From the comparison 
between those two insets, one can also see that, by setting $\mu_Q=50$~GeV, 
the merged result agrees almost perfectly (in shape and normalization) 
with the standalone $H+0j$ and $H+1j$ predictions (where relevant,
i.e. at small and large $\pt$'s respectively). Although this may suggest
a way towards the definition of an optimal matching scale, it is 
important to keep in mind that the capability of changing the merging
conditions is essential in order to assess the systematics that affects
the procedure. The dotted-plus-symbols green histograms in fig.~\ref{fig:1},
also presented in terms of ratios over the merged MC@NLO results
in the lower insets, are the \alpgen\ predictions; for consistency
with the $N=1$ case considered here, the $H+0$, $H+1$, and $H+2$
parton samples in \alpgen\ have been generated, and combined according
to the MLM prescription. Although in reasonable agreement, we see that the
$\pt(H)$ shape as given by \alpgen\ is harder than that of MC@NLO; we point 
out, however, that to some extent this difference might be due to the fact
that we have run \alpgen\ with its default scales, and with 
the LO version of the NLO PDFs we have used for MC@NLO.

In the bottom panels of fig.~\ref{fig:1} we present two observables
directly related to jet hardness, namely the $\pt$ of the leading jet,
and $d_1$. They display similar features, the most striking of which
is a kink at $\pt(j_1)\simeq d_1\simeq\mu_Q$. It is worth stressing
that the same kink appears in the \alpgen\ results. This emphasises
the fact that such a kink is a more general feature than being simply
an artifact of the merging prescriptions adopted in these plots 
(which are obviously very different from each other\footnote{Just as 
obviously, it is not implied that kinks will appear regardless 
of the merging technique employed; we shall show later how to get rid 
of them within the NLO approach proposed here. However, they are a 
persistent characteristic -- see e.g.~ref.~\cite{Mrenna:2003if}
for examples relevant to CKKW (although for processes different from
Higgs hadroproduction), and ref.~\cite{Hamilton:2009ne} for a possible
amendment, shown there to work in $e^+e^-$ collisions.}), since it is
ultimately caused by a significant mismatch 
between the Monte Carlo and matrix-element
descriptions in the region chosen for merging. As far as we know, the 
problem posed by this kink is very often ignored: in fact, if one
is interested in $1$-jet exclusive observables, a lower bound $\ptcut$
on the jet transverse momentum is imposed, and the issue of the
kink is simply bypassed, in the context of a merging procedure,
by choosing $\mu_Q<\ptcut$. With this, one is essentially assuming 
that the proper description for such a jet is a matrix-element one.
However, this is not necessarily the case; for example, a $40$-GeV-$\pt$
jet is presumably well described by matrix elements if produced
in association with a $80$~GeV Higgs, but likely much better modeled
by parton showers when the mass of the Higgs is $600$~GeV. Furthermore,
the mass scales involved in the problem are not the only relevant
factors: for example, in a $gg$-dominated process such as Higgs 
production, Monte Carlo effects will extend farther in $\pt$
in comparison with a process characterized by exactly the same
mass scales, but $q\bar{q}$-initiated (e.g., $Z^\prime$ production
with $m_{Z^\prime}=m_H$).
The bottom line is that, even for $J$-jet exclusive observables,
the choice $\mu_Q<\ptcut$ can be misleading, if anything because
it prevents one from assessing theoretical uncertainties in a complete
manner. Merging procedures do offer a systematic way of addressing this 
problem, provided that $\ptcut$ is not regarded as a natural upper bound
for the matching scale; jet transverse momenta (or related quantities)
must be studied in a range that includes the matching scale.
We conclude this discussion by mentioning that the kink that appears
in the bottom panels of fig.~\ref{fig:1} is not peculiar of the
$\kt$ algorithm with $R=0.6$ -- we have found the same feature
for all the jet algorithms we have considered.

\begin{figure}[htb!]
  \begin{center}
        \epsfig{file=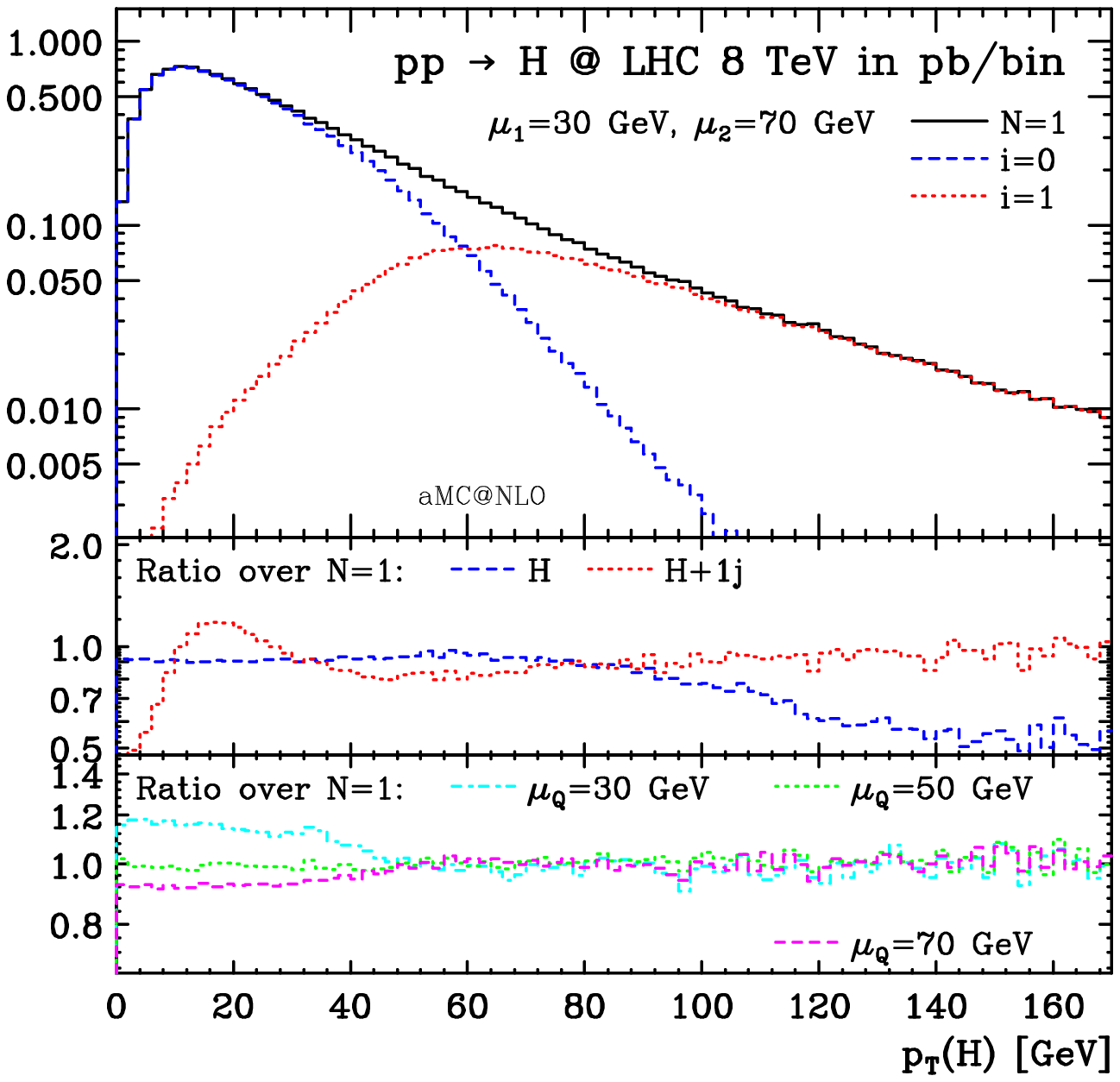, width=0.49\textwidth}
        \epsfig{file=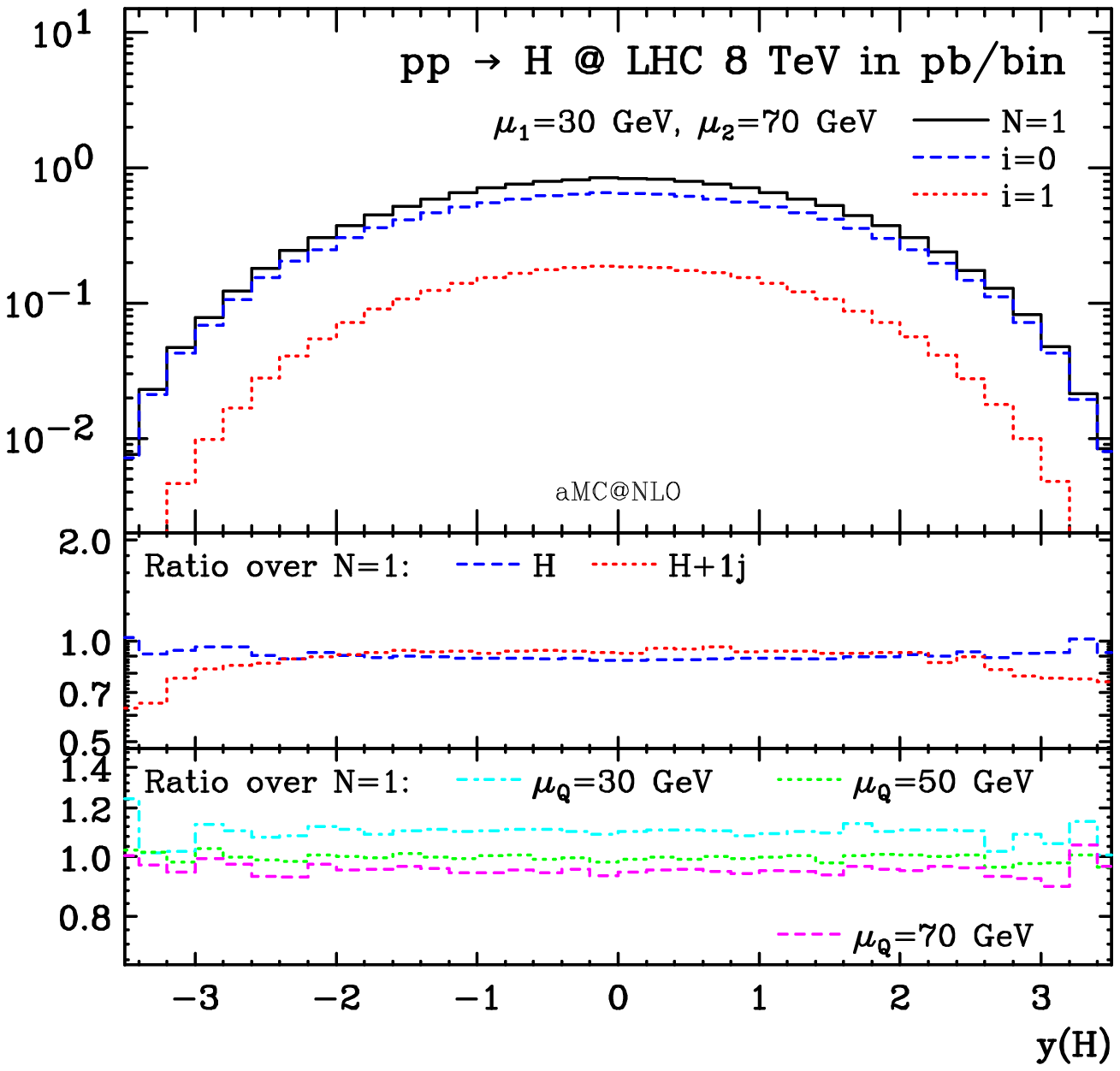, width=0.49\textwidth}
  \end{center}
  \begin{center}
        \epsfig{file=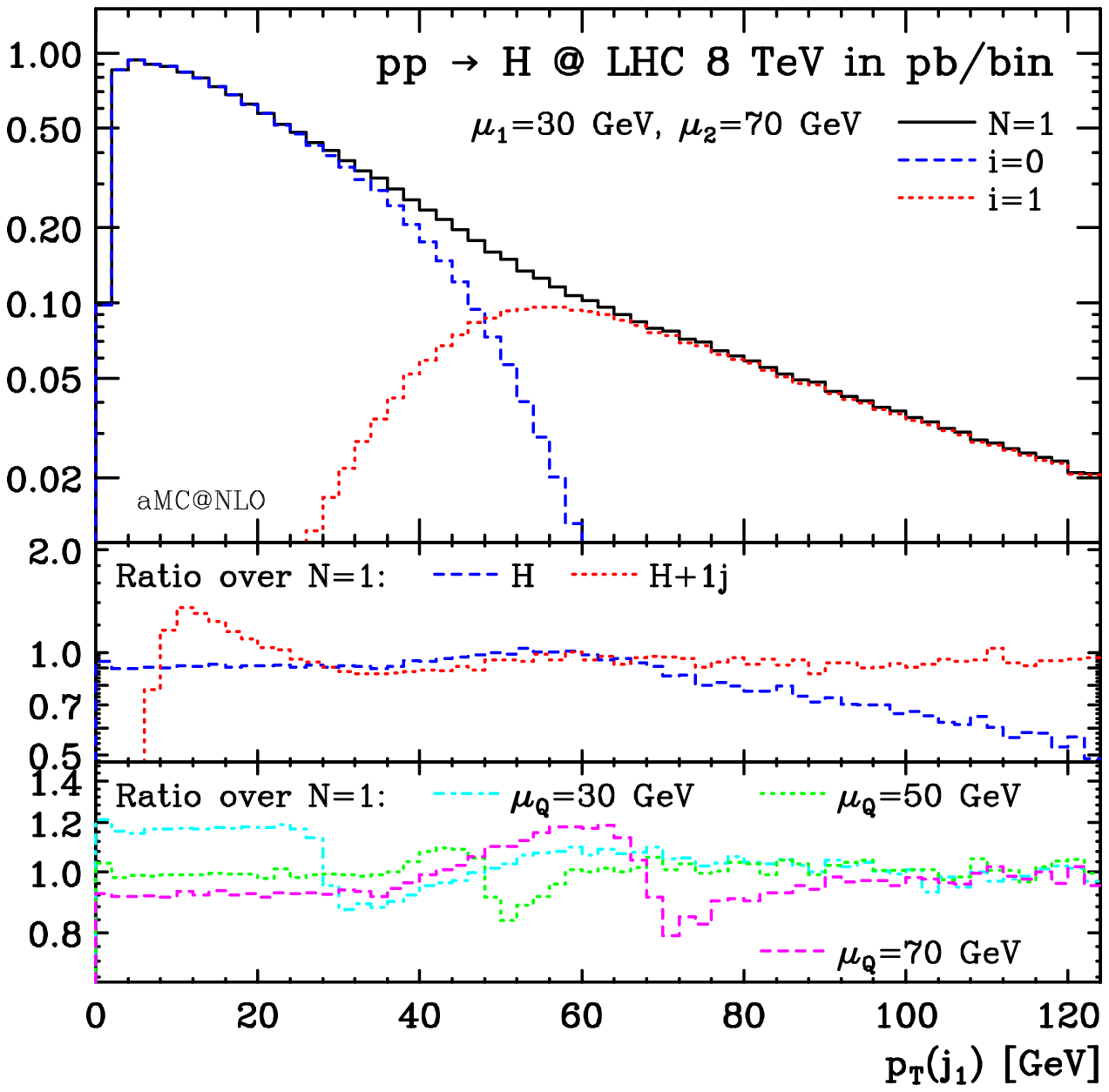, width=0.49\textwidth}
        \epsfig{file=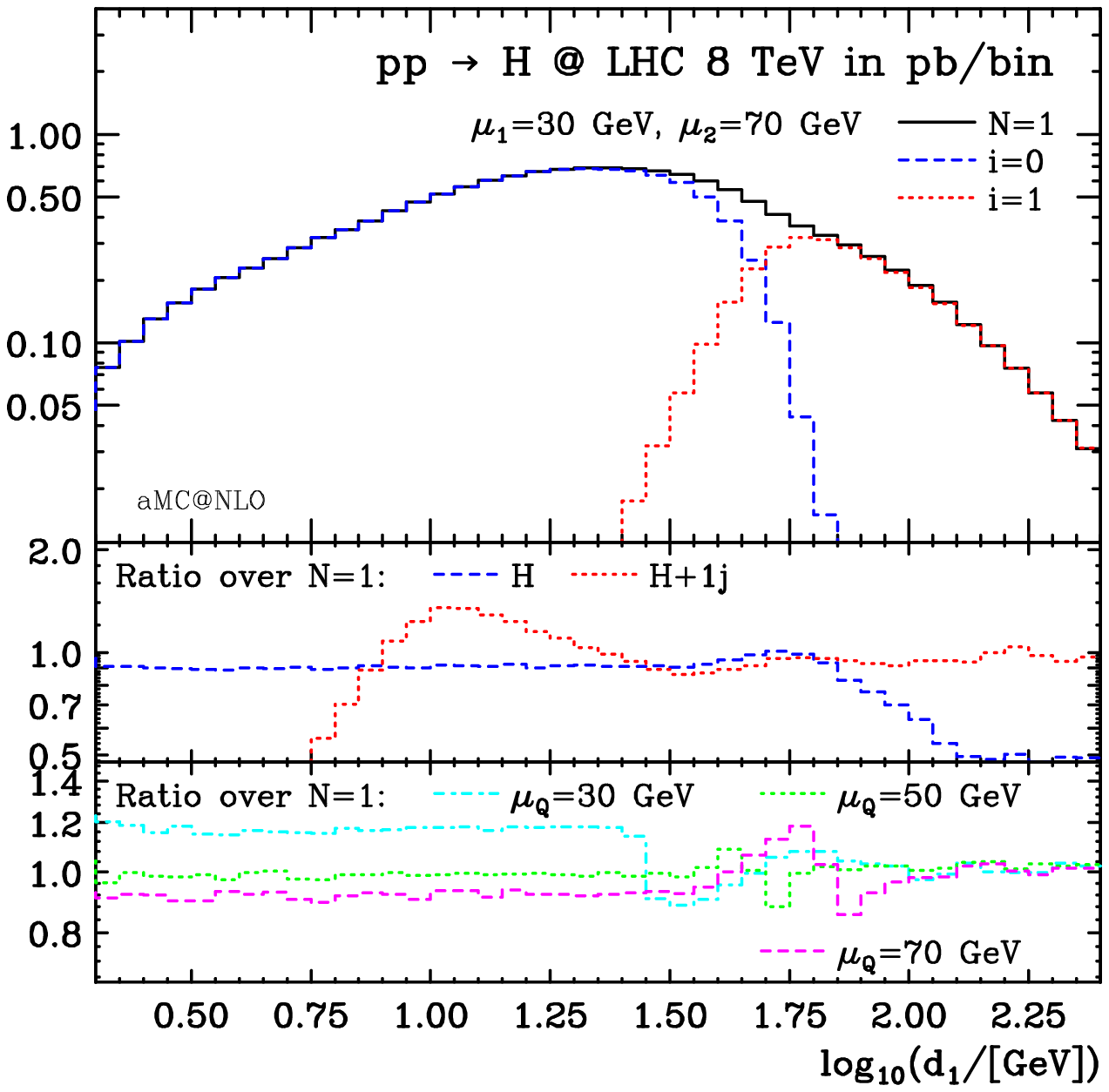, width=0.49\textwidth}
  \end{center}
  \vspace{-20pt}
  \caption{As in fig.~\protect\ref{fig:1}, with a smooth $D$ function.}
\label{fig:2}
\end{figure}
One interesting aspect of fig.~\ref{fig:1} is the general agreement
between MC@NLO and \alpgen\ for the basic features of the observables.
This may be surprising at first, given the fact that only in \alpgen\
a Sudakov suppression is implemented (effectively), which incorporates
information on the behaviour of the Monte Carlo in the merging scheme. 
In fact, such information is also included in MC@NLO via the matching 
procedure (in particular, in the MC subtractions), in a way that differs 
from that of \alpgen\ at relative ${\cal O}(\as^2)$. This confirms the naive
expectation that effects due to the mismatch between matrix elements 
and Monte Carlos, which are mitigated at the LO by merging, are largely
dealt with by matching at the NLO. It also leads one to expect that,
if similar merging procedures can be implemented at both the LO and
the NLO, the latter results will be better behaved than the former. 
We shall show later explicit examples that this is indeed the case.

\begin{figure}[htb!]
  \begin{center}
        \epsfig{file=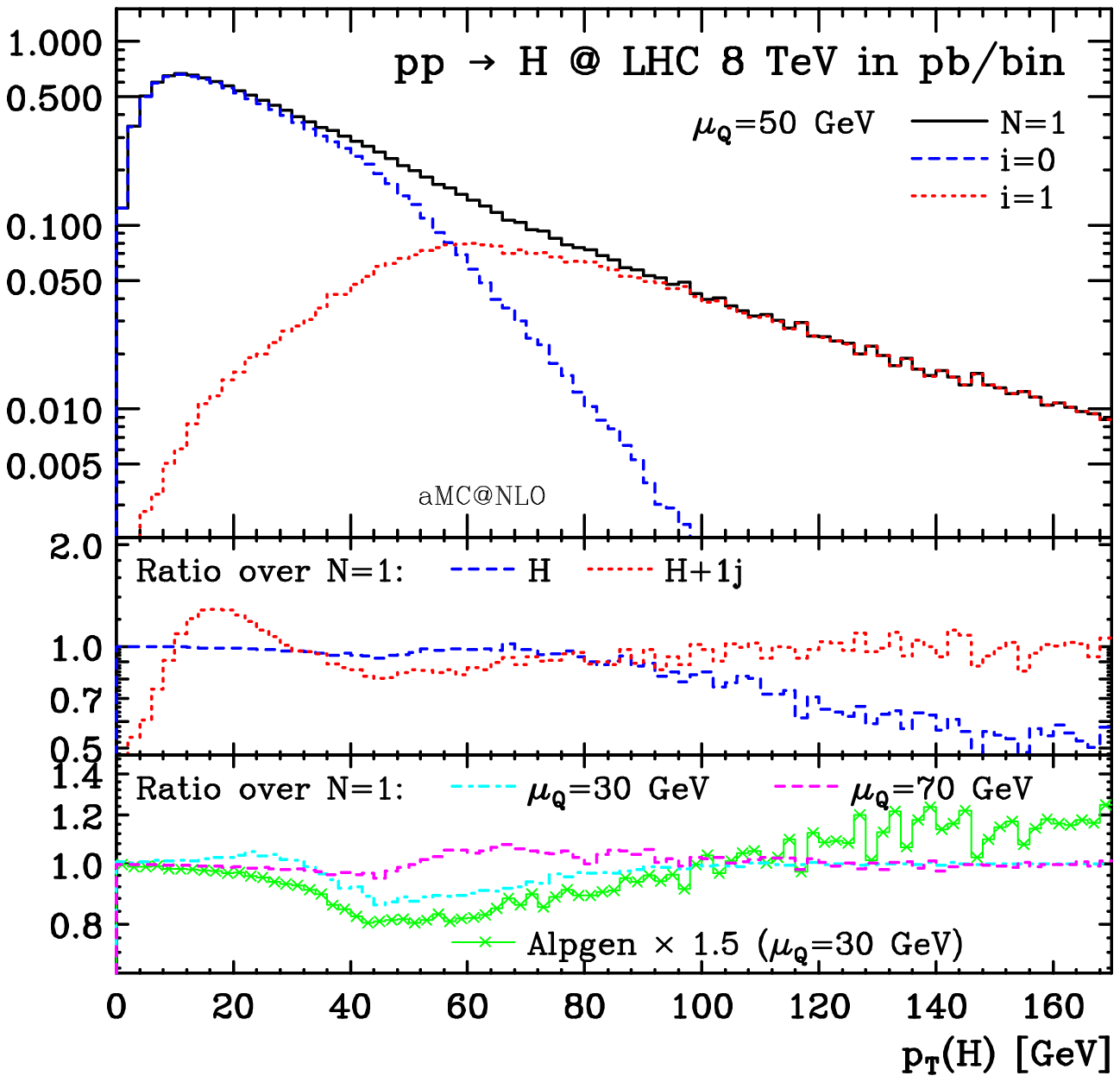, width=0.49\textwidth}
        \epsfig{file=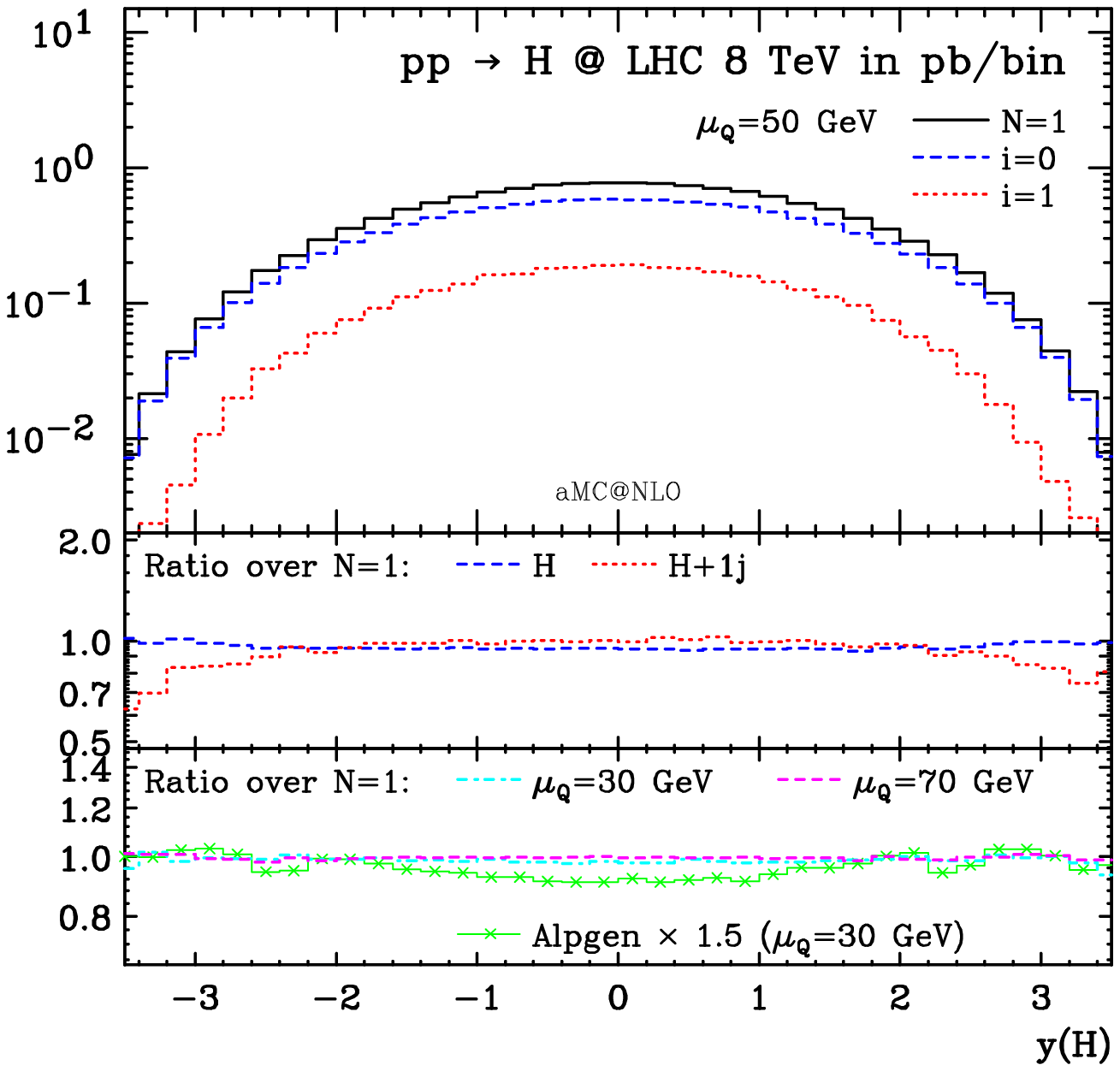, width=0.49\textwidth}
  \end{center}
  \begin{center}
        \epsfig{file=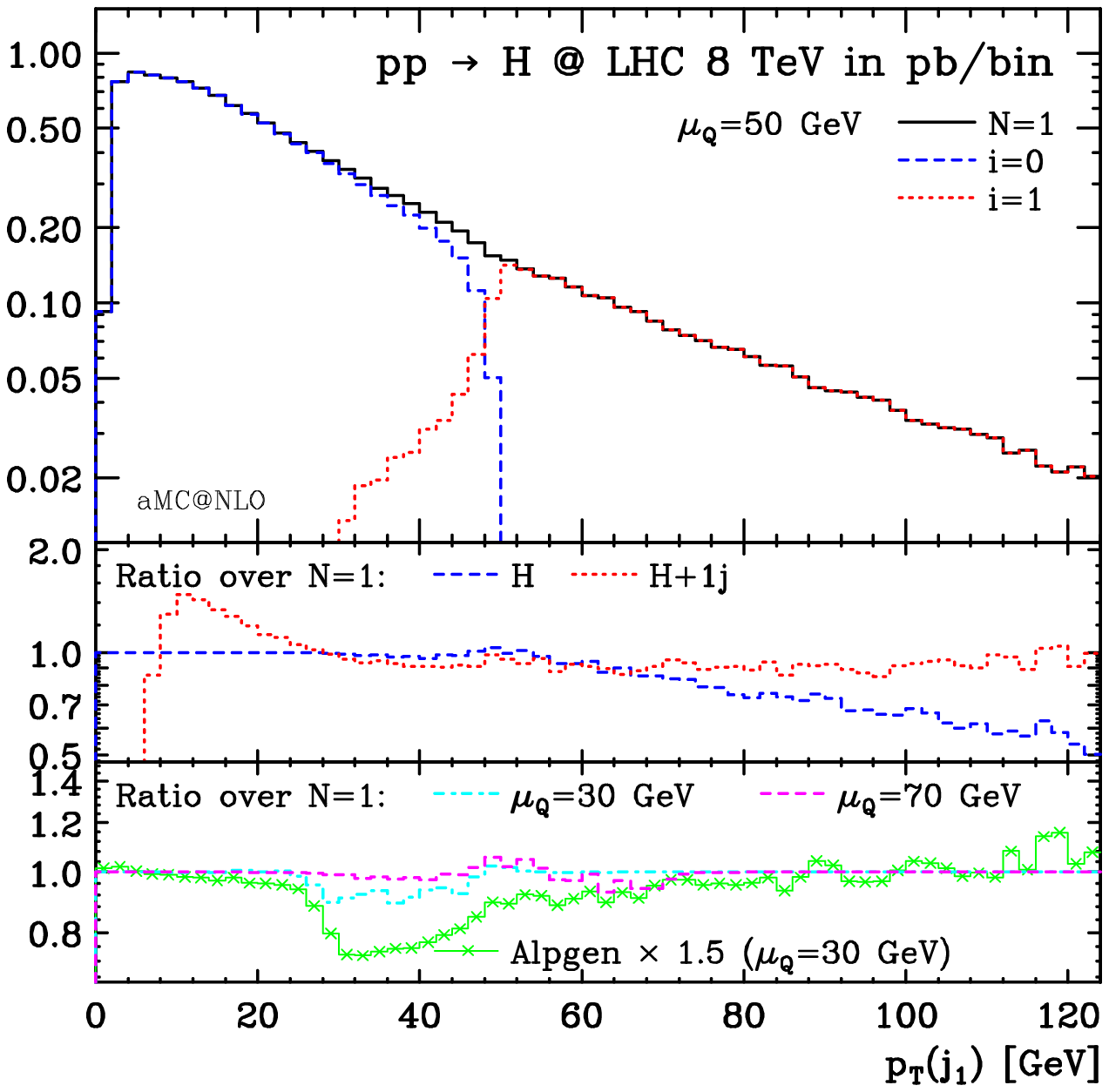, width=0.49\textwidth}
        \epsfig{file=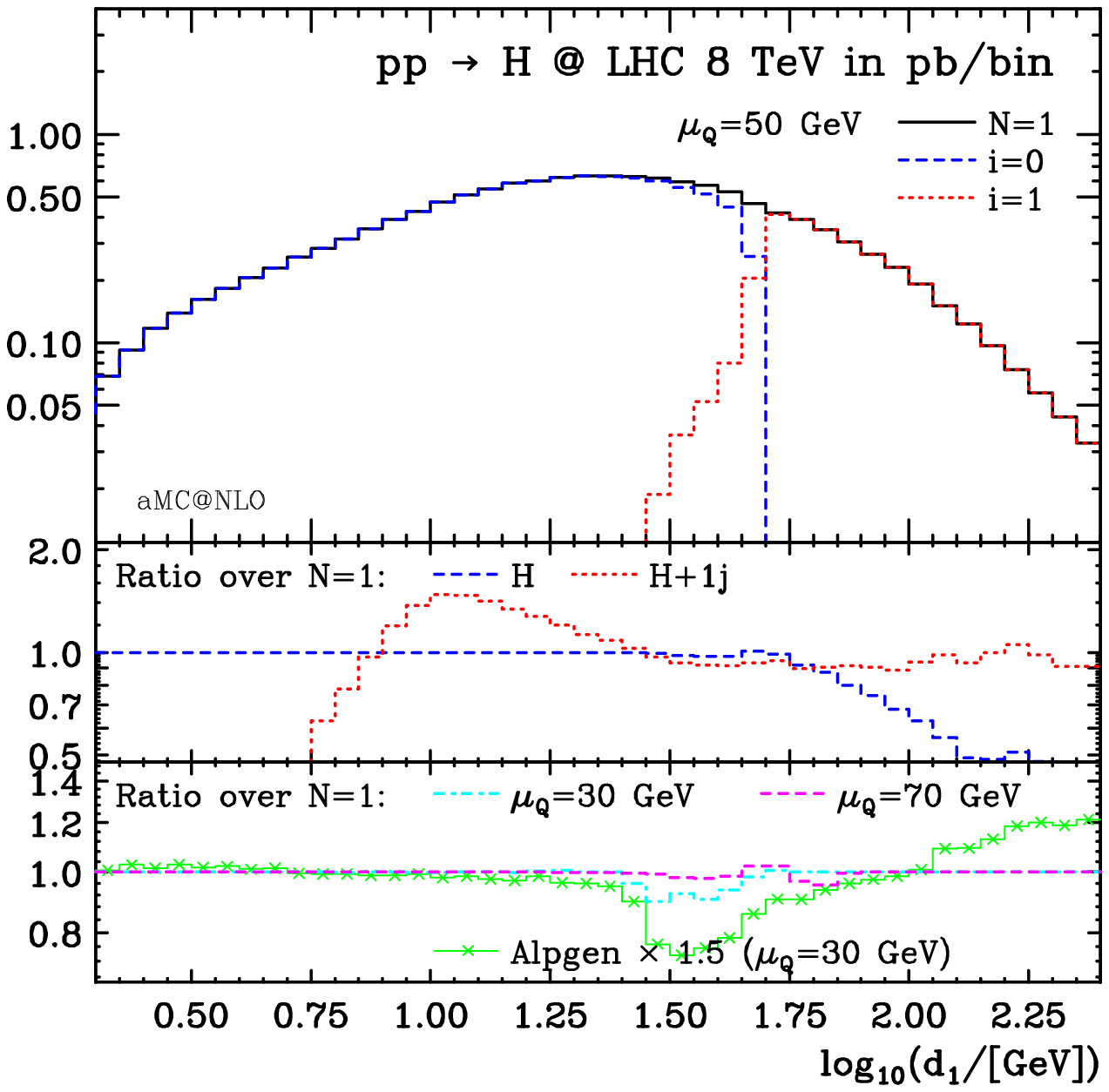, width=0.49\textwidth}
  \end{center}
  \vspace{-20pt}
  \caption{As in fig.~\protect\ref{fig:1}, with Sudakov reweighting.}
\label{fig:3}
\end{figure}
In fig.~\ref{fig:2} we plot the same observables as in fig.~\ref{fig:1},
computed with the same settings as before, except for the fact
that here the function $D$ is smooth, with $\mu_1=30$~GeV and
$\mu_2=70$~GeV. We have adopted the following functional form:
\beqn
D(\mu)&=&f(x(\mu))\,,\;\;\;\;\;\;\;\;\;\;
f(x)=\frac{(1-x)^{2\alpha}}{(1-x)^{2\alpha}+cx^{2\alpha}}\,,
\label{fdef}
\\
x(\mu)&=&\frac{\mu-\mu_1}{\mu_2-\mu_1}\stepf(\mu_2-\mu)\stepf(\mu-\mu_1)
+\stepf(\mu-\mu_2)\,,
\label{Dpheno}
\eeqn
with $\alpha=c=1$. The histograms shown in the lower insets are computed 
by taking the ratios of the predictions obtained with a sharp $D$ (for the 
three values of $\mu_Q$ previously considered), over those obtained with 
the smooth $D$ of eq.~(\ref{fdef}). Perhaps not surprisingly, the new 
results are in a very good overall agreement with those relevant to a sharp 
$D$ function with $\mu_Q=50$~GeV. The exception is the marked disagreement
\begin{figure}[htb!]
  \begin{center}
        \epsfig{file=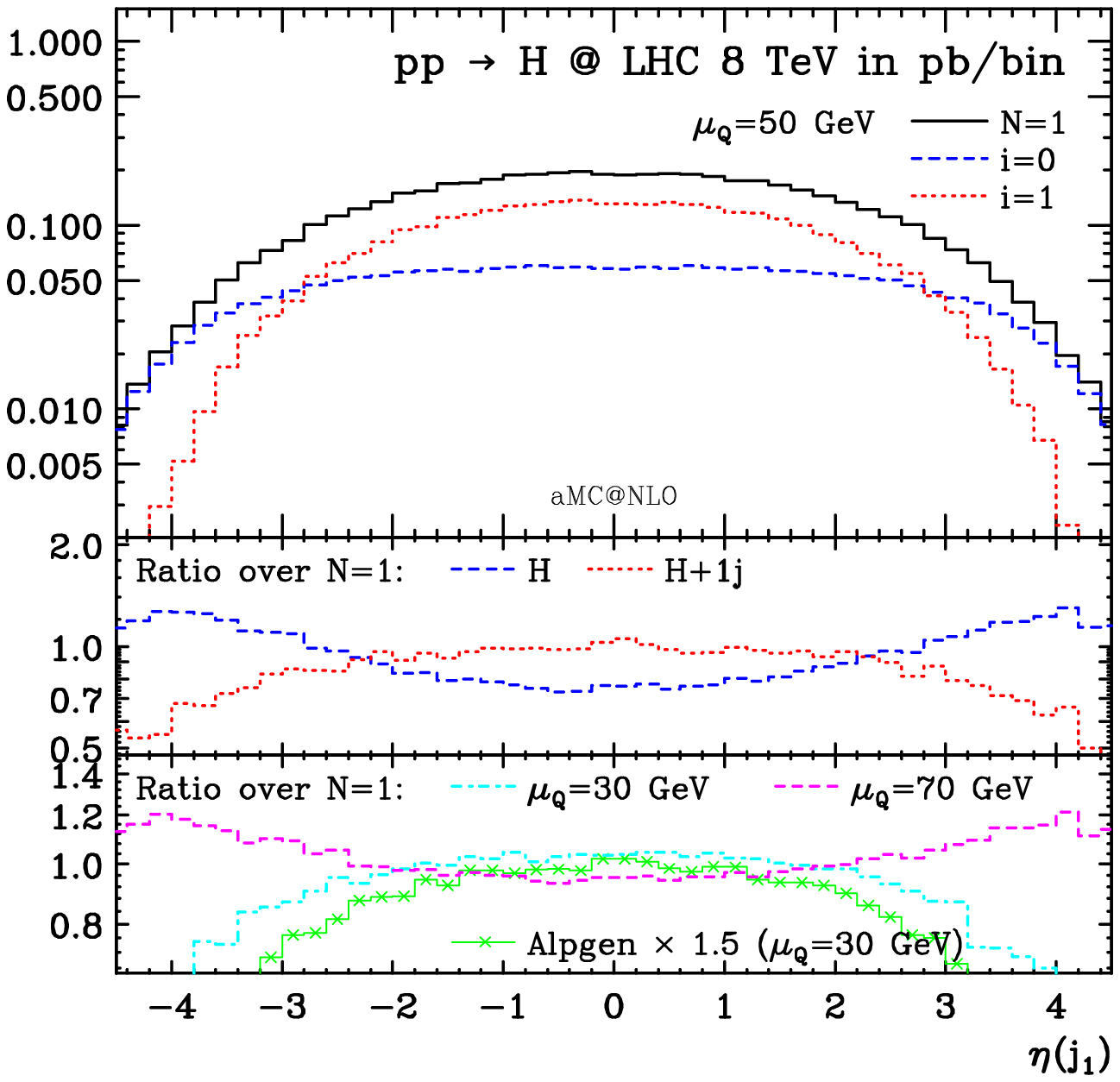, width=0.49\textwidth}
        \epsfig{file=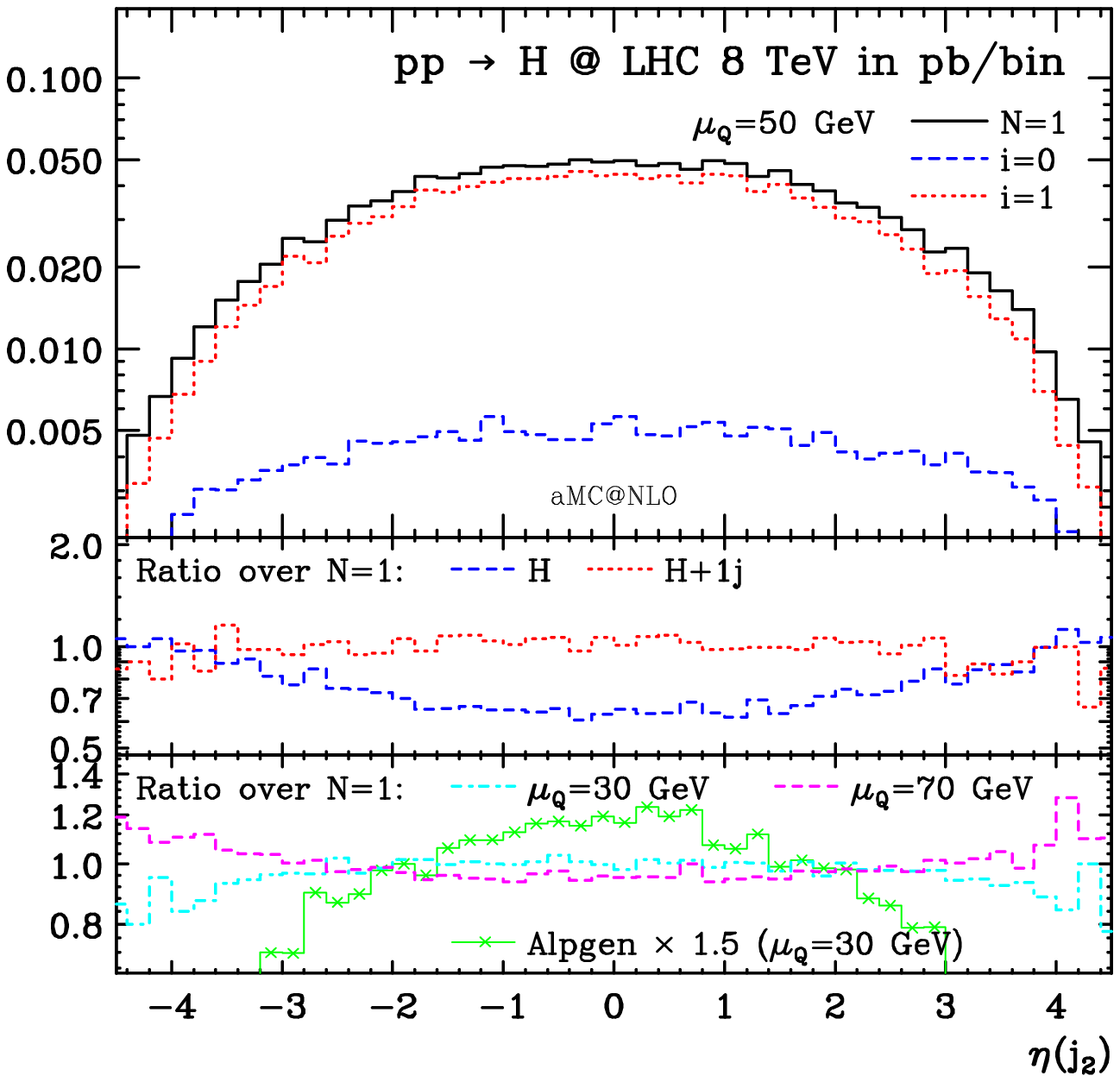, width=0.49\textwidth}
  \end{center}
  \begin{center}
        \epsfig{file=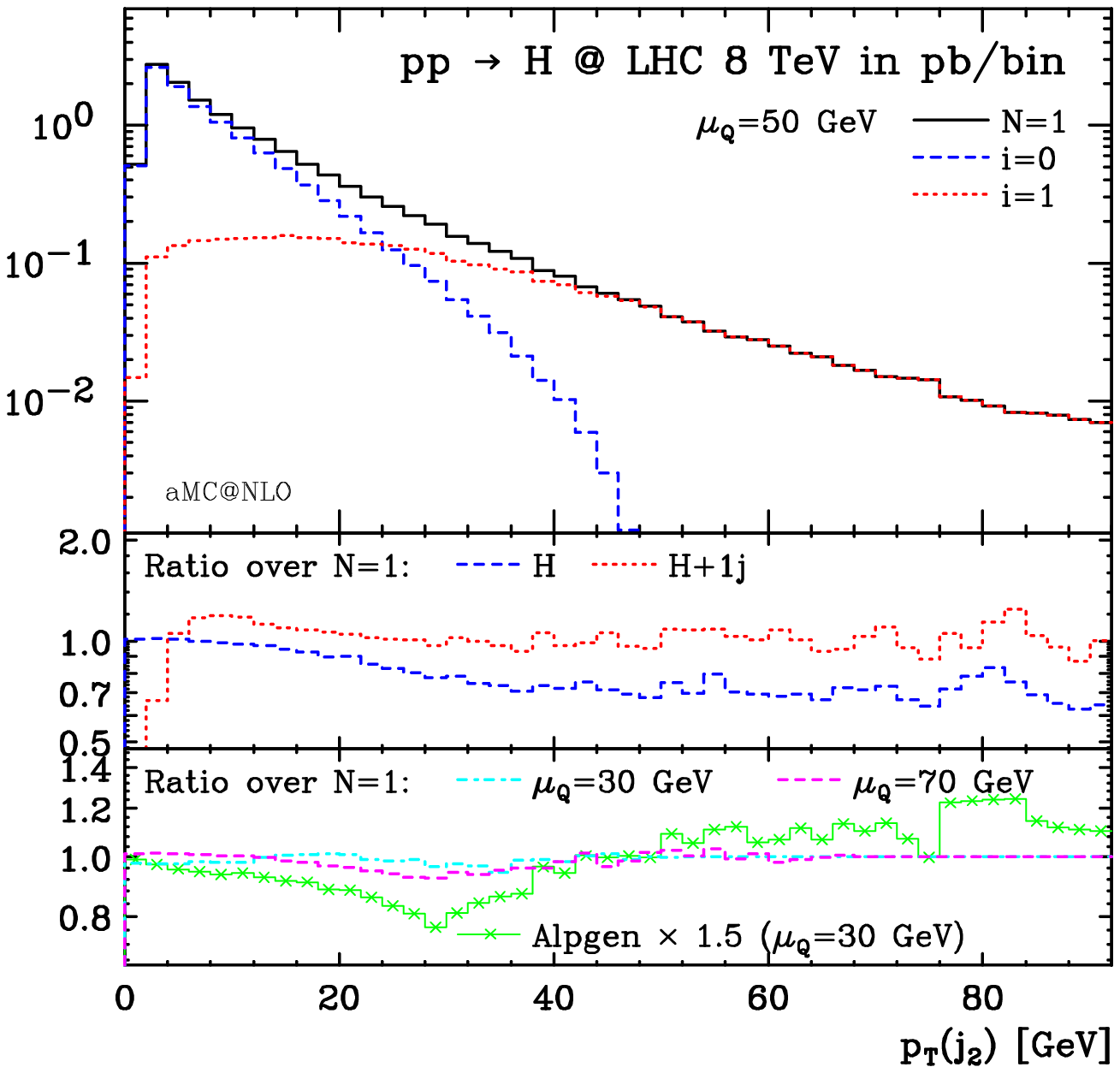, width=0.49\textwidth}
        \epsfig{file=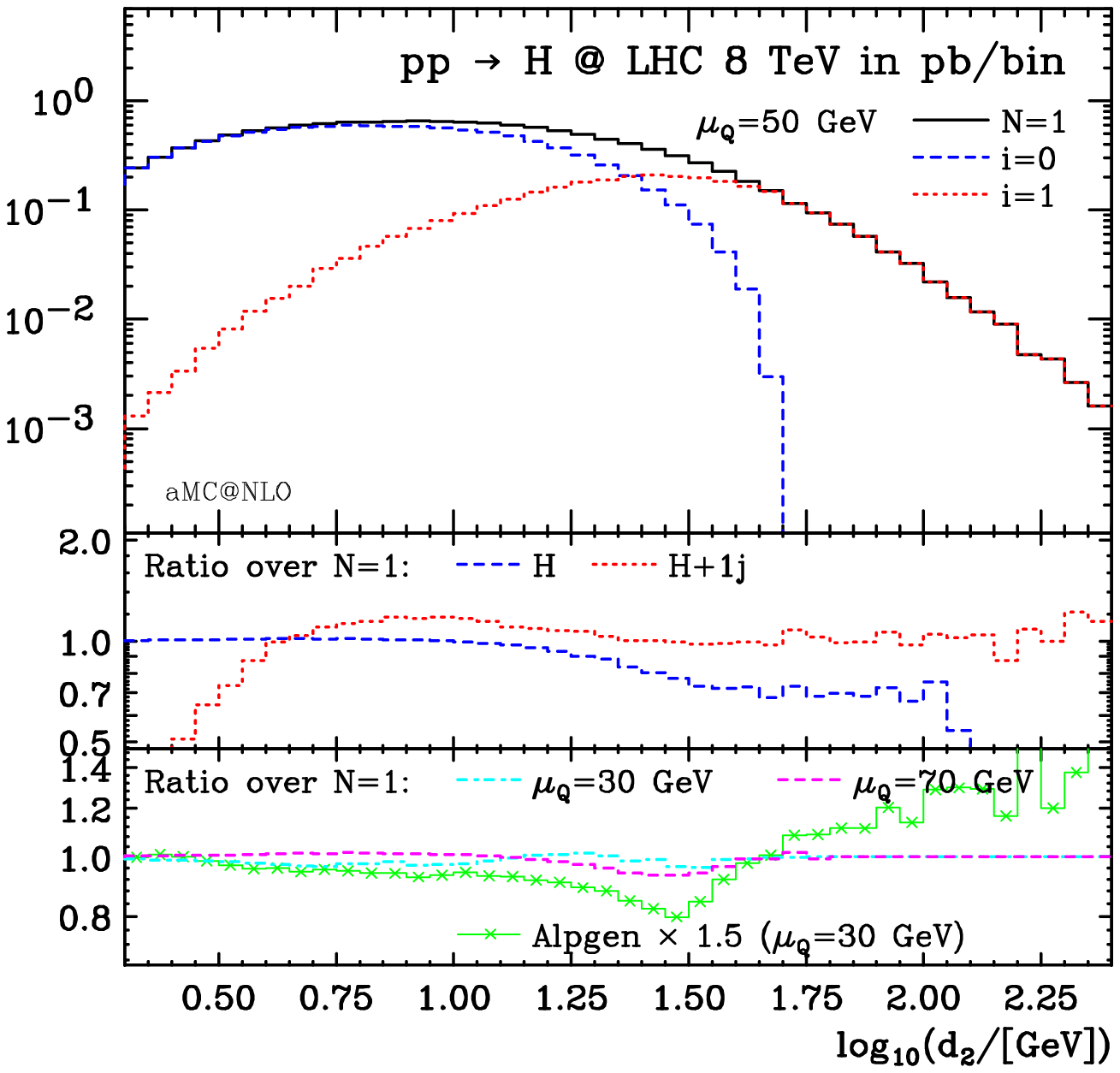, width=0.49\textwidth}
  \end{center}
  \vspace{-20pt}
  \caption{As in fig.~\protect\ref{fig:3}, for the pseudorapidity of the 
    hardest jet (upper left), the pseudorapidity (upper right) and 
    $\pt$ (lower left) of the second-hardest jet, and $d_2$ (lower right).
    In the case of $\eta(j_k)$, we have imposed a $\pt(j_k)\!>\!30$~GeV cut.}
\label{fig:4}
\end{figure}
around the $\pt(j_1)\simeq d_1\simeq\mu_Q$ regions, where the results
obtained with a smooth $D$ function do not present any kinks, and
are fairly regular. We thus conclude that the merging procedure 
with a smooth $D$ function (and without Sudakov reweighting) leads to 
satisfactory results\footnote{This suggests that it could be useful to
implement the CKKW or MLM merging procedures by randomly generating
the matching scale in pre-assigned ranges.}. Its drawback is that 
the assessment of the theoretical systematics is rendered more involved
because of the presence of two scales, that define the position and
width of a merging {\em range}. One possibility is that of taking the 
envelope of the predictions obtained with a sharp $D$, for values of $\mu_Q$ 
that span the merging range (which is essentially what is done in the lower
insets of fig.~\ref{fig:2}). On the one hand, this overestimates the
systematics, since the contributions due to scales close to the end-points
of the merging range are less important (in the effective average
performed by the smooth $D$ function) than those at its center.
On the other hand, this is not equivalent to assessing the effect
of changing the position and width of the merging range, which should
probably also be done. In any case, these appear to be pretty minor 
issues, given that the theoretical systematics associated with merging 
cannot be given a precise statistical meaning, and some degree of 
arbitrariness is always present.

\begin{figure}[htb!]
  \begin{center}
        \epsfig{file=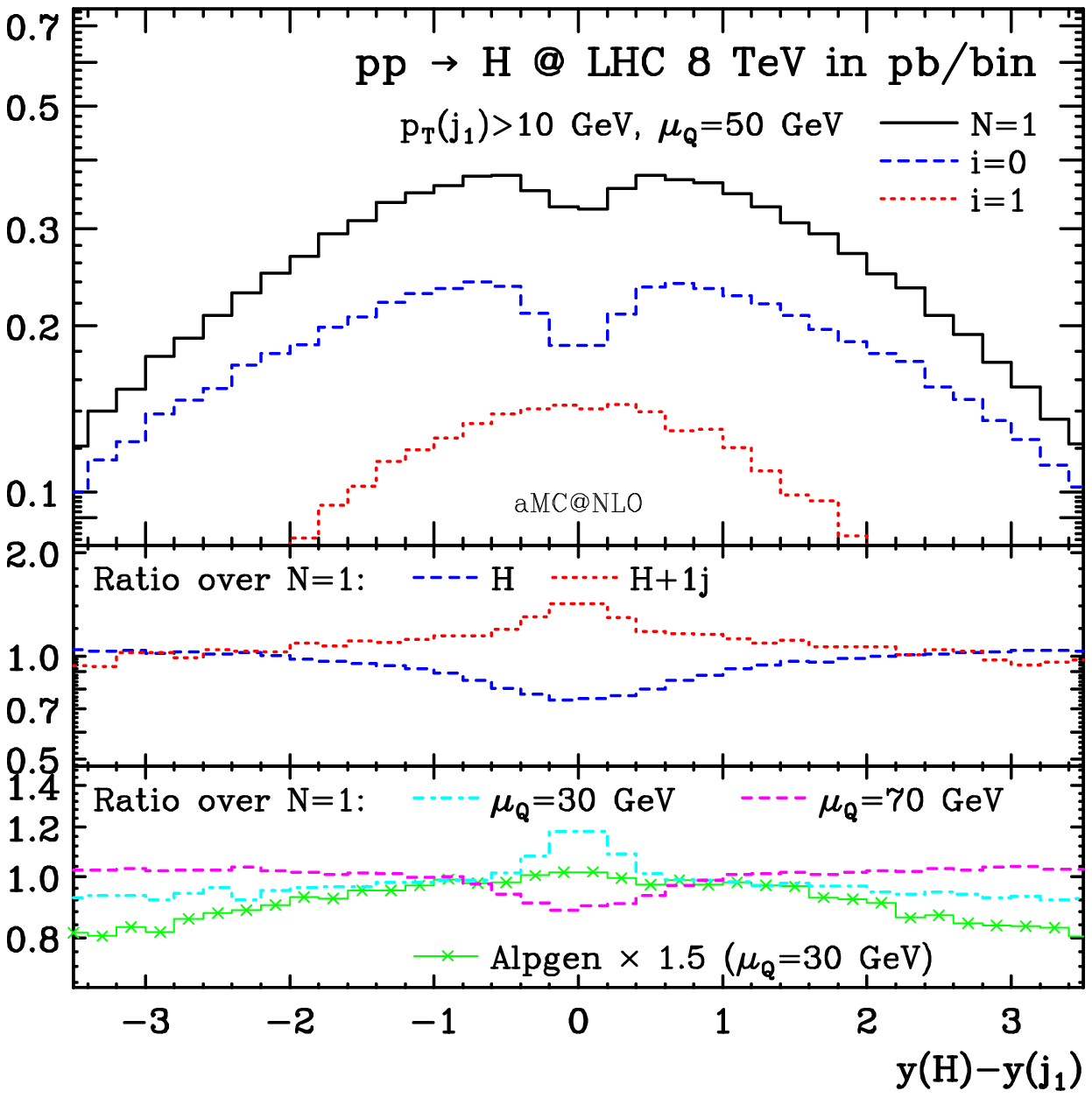, width=0.485\textwidth}
        \epsfig{file=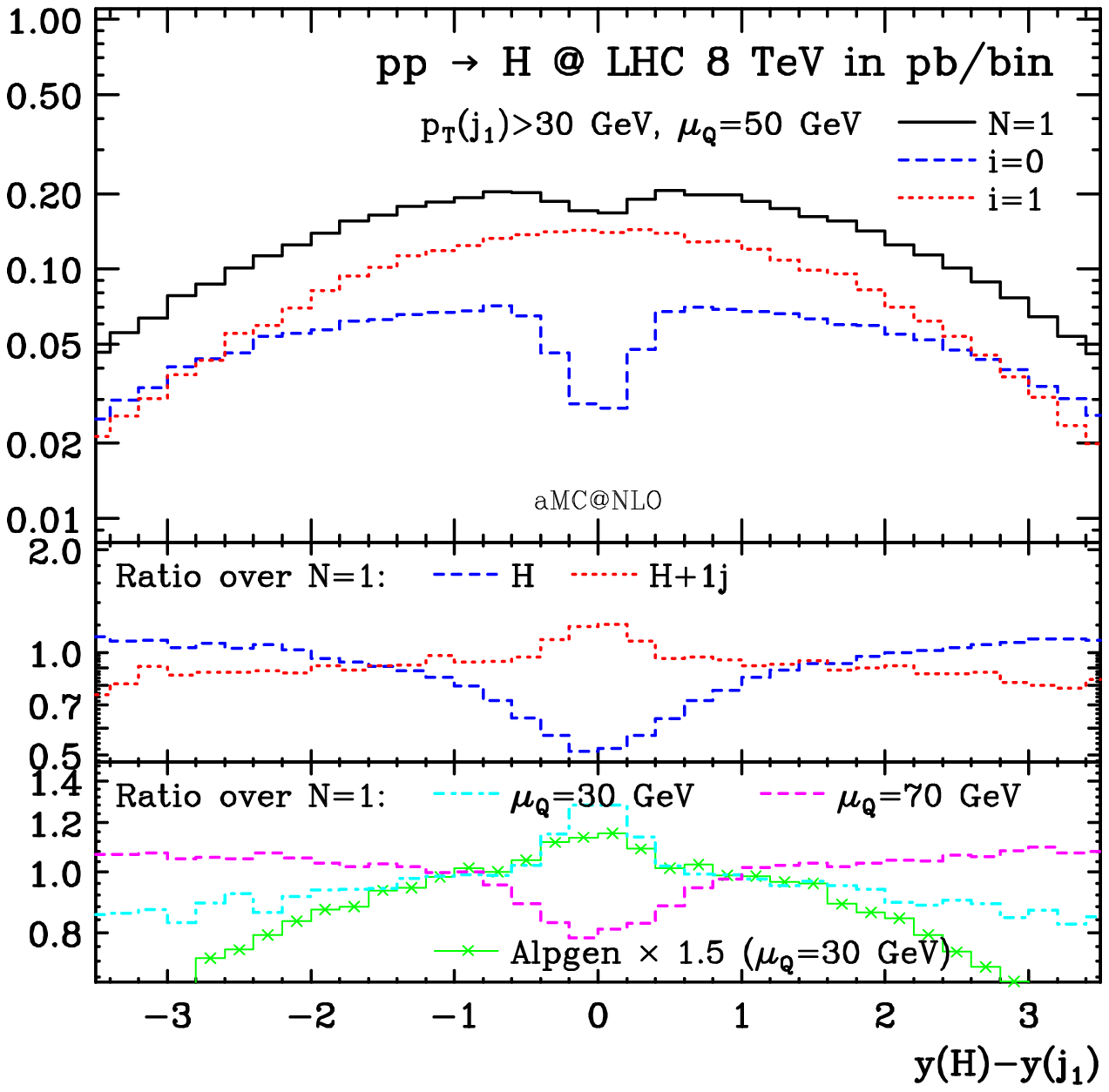, width=0.49\textwidth}
  \end{center}
  \begin{center}
        \epsfig{file=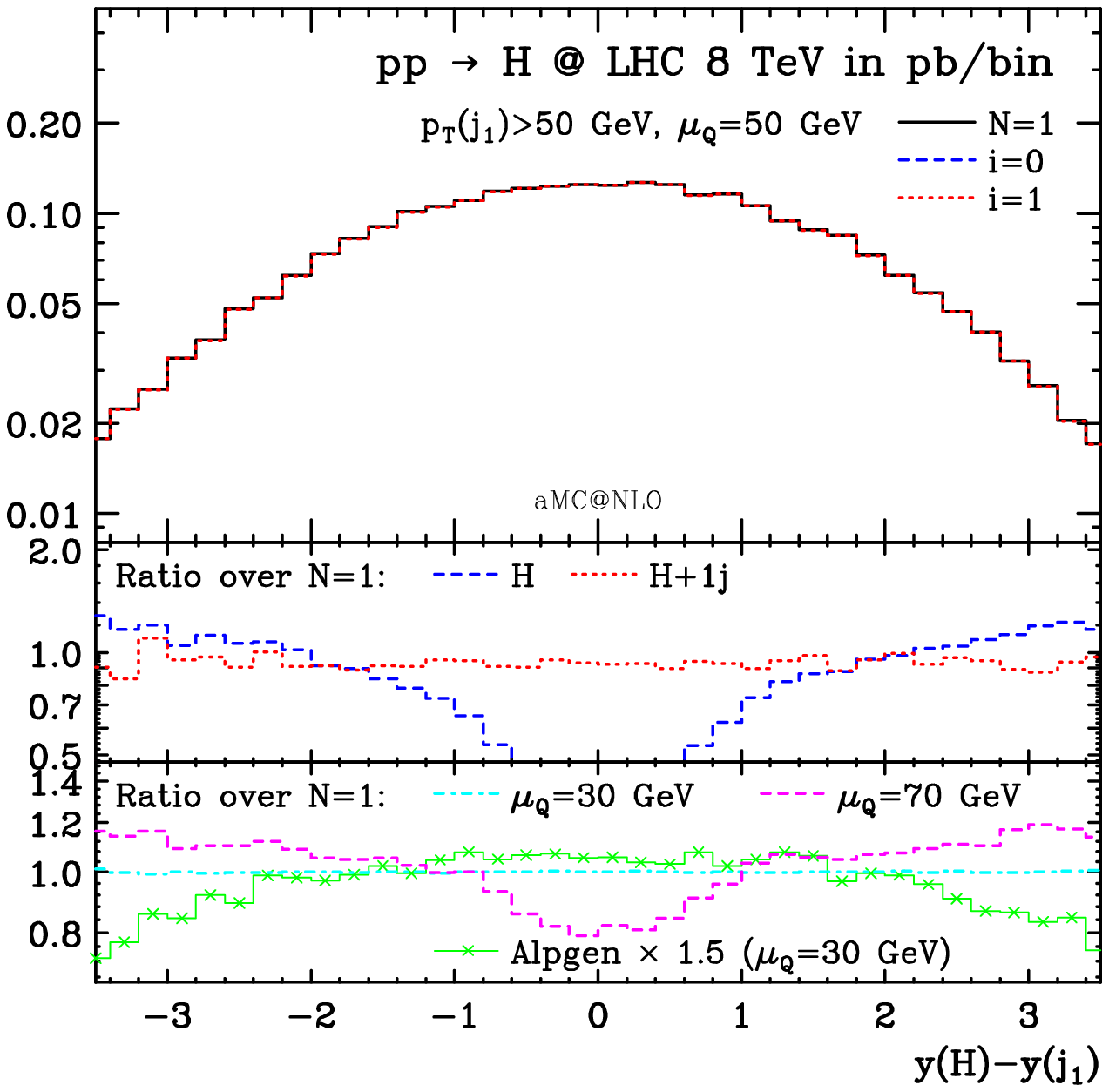, width=0.49\textwidth}
        \epsfig{file=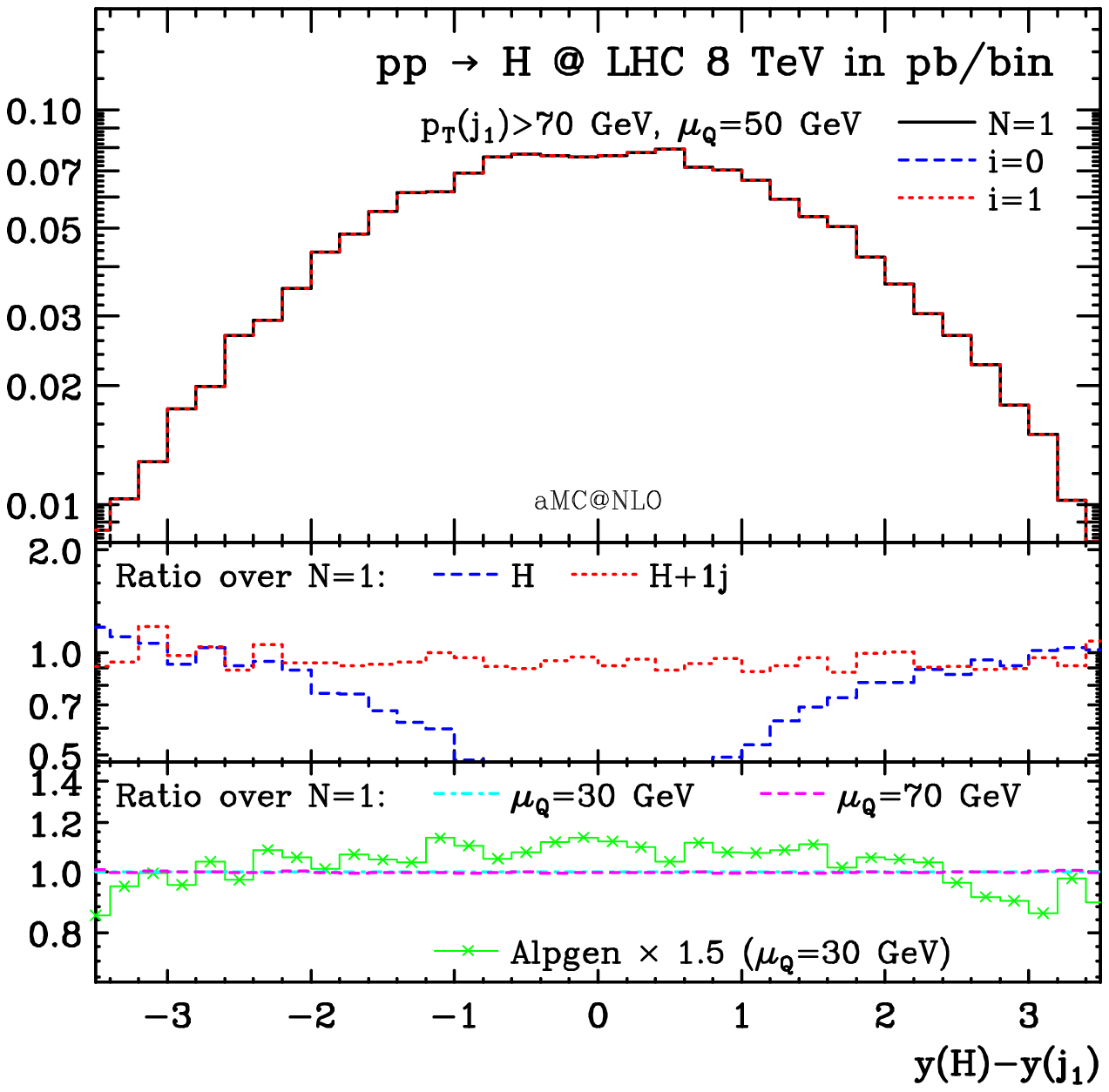, width=0.49\textwidth}
  \end{center}
  \vspace{-20pt}
  \caption{As in fig.~\protect\ref{fig:3}, for the difference in rapidity
    between the Higgs and the hardest jet, with four different $\pt$ cuts 
    on the latter.}
\label{fig:5}
\end{figure}

We now study the effect of the Sudakov reweighting, following the 
procedure described in sect.~\ref{sec:Sud}. We start by considering again 
the $N=1$ case, which we generate with a sharp $D$ function, and the 
three values $\mu_Q=30$, $50$, and $70$~GeV already employed. In 
fig.~\ref{fig:3} we plot the same observables as in fig.~\ref{fig:1} 
and~\ref{fig:2}; a few more jet-related observables are displayed
in figs.~\ref{fig:4} and~\ref{fig:5}. In all these figures, the main
frame presents the $\mu_Q=50$~GeV results, our ``central'' predictions
henceforth. The histograms in the lower insets are the ratios of the 
Sudakov-reweighted $\mu_Q=30$~GeV and $70$~GeV results over the 
central ones (in other words, there are no merged predictions in 
these plots that do not include the Sudakov reweighting). Also shown 
there are the ratios computed using \alpgen\ in the numerator,
over the central NLO-merged results.

The comparison of fig.~\ref{fig:3} with figs.~\ref{fig:1} 
and~\ref{fig:2} shows that the Sudakov reweighting on top of a sharp
$D$ function is as effective as the use of a smooth $D$ function
(without Sudakov reweighting) in removing the kinks.
There are quite small residual wiggles\footnote{These can be eliminated 
with a smooth $D$ function (plus Sudakov reweighting). We did not test 
this option here, since it appears to be just a phenomenological issue 
analogous to tuning. We reckon that a relatively small range $(\mu_1,\mu_2)$ 
will be sufficient; a further handle can be provided by the parameters 
$\alpha$ and $c$ that appear in eq.~(\ref{fdef}).}, which may be seen 
in some of the lower insets in the vicinity of the transition regions, 
that are in any case within a modest theoretical systematics; this
is smaller than 10\% everywhere, and in most regions quite negligible.
As one can infer from the comparison between the upper insets
of fig.~\ref{fig:3} with those of fig.~\ref{fig:2} (keeping in
mind that the standalone $H+0j$ and $H+1j$ MC@NLO predictions are
the same in these two figures) the sharp-$D$, Sudakov-reweighted
results are quite close to the smooth-$D$, non-Sudakov-reweighted ones,
with differences of the order of 5\% or smaller. The same holds true
for the sharp-$D$, non-Sudakov-reweighted results obtained with 
$\mu_Q=50$~GeV, in keeping with the idea that Sudakov effects in the
context of the present merging procedure are beyond NLO, hence small. 
Obviously, this smallness is true in a 
parametric sense, and numerically the case $\mu_Q=50$~GeV is
particularly fortunate, just because such a value happens to be 
quite suited for this process. When $\mu_Q=30$~GeV or $\mu_Q=70$~GeV
the differences induced by the Sudakov reweighting on top of a 
sharp $D$ function are numerically a bit larger, but there is
no difference of principle among the various scale choices.

The transverse momentum of the second-hardest jet, and $d_2$, presented
in the bottom panels of fig.~\ref{fig:4} show the same patterns
as $\pt(j_1)$ and $d_1$ (possibly with an even smaller theoretical
systematics). This also applies to the comparison with \alpgen,
since the latter has kinks at $\pt(j_2)\simeq d_2\simeq 30$~GeV,
analogous to those affecting $\pt(j_1)$ and $d_1$ (if a bit smaller).
The pseudorapidities of the hardest and second-hardest jets are presented
in the upper panels of fig.~\ref{fig:4}, for 
$\pt(j_k)\!>\!\ptcut\equiv 30$~GeV. 
In both cases, the merged predictions are more central than those obtained 
with the standalone $H+0j$ MC@NLO simulations, owing to the contribution 
of the $1$-parton sample. Still, since pseudorapidities receive the most 
important contributions from the region $\pt(j_k)\simeq\ptcut$, the 
$0$-parton sample will typically be dominant (except for very large 
$\ptcut$, which is not the case here), with the $1$-parton sample 
providing a subleading correction. Hence, the onset of the $1$-parton 
sample regime determines directly the amount of migration 
towards central $\eta$ values w.r.t.~the
standalone $H+0j$ results. In turn, this onset is controlled by the
matching scale; this explains why the systematics affecting $\eta(j_1)$
and $\eta(j_2)$ is larger than for other observables. In the case of
$\eta(j_1)$, the \alpgen\ result is in fact quite close to the NLO-merged 
prediction obtained with the same matching scale (30~GeV). On the
other hand, for $\eta(j_2)$ \alpgen\ is significantly more central
than MC@NLO, even with the same matching scale. This can be understood
as follows. Let us consider the $H+2$~parton matrix element which,
if surviving the merging ``cuts'' in \alpgen, will result into two hard
jets. This is not quite the case in MC@NLO (except in the large-$\pt$
region, which is not important here), where its contribution to the
$\clH$ events of the $1$-parton sample is partly compensated by the
MC subtractions. This is what underscores the intuitive picture of
the $1$-parton sample being {\em kinematically} a $1$-jet cross section,
plus (small) corrections, and is consistent with the fact that NLO 
computations must be inclusive to a certain degree. Hence, despite
receiving contributions from the same tree level matrix elements,
\alpgen\ results will be more matrix-element driven than MC@NLO ones,
for observables sensitive to the largest multiplicity\footnote{We also 
note that the one-loop contribution to the $N$-parton sample has also 
an $N$-jet kinematics, which will give a further MC-driven
contribution to $(N+1)$-jet observables, not present in \alpgen.}. 
Indeed, we shall show later that the inclusion of the $2$-parton sample 
in MC@NLO results in a more central $\eta(j_2)$ distribution.

As the final example for the $N=1$, sharp-$D$ function, Sudakov-reweighted
merging, we present in fig.~\ref{fig:5} the difference in rapidity
between the Higgs and the hardest jet, by imposing that the $\pt$ of
the latter be larger than 10, 30, 50, and 70~GeV. This observable has
attracted some attention in the past, because of the presence in
the standalone $H+0j$ MC@NLO results of a dip in the central region
(analogous features can be found in \mbox{$y(S)-y(j_1)$} or $y(j_1)$ for 
standalone $S+0j$ MC@NLO runs -- see e.g.~ref.~\cite{Mangano:2006rw} for a 
discussion of the case $S=t\tb$, to which we shall return later). It should 
be pretty clear that the dip is inherited by MC@NLO from the underlying Monte 
Carlo, which has in fact a deeper dip, partly filled in MC@NLO by the 
$\clH$-event contribution. This is documented in ref.~\cite{Torrielli:2010aw},
where MC@NLO was matched with $Q^2$-ordered \PY; since \PY,
at variance with \HWs\ or \HWpp, does not have a dip (at 
least in the low-$\pt$ region), MC@NLO does not have a dip.
Having clarified this, the natural question is the following:
if, for a given $\pt(j_1)\!>\!\ptcut$, the underlying Monte Carlo
has a dip in the rapidity difference, why MC@NLO does not remove
it completely? The answer is pretty simple: as mentioned above, MC@NLO 
fills the dip through the $\clH$-event contribution, which is 
matrix-element driven and of relative ${\cal O}(\as)$. Such contribution 
will thus be overwhelmed by Monte Carlo effects (from the first emission
onwards), when radiation of ${\cal O}(\ptcut)$ can be easily achieved by 
parton showers\footnote{Which also explains why, if $\ptcut$ is 
sufficiently large, MC@NLO does fill the dip completely.}. In other
words, in spite of the fact that $\ptcut$ may seem to define a
hard scale, $\pt(j_1)\simeq\ptcut$ could still be in an MC-dominated 
region (which is a process- and MC-dependent statement). Hence, in such
a case MC@NLO will not ``assume'' that a matrix element description is 
correct, but will rather follow the pattern of the underlying Monte Carlo.
The implication is that, if the dip is phenomenologically untenable,
a solution has to be found at the Monte Carlo level, and not in the
matching procedure.

An alternative point of view, which will however lead one to 
the same conclusions, is the following. Since $y(S)-y(j_1)$ is
one-jet exclusive, standalone $S+0j$ MC@NLO will give an LO-accurate
description at best, so we are much better off by considering
standalone $S+1j$ MC@NLO instead. While this is true, it is indeed 
equivalent to {\em assuming} that for $\pt(j_1)\simeq\ptcut$
a matrix-element description is correct. This is obviously the 
same problem as that of $\pt(j_1)$, discussed in the context of 
fig.~\ref{fig:1}. Therefore, as in that case, a merging procedure
will certainly constitute an improvement over standalone results,
provided that the merging systematics is correctly assessed
(i.e., $\mu_Q$ must be chosen and varied independently of $\ptcut$).
As can be seen from the four panels of fig.~\ref{fig:5}, the
NLO-merged results are in much better agreement with what one expects 
from matrix elements (equivalent to the standalone $H+1j$ histograms
here) than with the standalone $H+0j$ predictions.
There are, however, significant differences among the results
obtained with different $\mu_Q$'s, since the dip is typically present only
when $\ptcut<\mu_Q$ (but not necessarily so: e.g.~when $\mu_Q=30$~GeV,
there is no dip even for $\ptcut=10$~GeV). For example, when 
$\ptcut=30$~GeV, the NLO-merged result has a dip when $\mu_Q=50$ 
and $70$~GeV, while it has no dip when $\mu_Q=30$~GeV. It is interesting 
to observe that \alpgen\ does have dips when $\ptcut\le\mu_Q=30$~GeV,
while the NLO-merged prediction obtained with the same matching scale 
does not, as mentioned above. This is not surprising, on the 
basis of what was discussed before about the inheritance of
this feature from the underlying Monte Carlo. The conclusion is
that, in the context of merging at both the LO and the NLO accuracy, 
the dip can be tuned away by a suitable choice of merging parameters;
this is acceptable only if the merging systematics is exhaustively
assessed. We expect that, for small $\ptcut$ (where ``small'' can
be precisely defined given the production process), such systematics
will be large if the underlying Monte Carlo features the dip,
\begin{figure}[htb!]
  \begin{center}
        \epsfig{file=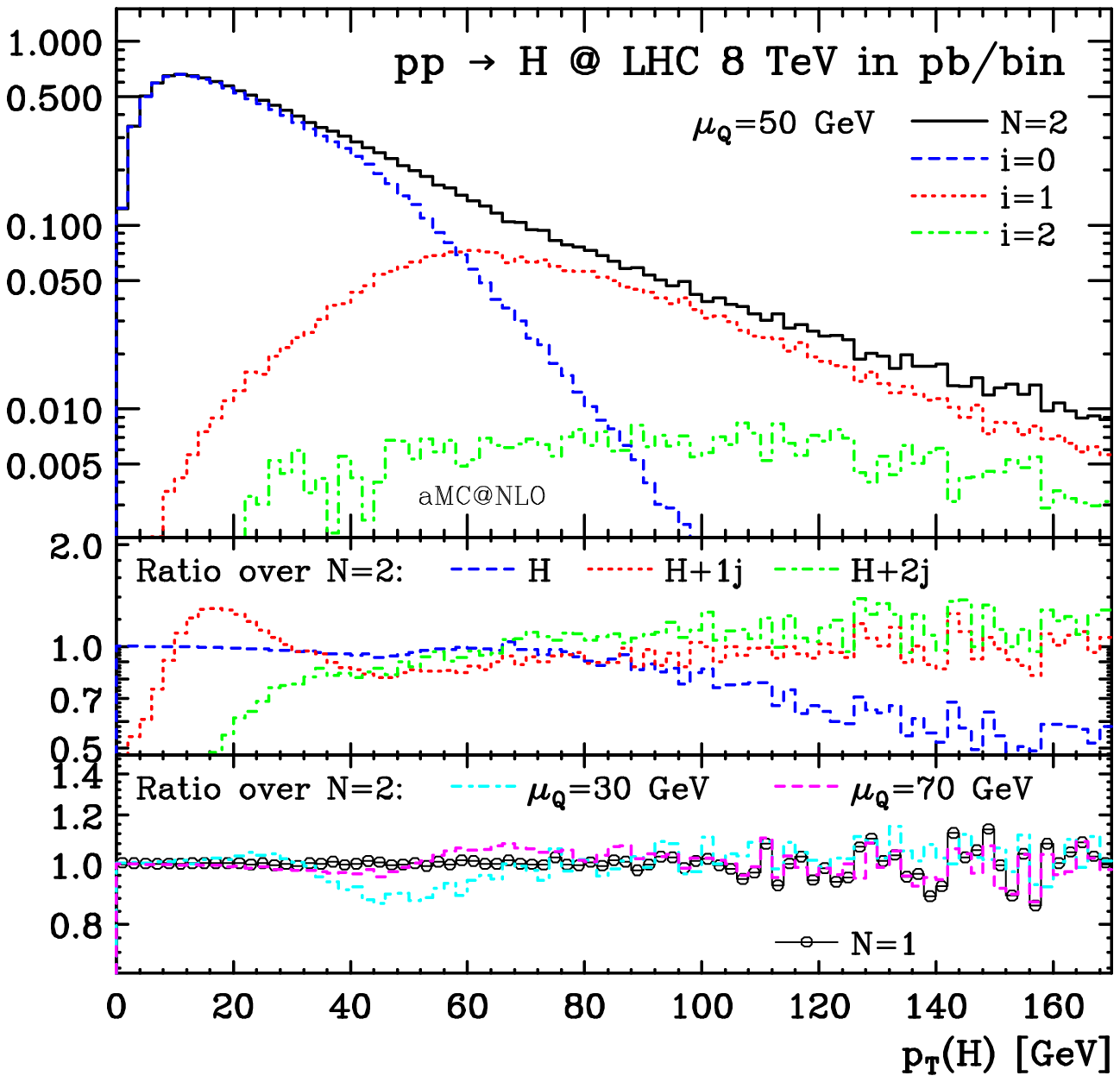, width=0.485\textwidth}
        \epsfig{file=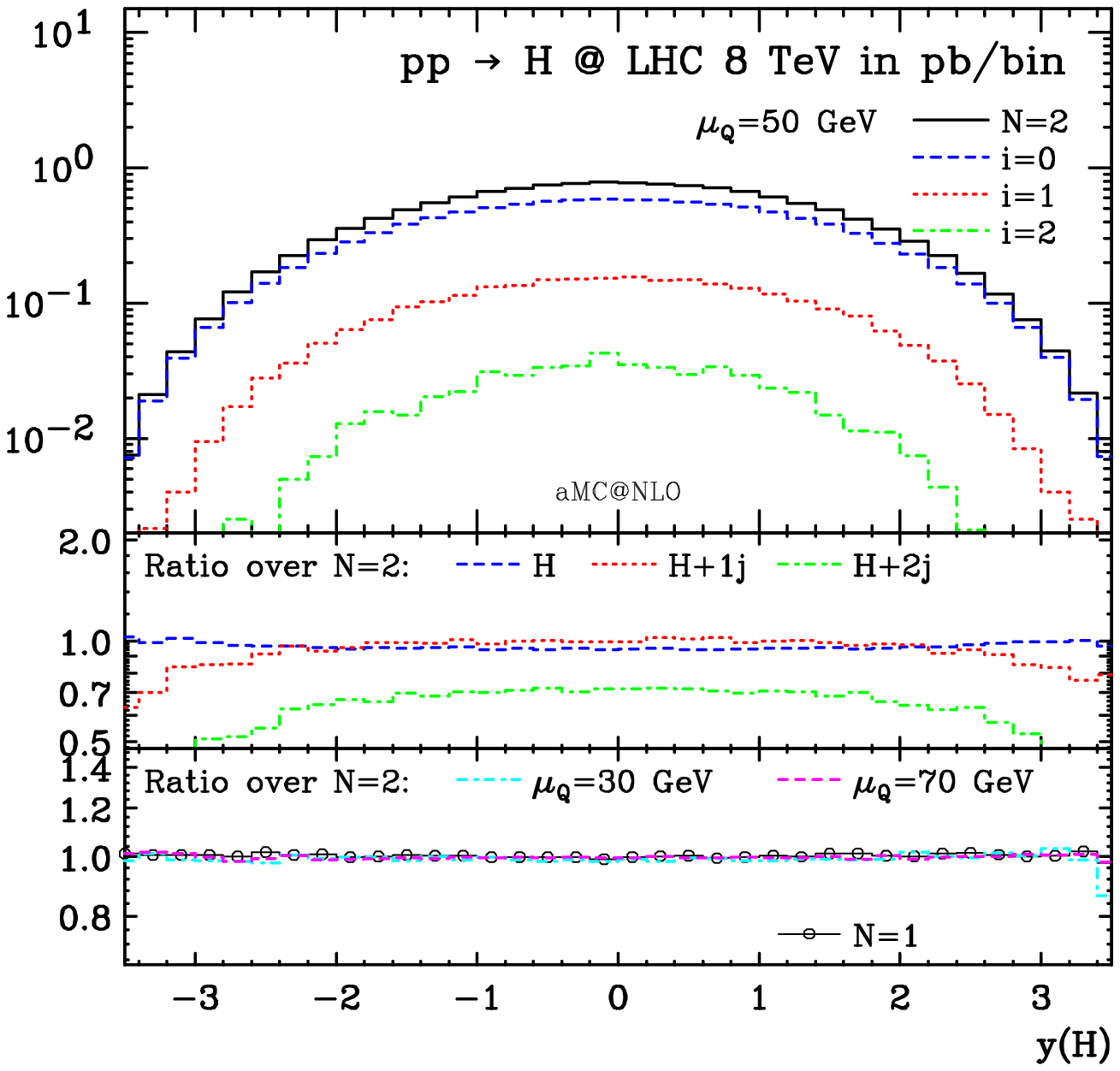, width=0.49\textwidth}
  \end{center}
  \begin{center}
        \epsfig{file=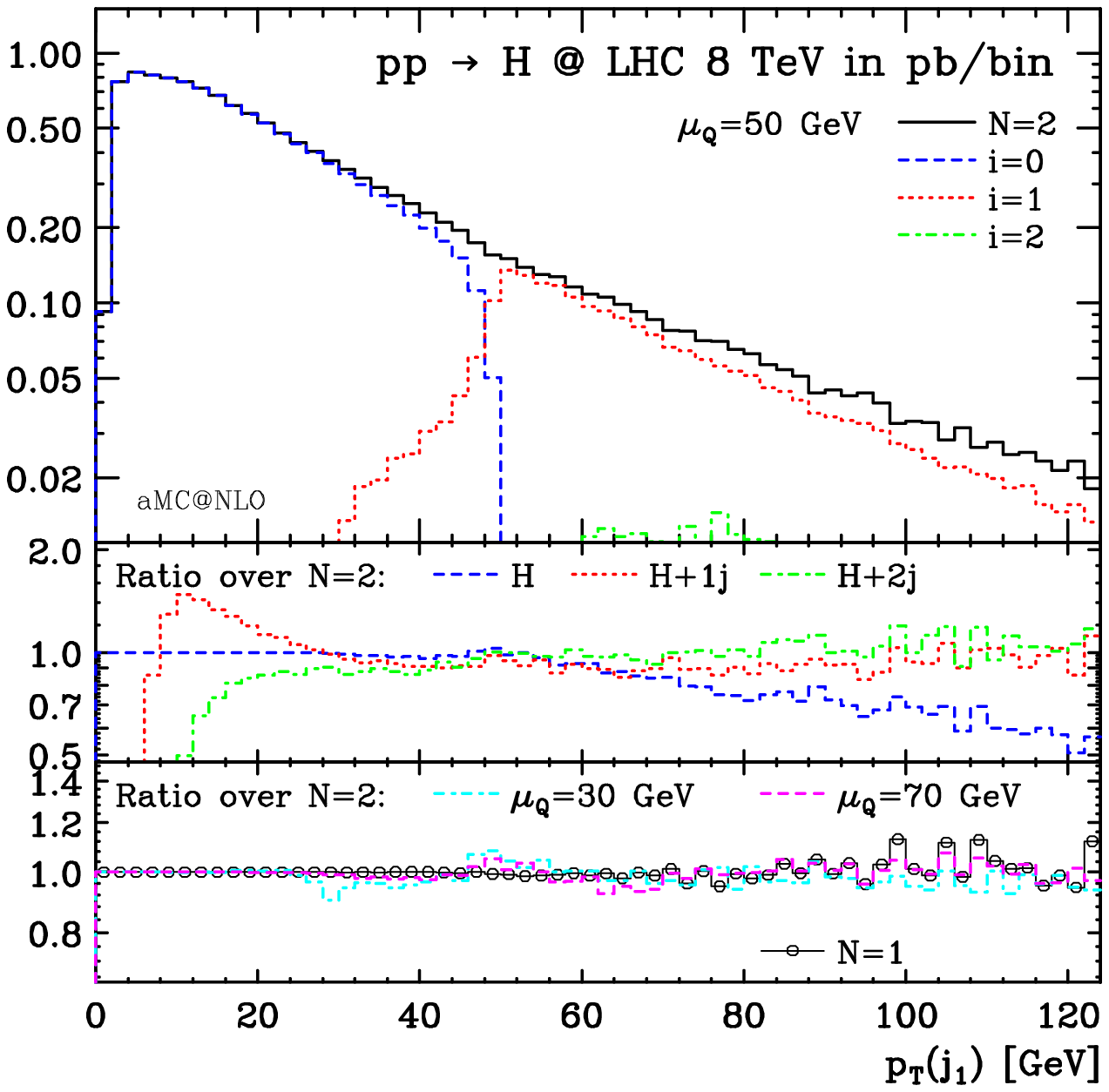, width=0.49\textwidth}
        \epsfig{file=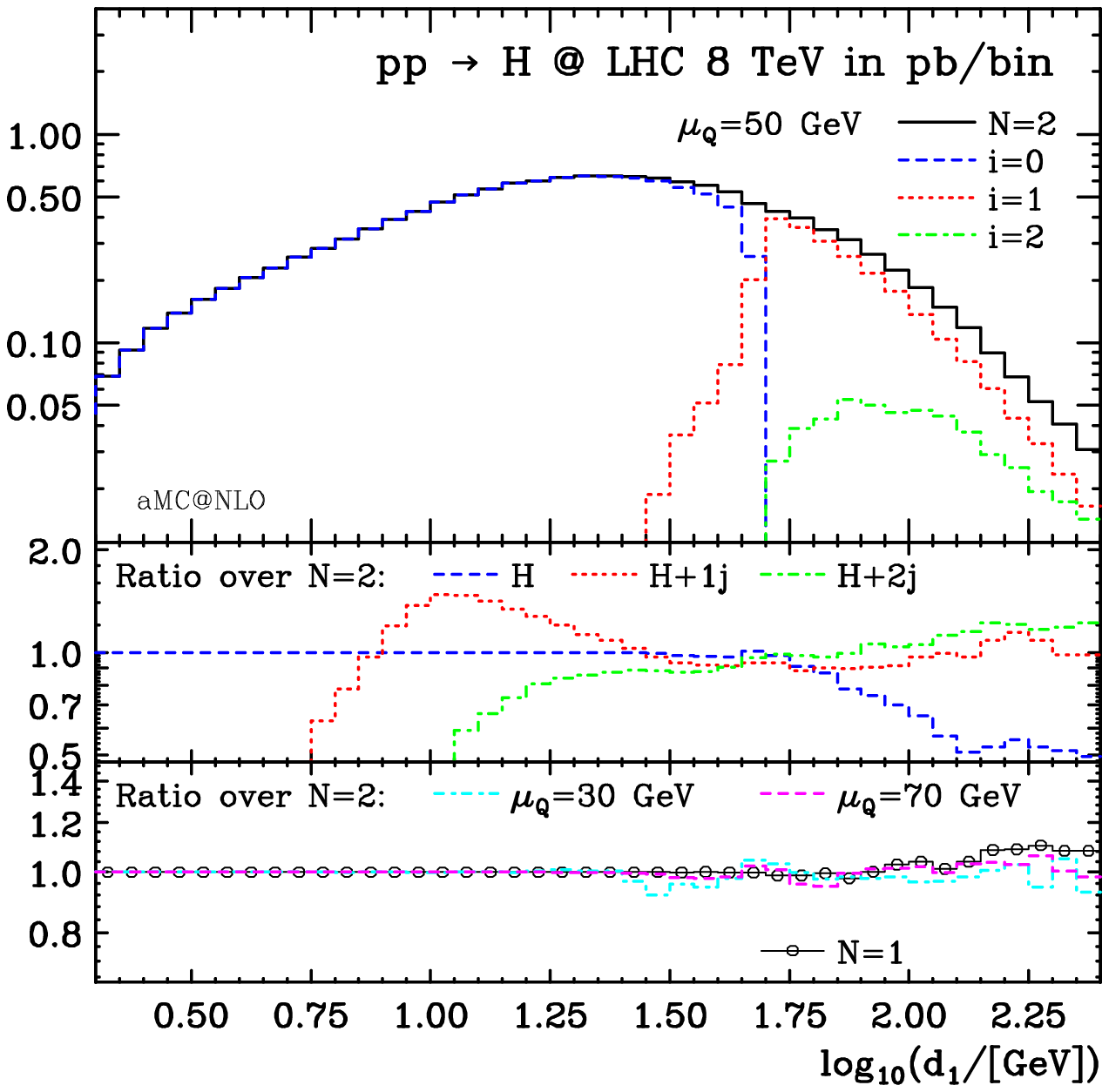, width=0.49\textwidth}
  \end{center}
  \vspace{-20pt}
  \caption{As in fig.~\protect\ref{fig:3}, with $N=2$.}
\label{fig:6}
\end{figure}
and much smaller otherwise -- hence, theoretical uncertainties
will be strictly MC-dependent. This dependence is bound to disappear,
and the merging-parameter dependence reduced, when $\ptcut$ becomes large.

\begin{figure}[htb!]
  \begin{center}
        \epsfig{file=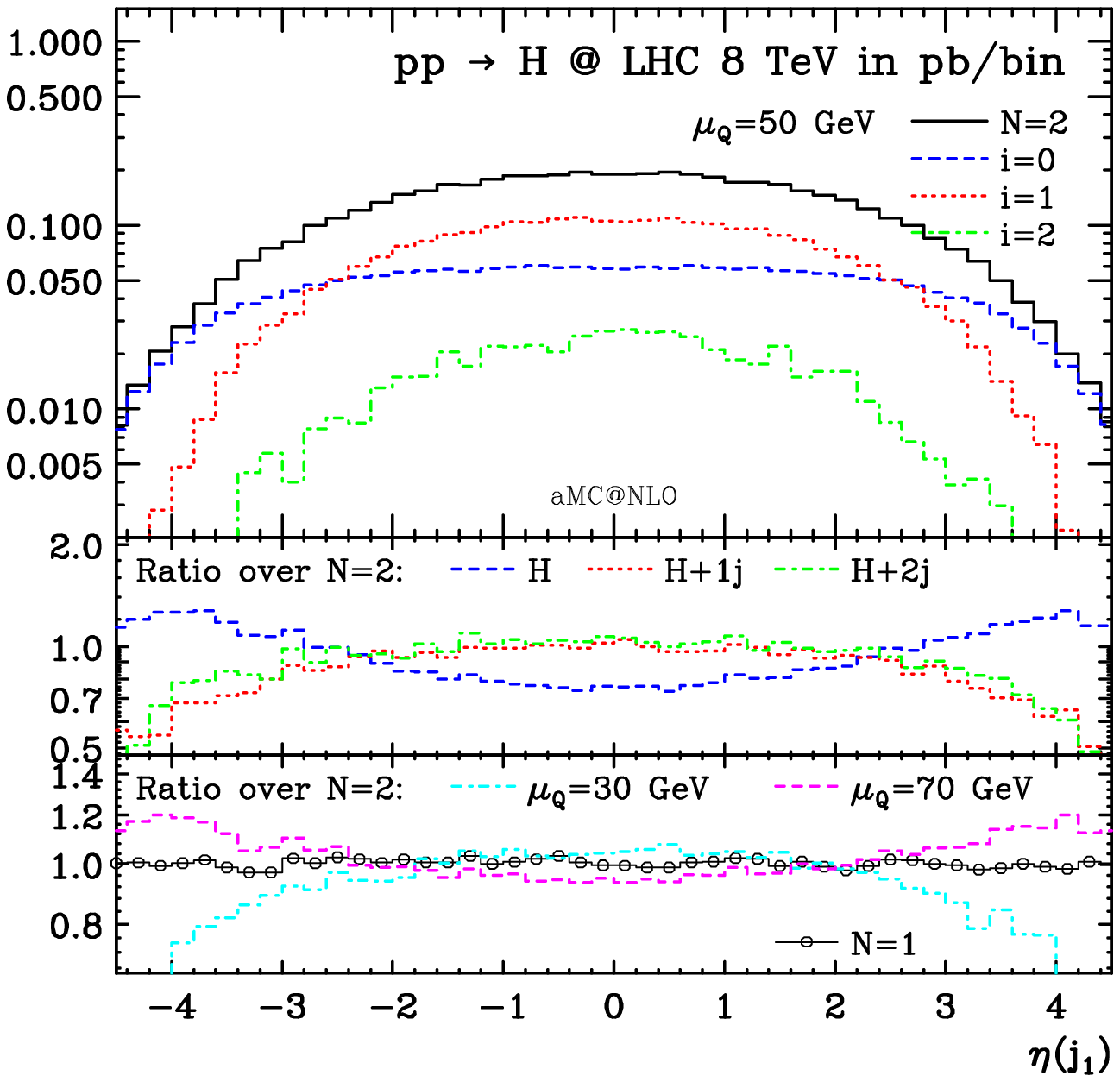, width=0.49\textwidth}
        \epsfig{file=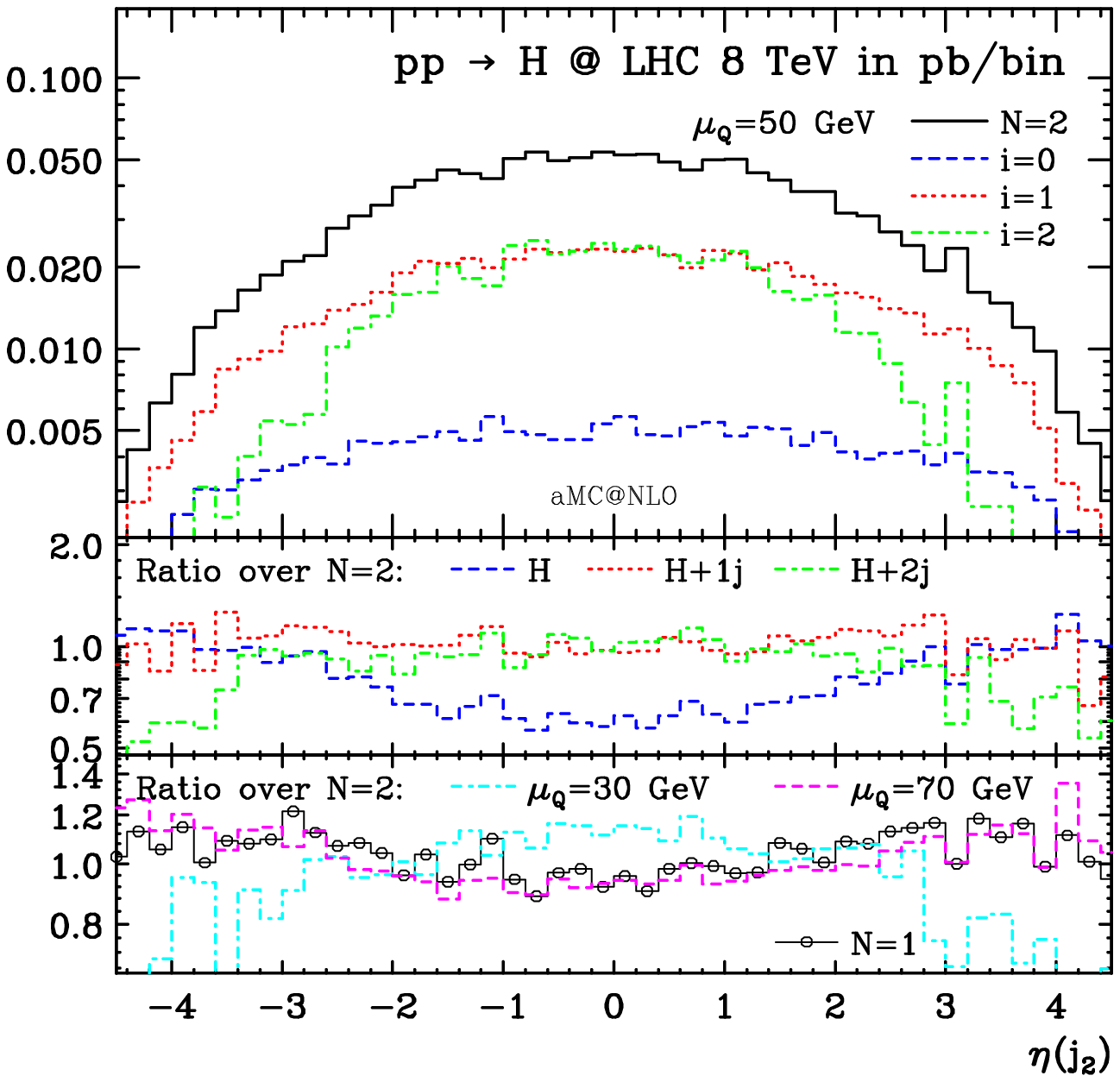, width=0.49\textwidth}
  \end{center}
  \begin{center}
        \epsfig{file=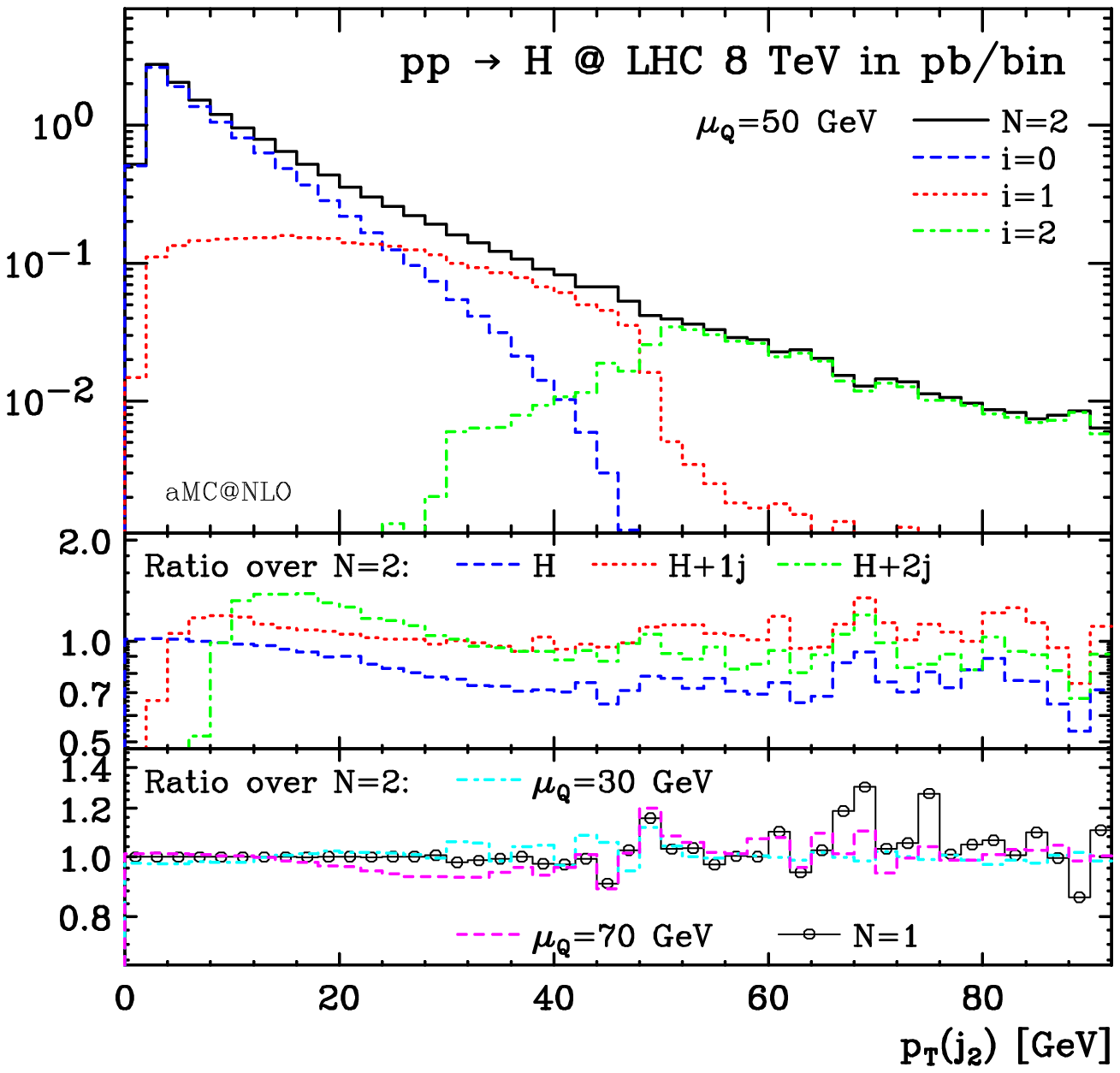, width=0.49\textwidth}
        \epsfig{file=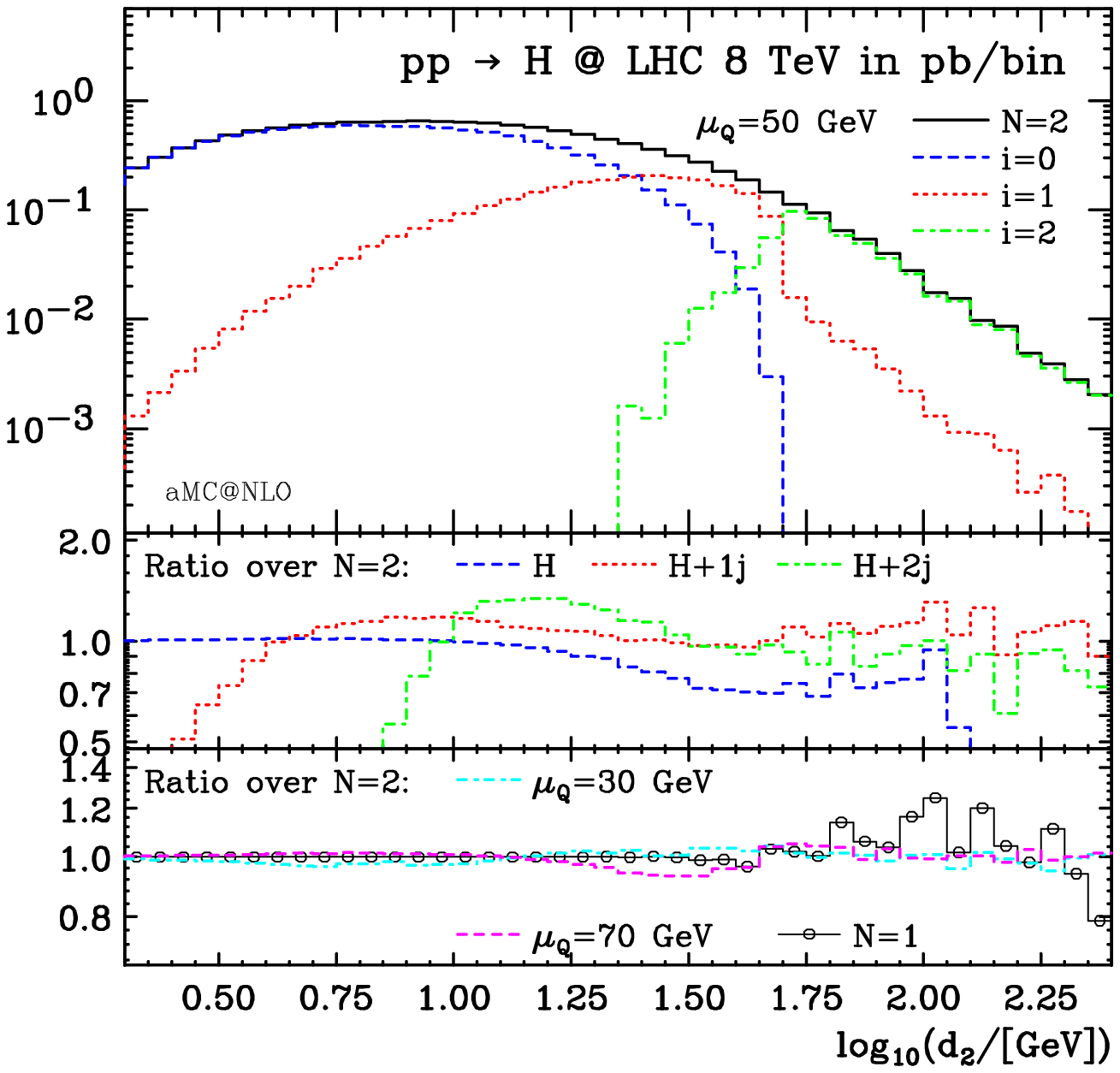, width=0.49\textwidth}
  \end{center}
  \vspace{-20pt}
  \caption{As in fig.~\protect\ref{fig:4}, with $N=2$.}
\label{fig:7}
\end{figure}
\begin{figure}[htb!]
  \begin{center}
        \epsfig{file=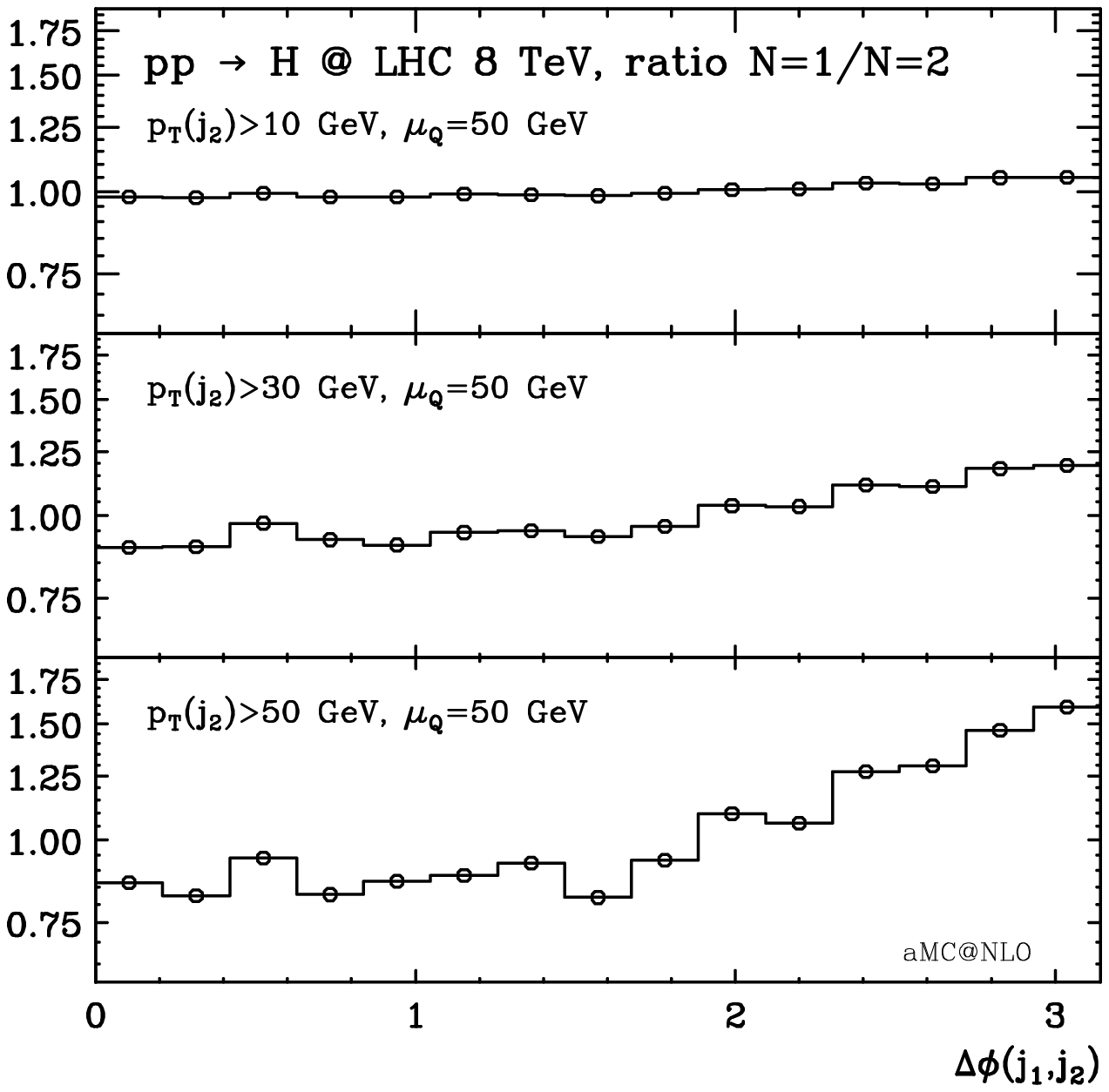, width=0.49\textwidth}
        \epsfig{file=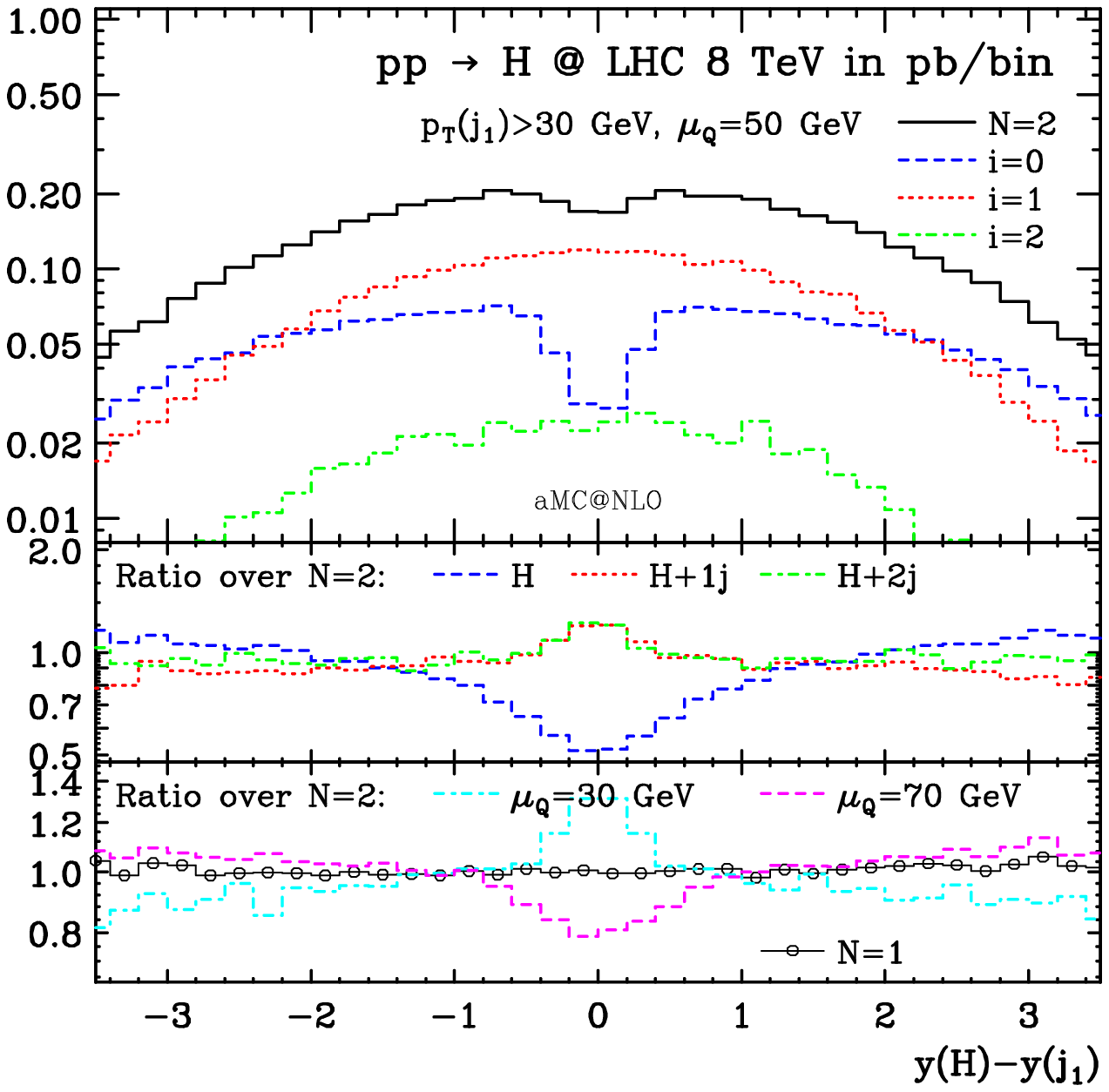, width=0.49\textwidth}
  \end{center}
  \begin{center}
        \epsfig{file=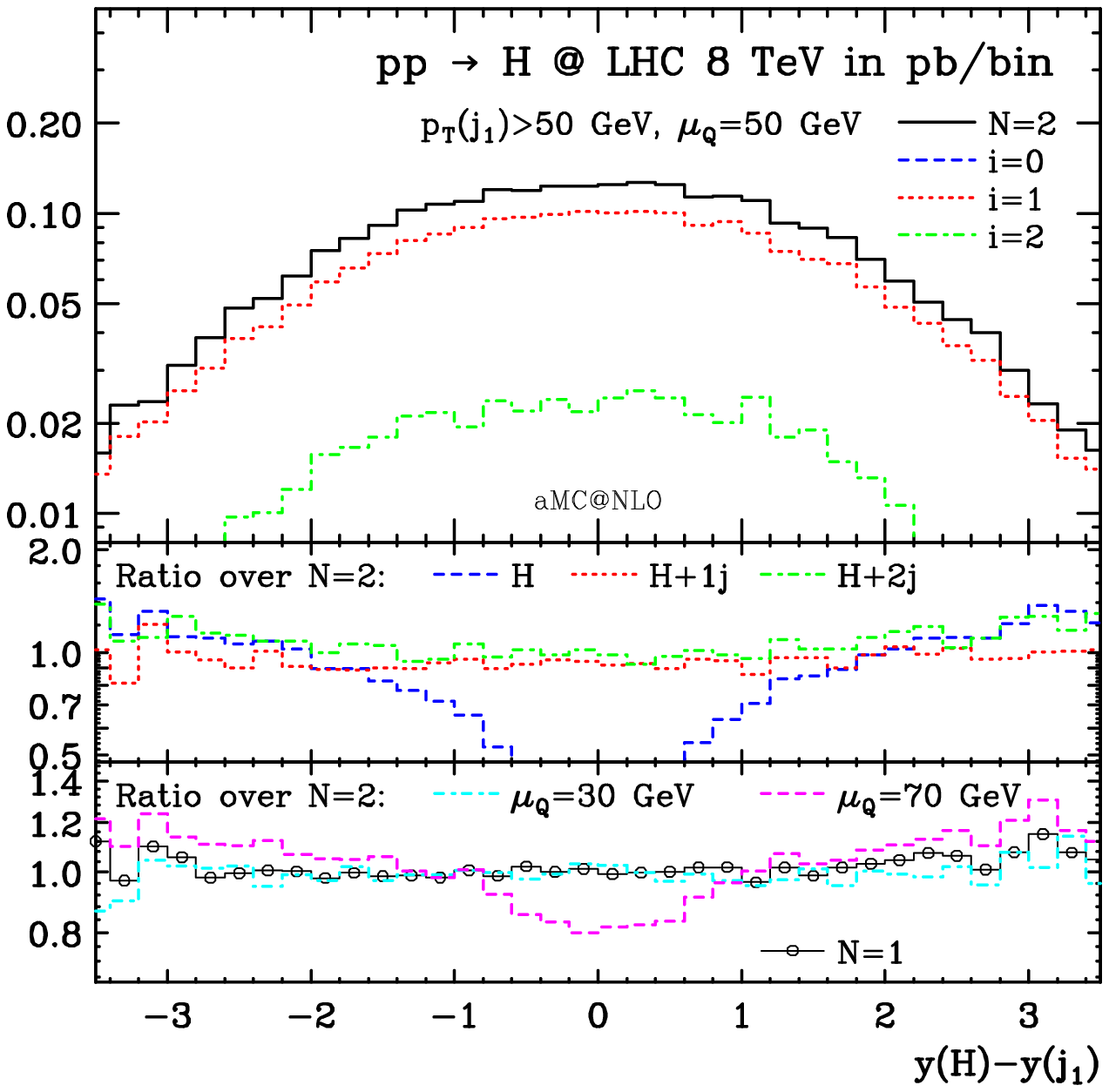, width=0.49\textwidth}
        \epsfig{file=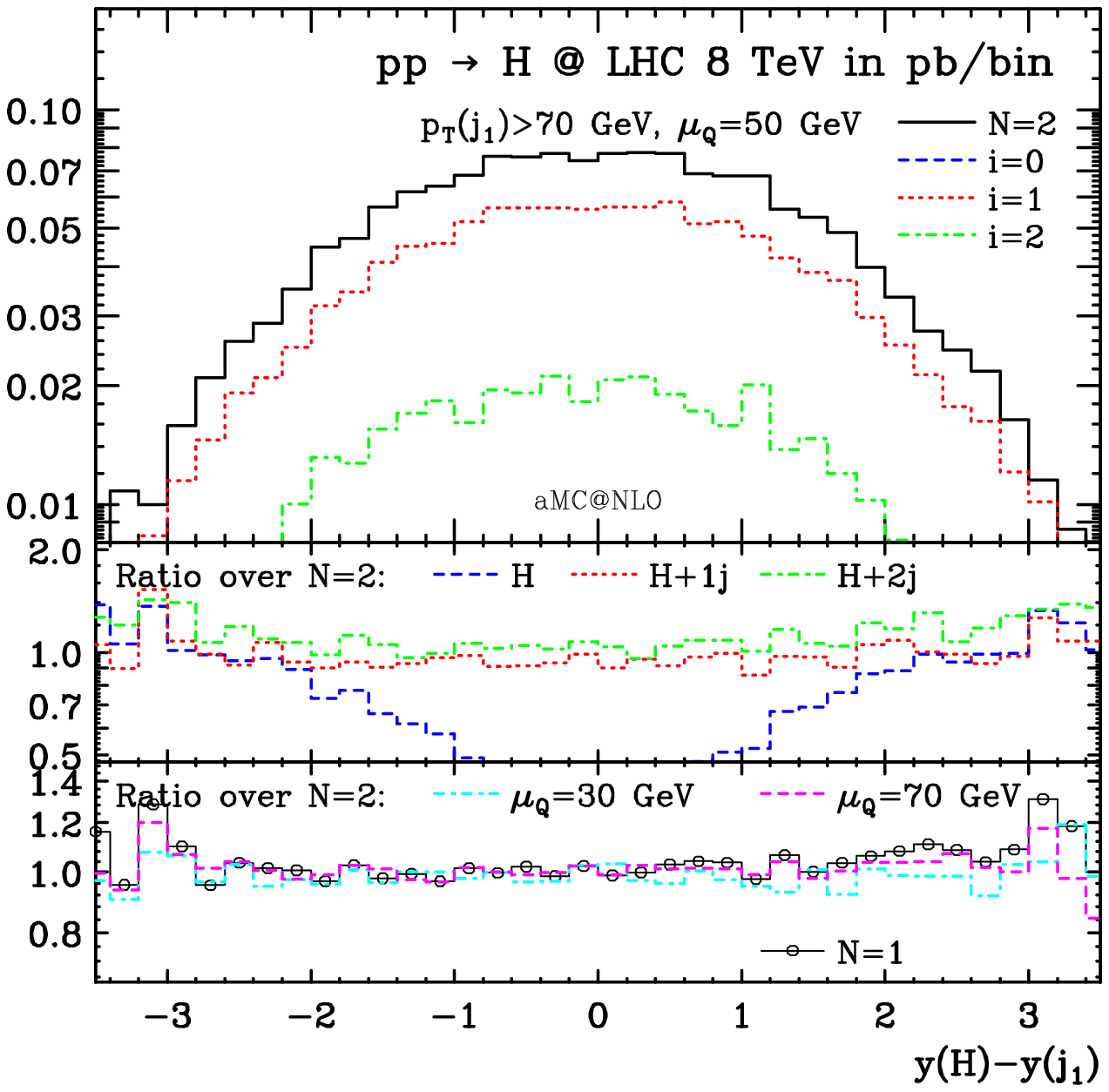, width=0.49\textwidth}
  \end{center}
  \vspace{-20pt}
  \caption{As in fig.~\protect\ref{fig:5}, with $N=2$. The $y(H)-y(j_1)$
    observable with $\pt(j_1)>10$~GeV of the upper left panel of
    fig.~\protect\ref{fig:5} has been replaced here by a plot presenting the 
    ratios of the $N=1$ results over the $N=2$
    results for the azimuthal separation between the two hardest jets, with
    three different $\pt$ cuts on the second-hardest jet.}
\label{fig:8}
\end{figure}
We finally turn to discussing the case of the $N=2$, sharp-$D$ function,
Sudakov-reweighted merging; that is, we increase the largest multiplicity
by one unit w.r.t.~what was done before. The settings are the same
as in the $N=1$ case, and figs.~\ref{fig:6}, \ref{fig:7}, and~\ref{fig:8}
are the analogues of figs.~\ref{fig:3}, \ref{fig:4}, and~\ref{fig:5}
respectively (with the exception of one panel in fig.~\ref{fig:8}).
The numerators of the ratios that appear in the upper insets
are the same as before for the $H+0j$ and $H+1j$ cases; that for
$H+2j$ is obviously specific to $N=2$. In the lower insets,
together with the ratios that allow one to assess the merging systematics,
we have plotted (as histograms overlaid with open circles) the ratios 
of the $N=1$ results over the $N=2$ ones, both for $\mu_Q=50$~GeV.
We have also recomputed the \alpgen\ predictions, by adding the 
$H+3$ parton sample, for consistency with $N=2$. The corresponding
results will not be shown in the plots, since these are already quite
busy, and there is no difference at all in the patterns discussed above,
except in a very few cases which we shall comment upon when appropriate.

The common feature of all but one of the observables presented
in figs.~\ref{fig:6}--\ref{fig:8} is that they are extremely close,
in both shape and normalization, to their $N=1$ counterparts
of figs.~\ref{fig:3}--\ref{fig:5}.
This is highly non-trivial, since the individual $i$-parton 
contributions are different in the two cases. The exception
is the pseudorapidity of the second-hardest jet (upper right
panel of fig.~\ref{fig:7}), which the inclusion of the $2$-parton 
sample turns into a more central distribution, as anticipated
in the discussion relevant to fig.~\ref{fig:4}, and brings it
very close to the \alpgen\ result obtained with the same $\mu_Q$.

The small impact of the increase of the largest multiplicity is also 
generally in agreement with what is found in \alpgen, where the inclusion of
the $H+3$ parton contribution changes the fully-inclusive rate by $+0.3$\%.
The effects on differential observables are also comparably small,
growing to only a few percent in the tail of $\pt$ distributions.
The exceptions are $\pt(j_2)$ and $\pt(j_3)$, which are significantly
harder in \alpgen\ after the inclusion of the $H+3$ parton sample
(a 20\% and 30\% effect respectively). However, this may be related
to the fact that such an inclusion also leads to larger kinks 
(of which we see no trace in the NLO-merged results) at
$\pt\simeq\mu_Q$ for these two observables; thus, the hardening may be
partly an artifact of the LO-merging procedure; in order to further
this point, it will be useful to study the merging systematics 
affecting $\pt(j_2)$ and $\pt(j_3)$ in \alpgen.
It is obvious that the relative stability of the NLO-merged results
against the inclusion of higher multiplicities is observable-dependent.
$\eta(j_2)$ gives a counterexample, but more spectacular ones can
be found by looking at correlations. As an example of this, we
show in the upper left panel of fig.~\ref{fig:8} the azimuthal
distance $\Delta\phi(j_1,j_2)$ between the two hardest jets, 
for three different cuts $\pt(j_2)>\ptcut$ on the subleading 
jet\footnote{For reasons of space, this plot has replaced that of
the rapidity difference with $\pt(j_1)>10$~GeV. The latter is in
fact not particularly illuminating since, exactly as its analogues
shown in fig.~\ref{fig:8} for larger $\pt$ cuts, the merged $N=2$ result
is identical to the $N=1$ one shown in fig.~\ref{fig:5}.}. The plots 
are presented in the form of ratios of the $N=1$ results over the
$N=2$ ones. When $\ptcut$ is well below the matching scale (10~GeV
vs 50~GeV, see the upper inset), $\Delta\phi(j_1,j_2)$ is MC-dominated,
and the inclusion of the $2$-parton sample is irrelevant. However,
matrix-element effects start to be felt when $\ptcut$ is increased
(middle inset), to become quite important at $\ptcut=50$~GeV (lower inset).
This clearly shows the impact of the largest matrix-element multiplicity 
on $\Delta\phi(j_1,j_2)$, and demonstrates that for the predictions of 
observables exclusive in up to $J$ jets is best to have $N\ge J$.
We conclude this section by remarking that in general the merging 
systematics is smaller (although for  some observables marginally so) 
when $N=2$ than in the $N=1$ case, which is exactly what one expects in 
a ``converging'' procedure, where the $N=2$ results are better 
(i.e., more accurate) than the $N=1$ ones.

\subsection{$e^+\nu_e$ production\label{sec:resW}}
In this section, we present the results for $e^+\nu_e$ production,
limiting ourselves to the case $N=1$, sharp-$D$ function, and
Sudakov-reweighted merging. We have treated the electron as massless,
and have set $m_W=80.419$~GeV, $\Gamma_W=2.0467$~GeV, and
\beq
\muME=\sqrt{M(e\nu)^2+\pt(e\nu)^2}\,,
\label{muMEW}
\eeq
where $M(e\nu)$ and $\pt(e\nu)$ are the invariant mass and transverse 
momentum of the lepton pair respectively; the former is also required 
to obey the following constraints:
\beq
m_W-30\,\Gamma_W\le M(e\nu)\le m_W+30\,\Gamma_W\,.
\eeq
The one-loop matrix elements are computed with \madloop~\cite{Hirschi:2011pa}.
We have considered three values for the matching scale, $\mu_Q=35$
(our default), $20$, and $50$~GeV. As a shorthand notation, we
may denote by $W$ the $e^+\nu_e$ pair in the following and in the
labels of the plots.
\begin{figure}[htb!]
  \begin{center}
        \epsfig{file=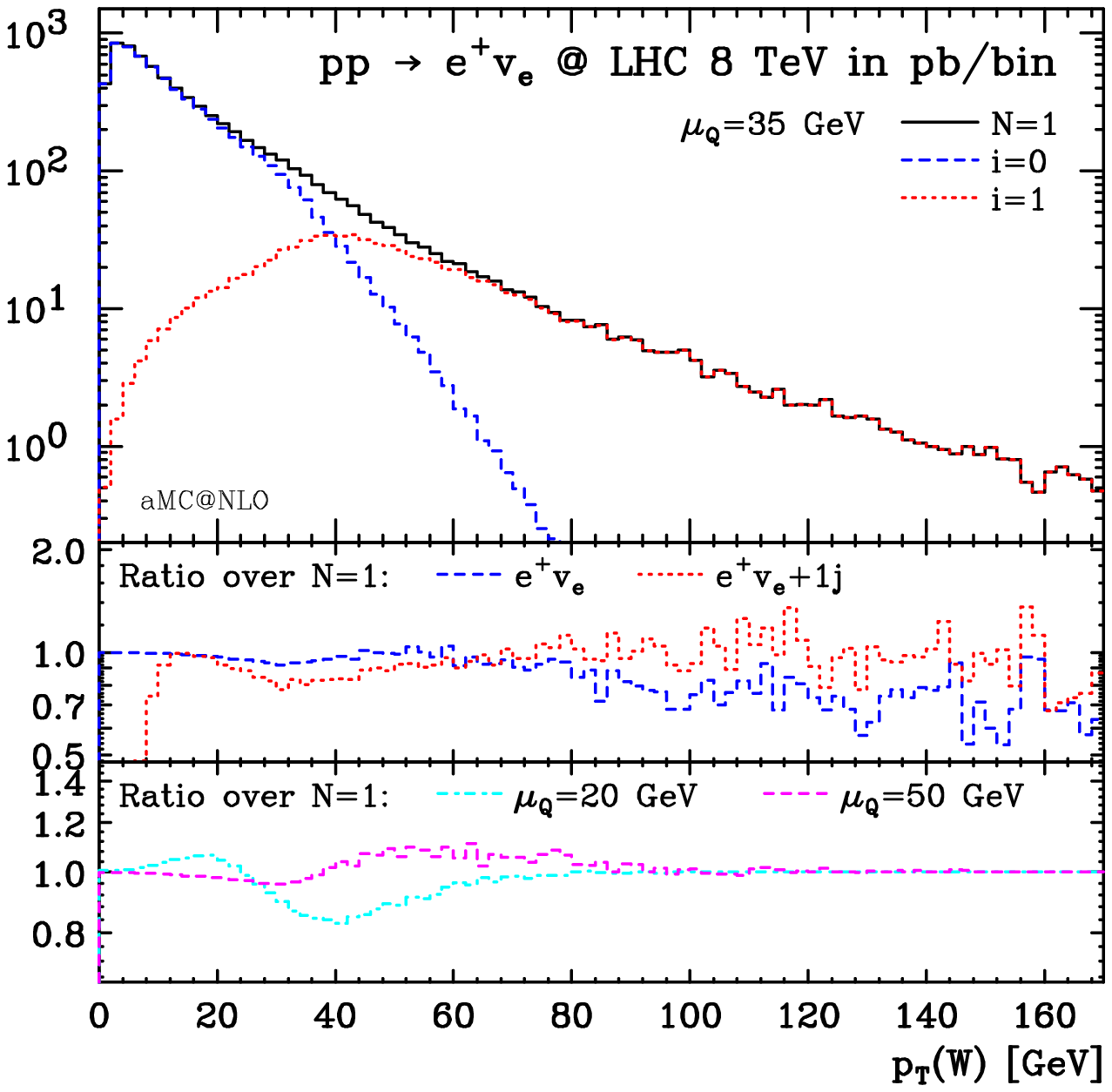, width=0.48\textwidth}
        \epsfig{file=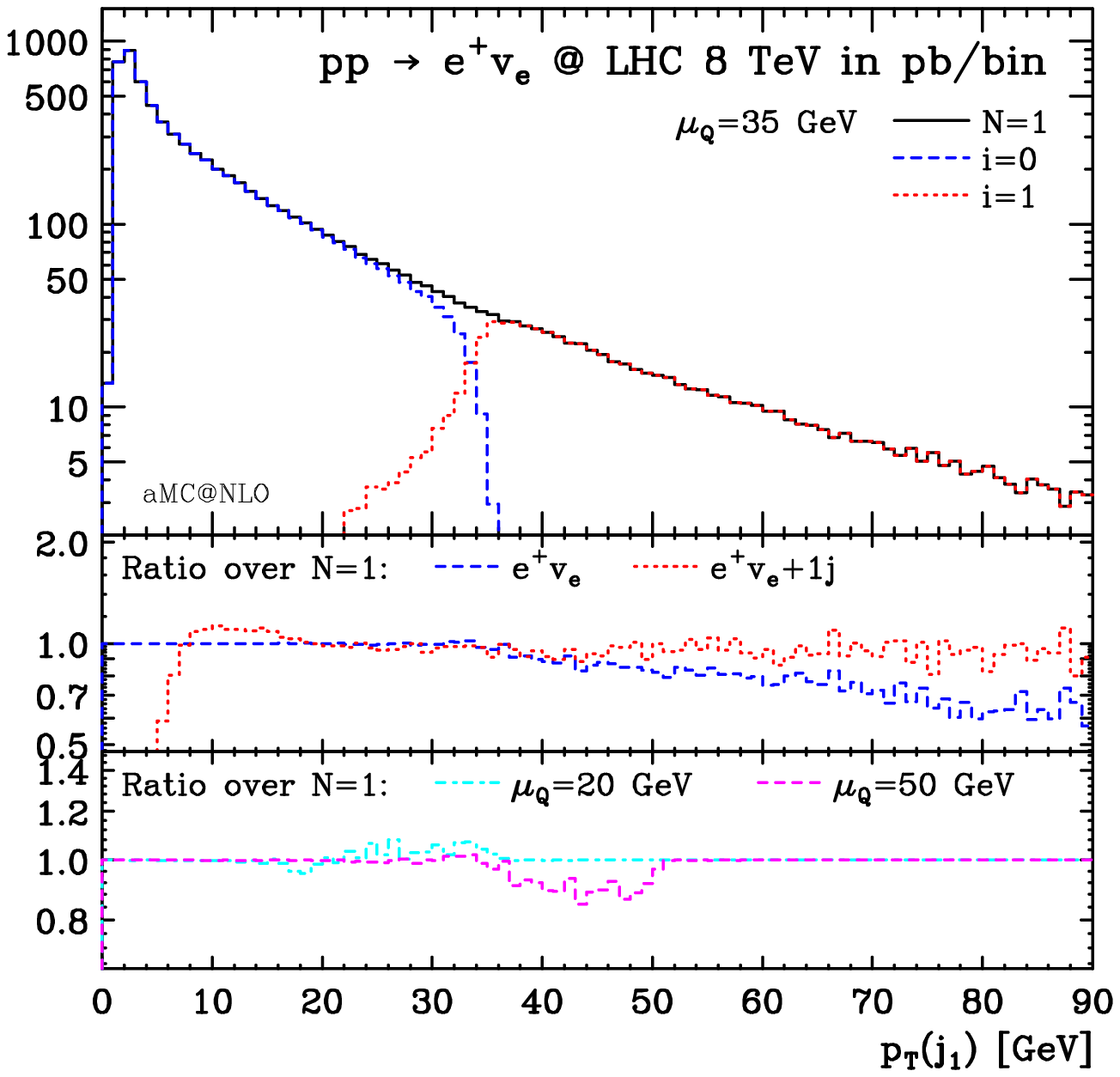, width=0.49\textwidth}
  \end{center}
  \begin{center}
        \epsfig{file=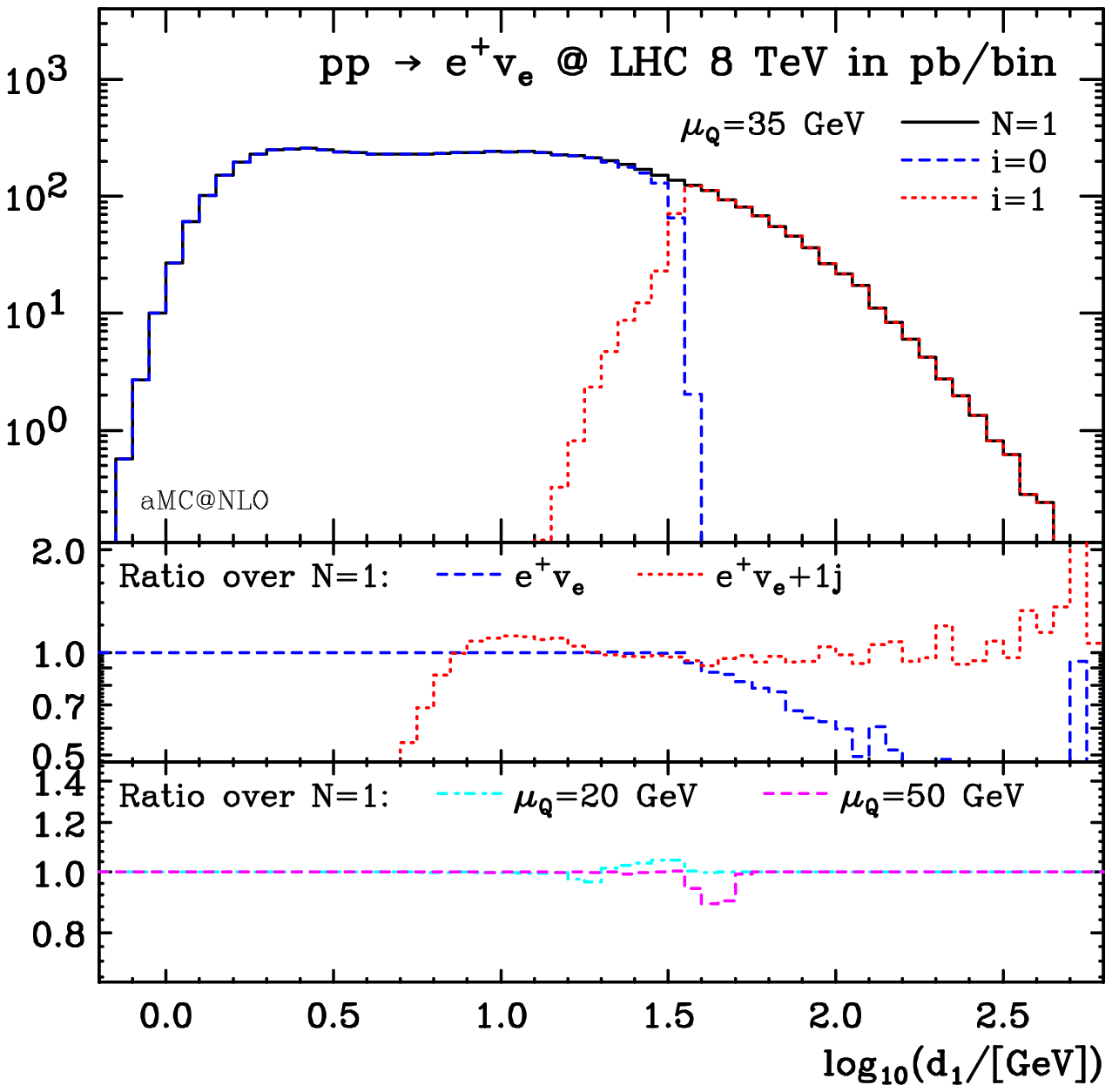, width=0.48\textwidth}
        \epsfig{file=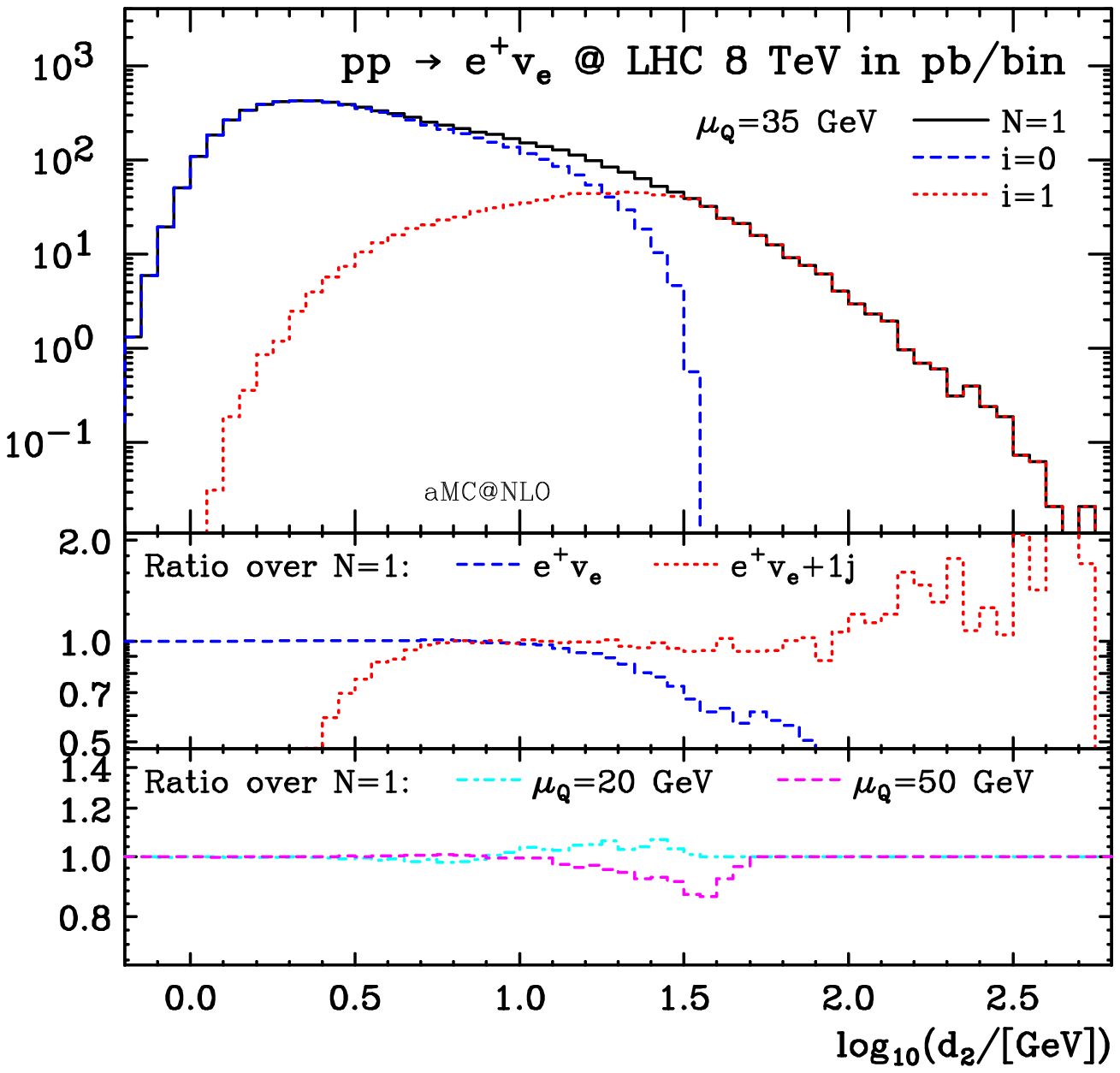, width=0.49\textwidth}
  \end{center}
  \vspace{-20pt}
  \caption{$e^+\nu_e$ production, with $N=1$, sharp $D$ function, and 
    Sudakov reweighting. Pair $\pt$ (upper left), hardest-jet $\pt$ 
    (upper right), $d_1$ (lower left), and $d_2$
    (lower right) are shown.}
\label{fig:9}
\end{figure}
\begin{figure}[htb!]
  \begin{center}
        \epsfig{file=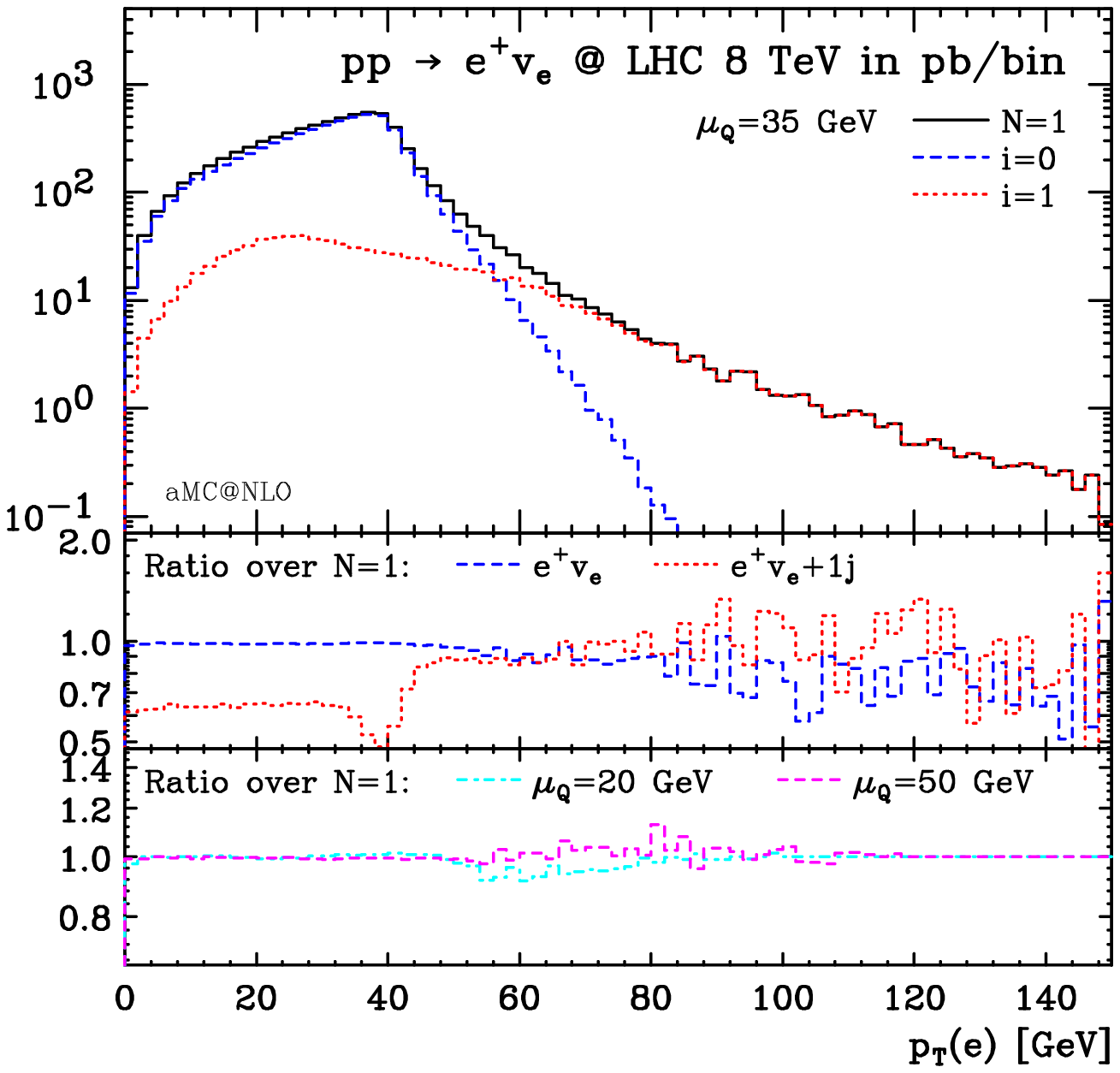, width=0.49\textwidth}
        \epsfig{file=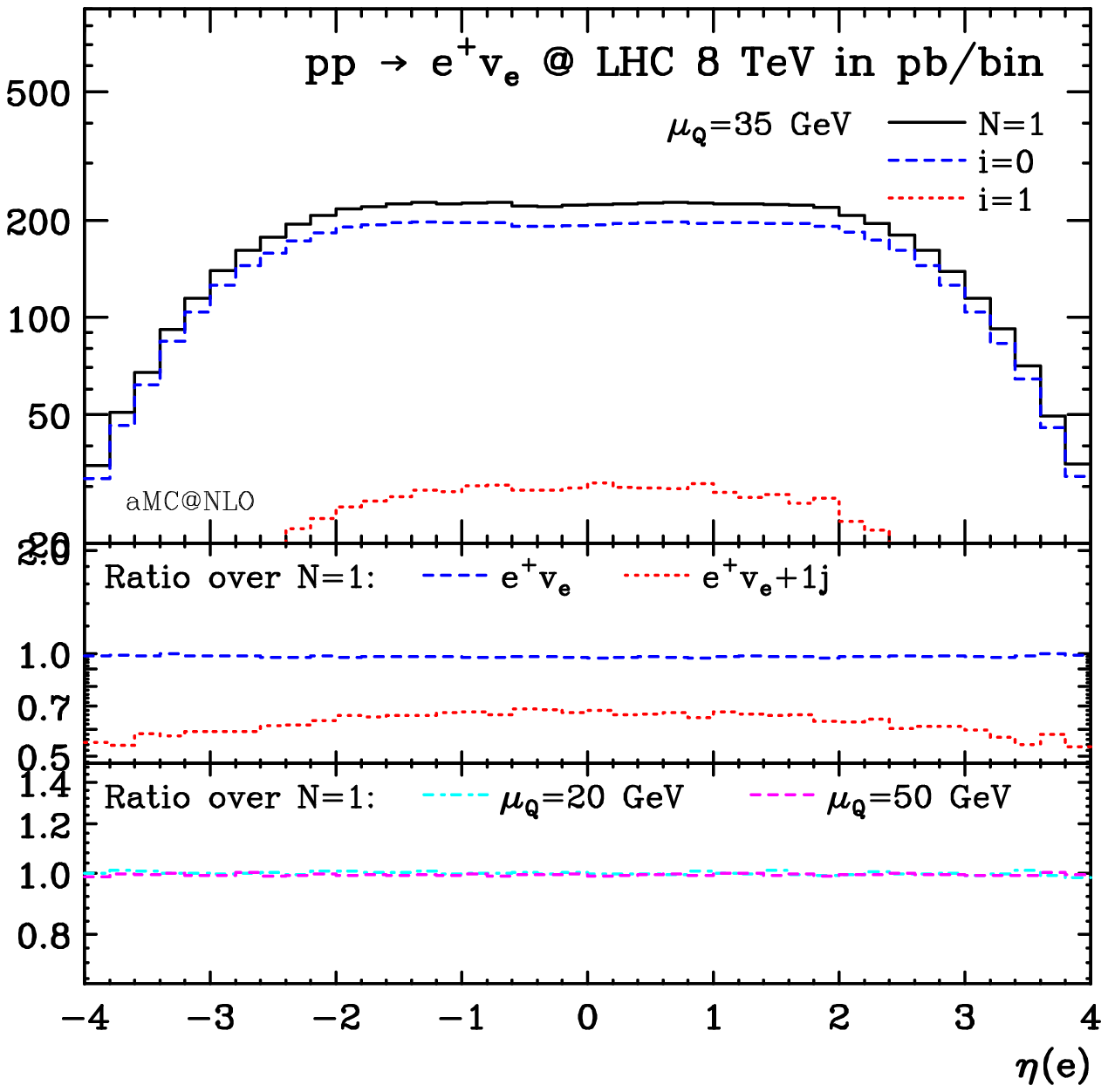, width=0.48\textwidth}
  \end{center}
  \begin{center}
        \epsfig{file=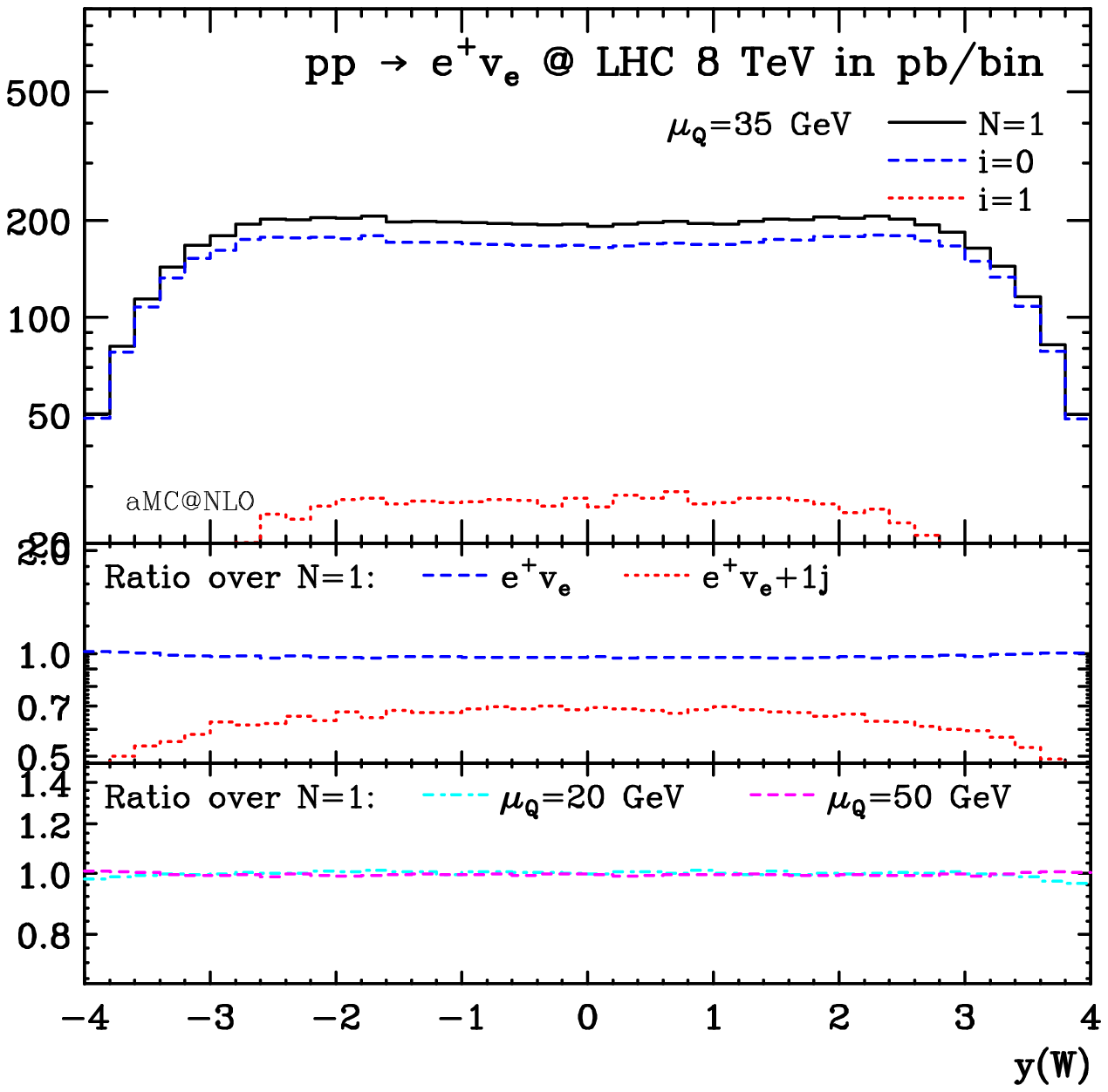, width=0.49\textwidth}
        \epsfig{file=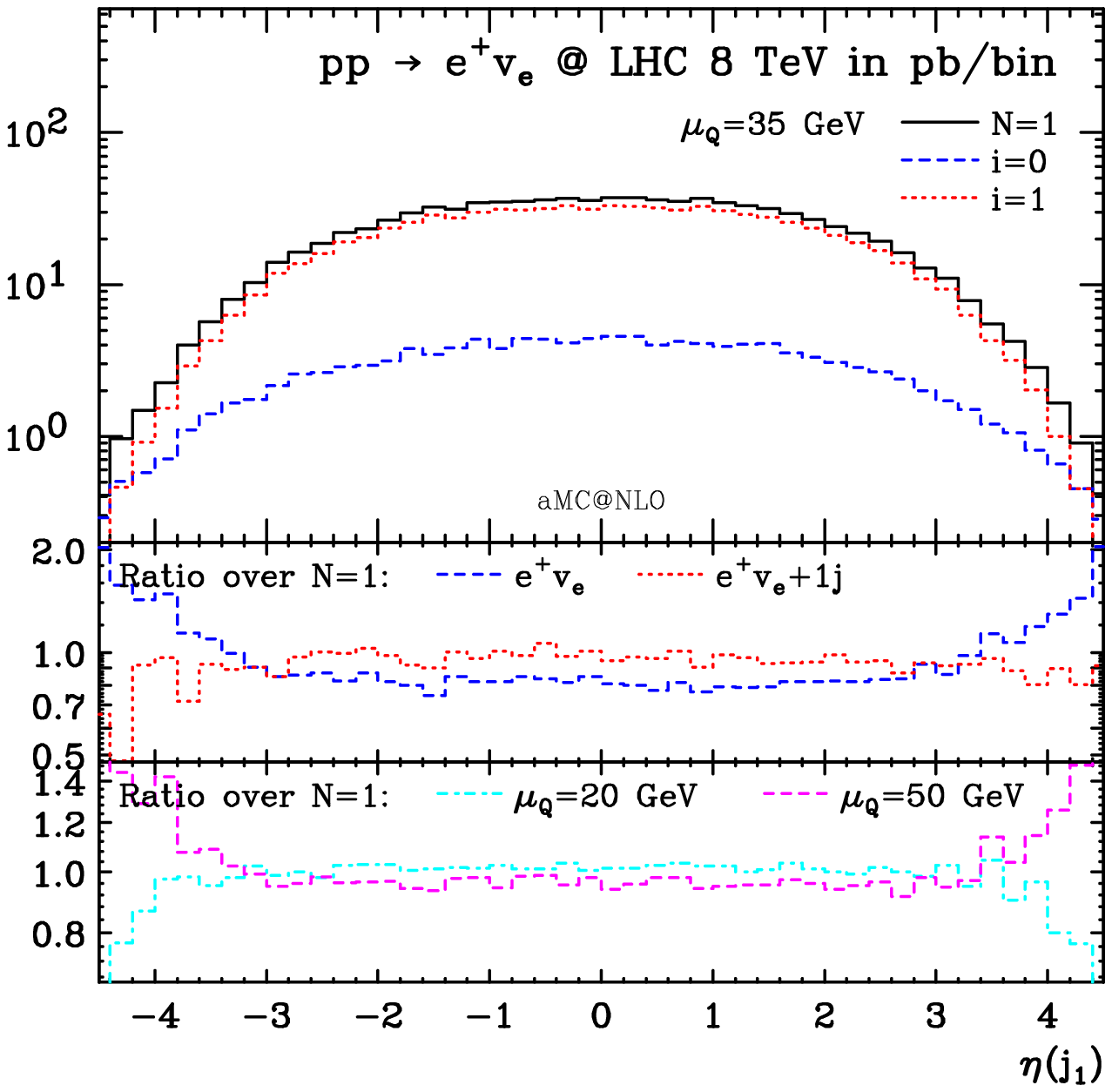, width=0.49\textwidth}
  \end{center}
  \vspace{-20pt}
  \caption{As in fig.~\protect\ref{fig:9}, for the electron $\pt$
    (upper left) and pseudorapidity (upper right), the pair rapidity
    (lower left), and the hardest-jet pseudorapidity (lower right).
    The latter observable is obtained with a $\pt(j_1)>30$~GeV cut.}
\label{fig:10}
\end{figure}
Our predictions are presented in figs.~\ref{fig:9}--\ref{fig:11},
where we have used the same layout and patterns as e.g.~in fig.~\ref{fig:3}
(except for the \alpgen\ results, which we did not generate for this
process). As in the case of Higgs production, the merged results
are fairly smooth, and affected by merging systematics which are
at most a 15\% effect, but typically much smaller than that.
Similarly to what happens for Higgs, some of the $\pt$ or $d$
distributions display wiggles at the transition between regions
dominated by different $i$-parton samples. While this is just a 
manifestation of the merging systematics, in phenomenology-oriented
applications one may consider using a smooth $D$ function, as was
already suggested in sect.~\ref{sec:resH}. As is expected, the inclusion
of the $1$-parton sample induces a hardening in the tails of $\pt$
distributions w.r.t.~the standalone $W+0j$ results (see the dashed blue
histograms in the upper insets), while affecting the fully-inclusive
rates only in a marginal manner.
\begin{figure}[htb!]
  \begin{center}
        \epsfig{file=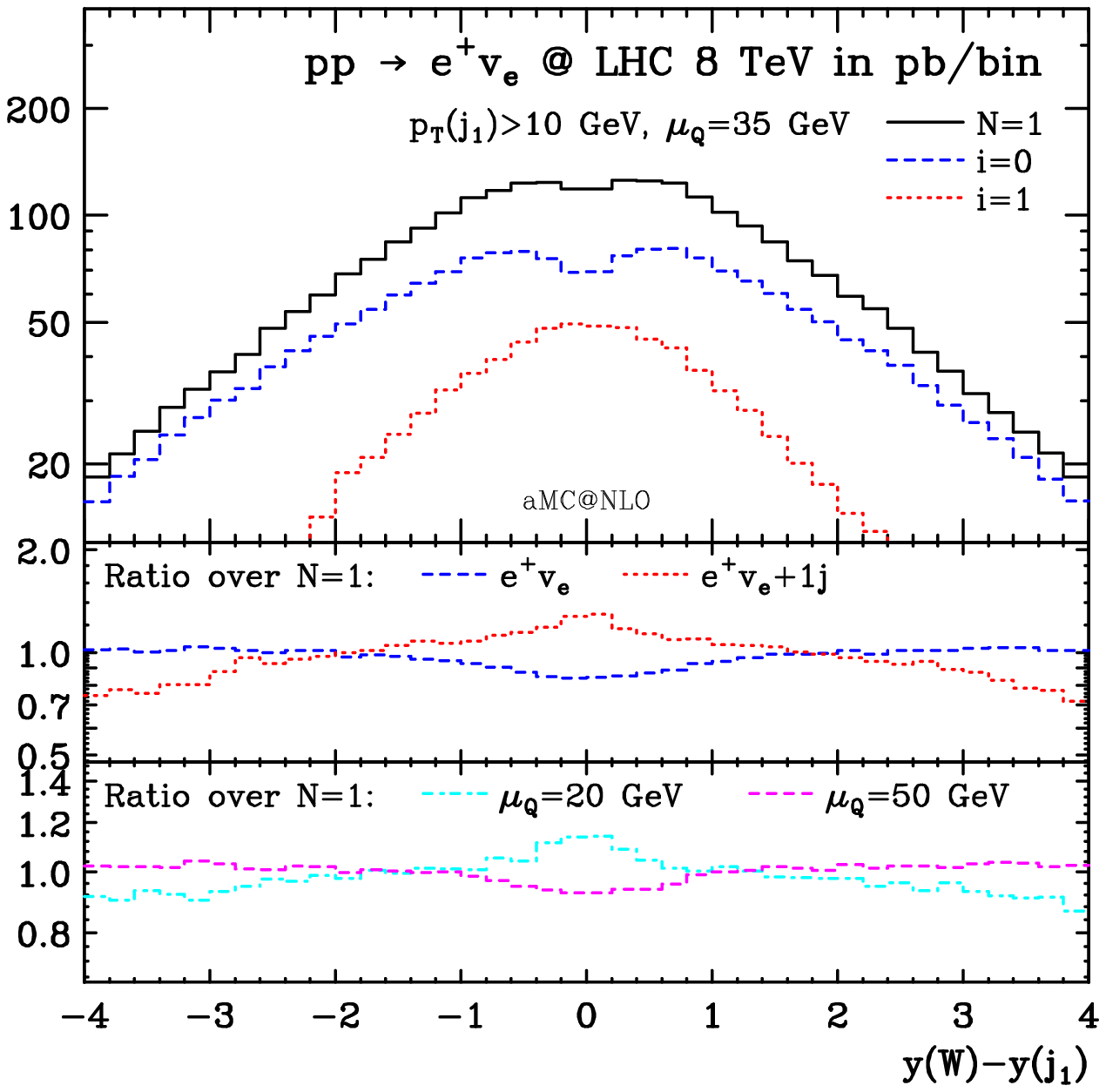, width=0.49\textwidth}
        \epsfig{file=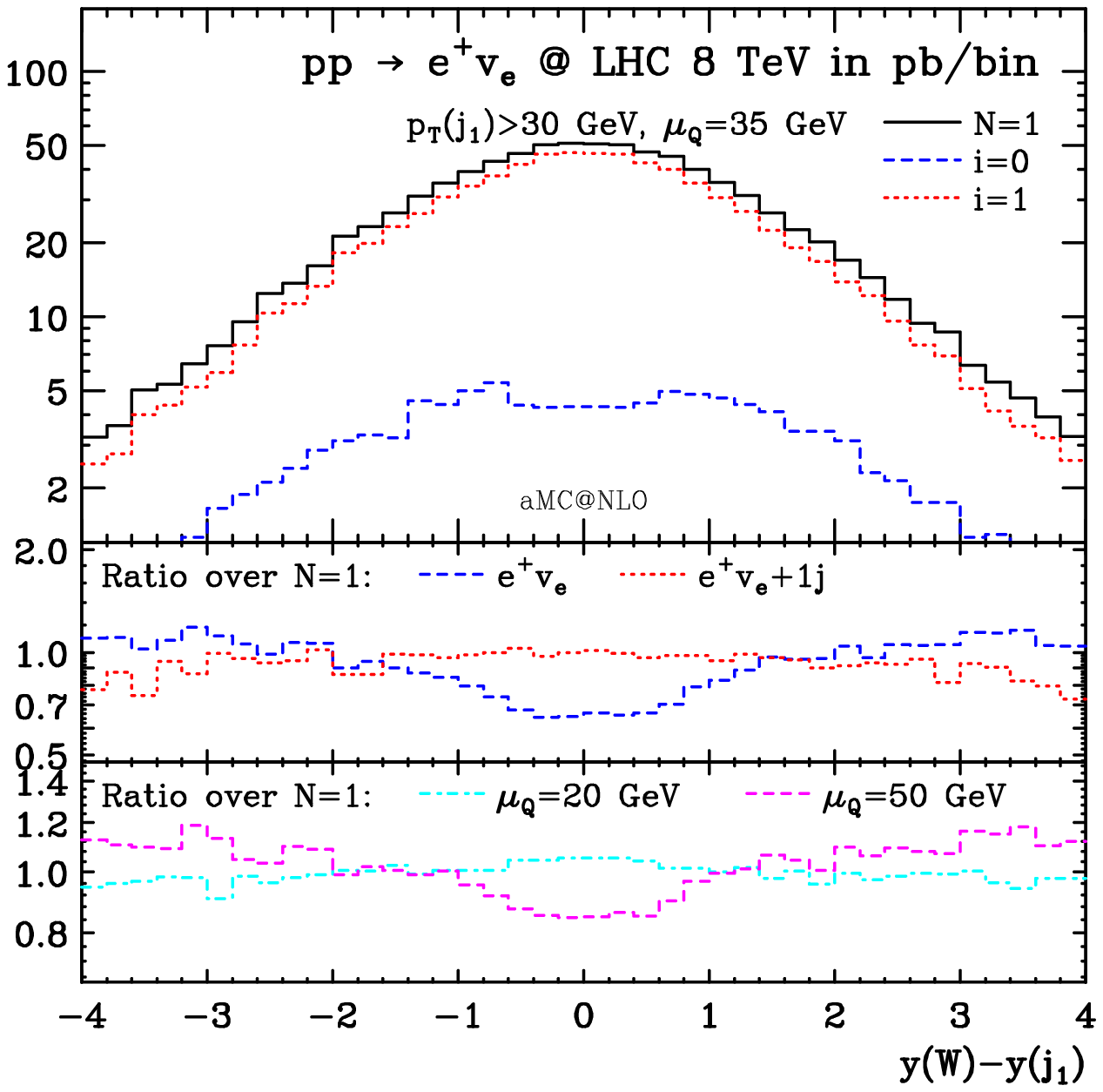, width=0.49\textwidth}
  \end{center}
  \begin{center}
        \epsfig{file=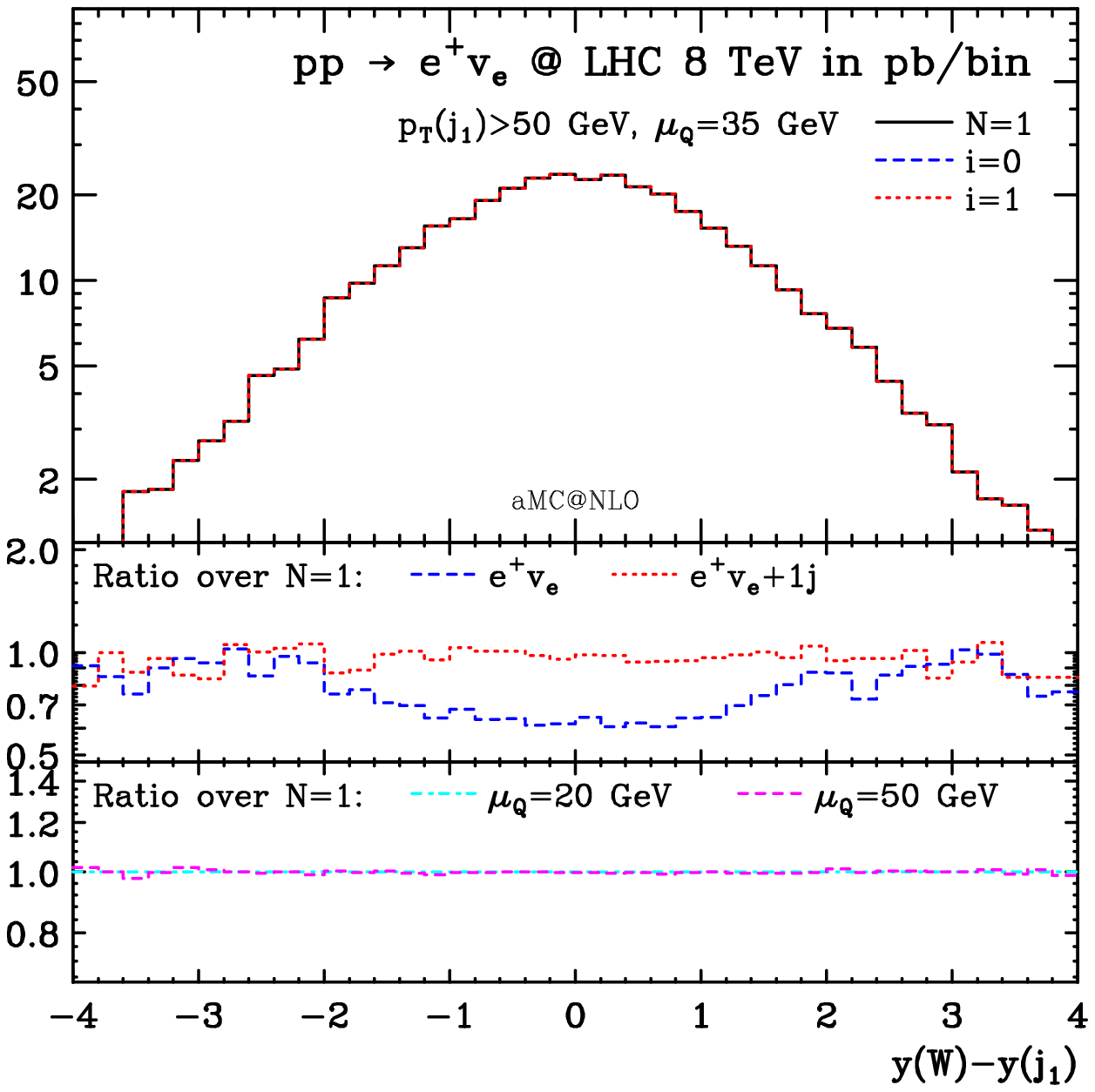, width=0.485\textwidth}
        \epsfig{file=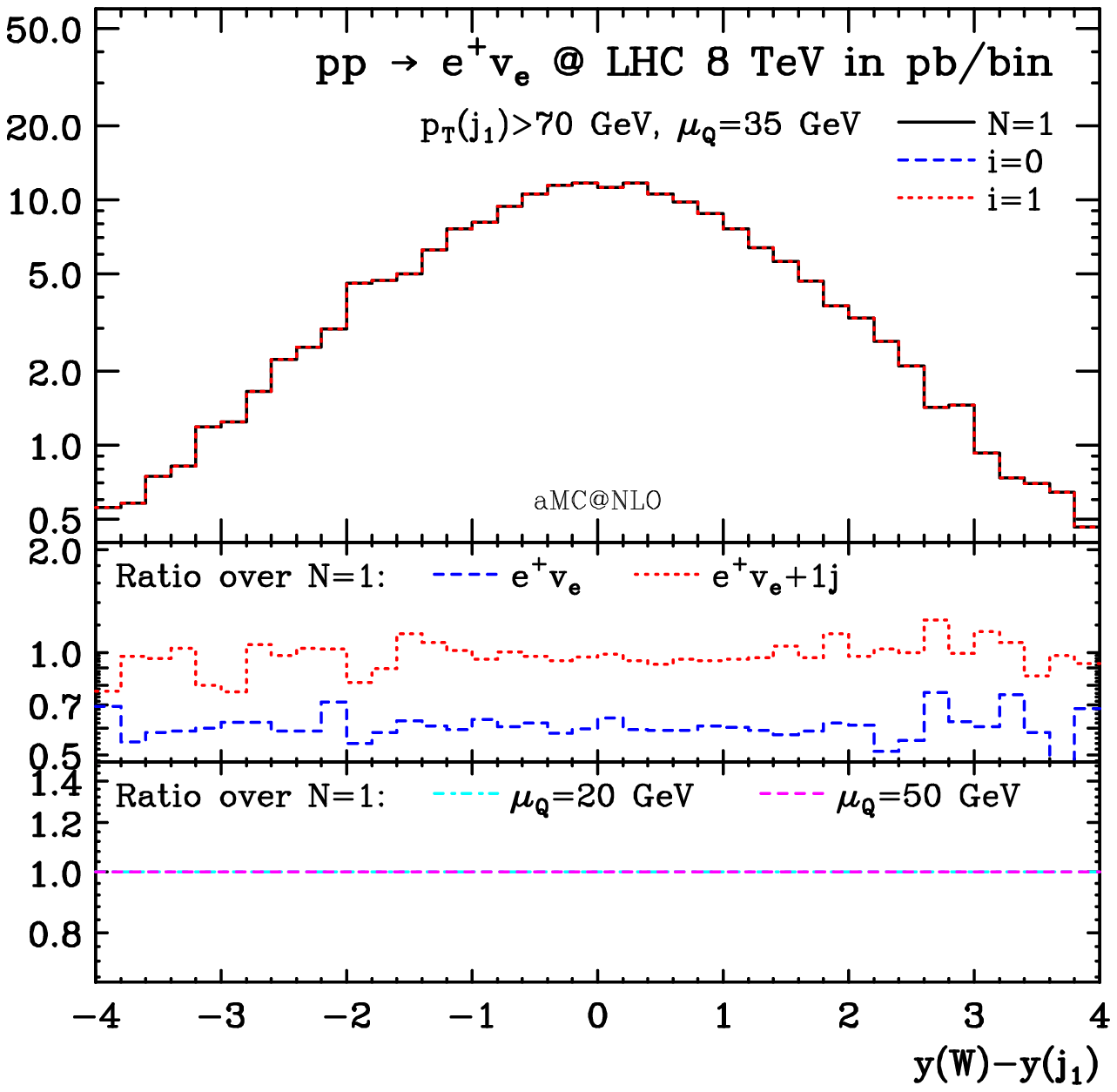, width=0.49\textwidth}
  \end{center}
  \vspace{-20pt}
  \caption{As in fig.~\protect\ref{fig:9}, for the difference in rapidity
    between the $e^+\nu_e$ pair and the hardest jet, for four different 
    $\pt$ cuts on the latter.}
\label{fig:11}
\end{figure}

In fig.~\ref{fig:11} we show the rapidity difference between the
lepton pair and the hardest jet, for four different $\pt$ cuts on 
the latter. This is in fully analogy with the case of fig.~\ref{fig:5},
and it is immediate to see that the general discussion given there applies 
to $e^+\nu_e$ production as well. In particular, the pattern of the 
presence or absence of the dip is exactly the same, while the
specific behaviour at a given $\ptcut$ is different because of
the differences between the two processes (i.e., $m_W$ vs $m_H$
and $q\bar{q}$ vs $gg$). We also point out that fig.~\ref{fig:11} 
can be quite directly compared with figs.~5 and~6 of 
ref.~\cite{Torrielli:2010aw}, where the same observable is computed
with standalone MC@NLO matched with $Q^2$-ordered \PYs; this underlines 
again the MC-dependence of these distributions for small $\ptcut$.

\subsection{$t\tb$ production\label{sec:restt}}
We conclude this phenomenology section by presenting our predictions 
for $t\tb$ production, again limiting ourselves to the case $N=1$, 
sharp $D$ function, and Sudakov-reweighted merging. The top quarks 
are produced on the mass shell; they are decayed leptonically in the 
shower phase, in order to limit the contamination of the hadronic 
activity of the events. Furthermore, the $b$ quarks emerging from
the top decays are not included in the jets whose distributions we 
present below. We have set $m_t=172.5$~GeV, and
\beq
\muME=\max\left(\mt(t),\mt(\tb)\right)\,,
\label{muMEtt}
\eeq
where we have denoted by $\mt(X)$ the transverse mass of $X$.
\begin{figure}[htb!]
  \begin{center}
        \epsfig{file=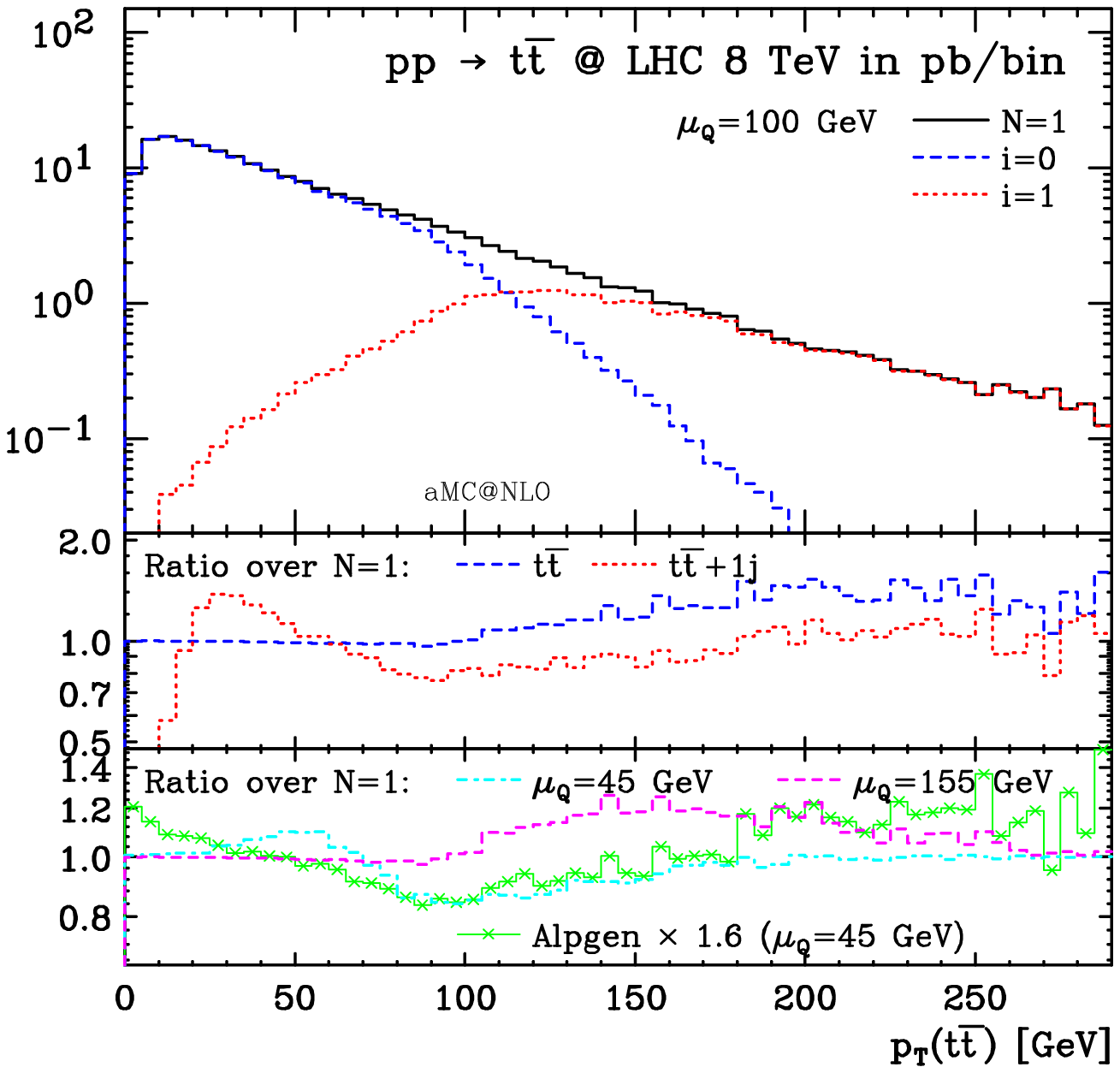, width=0.49\textwidth}
        \epsfig{file=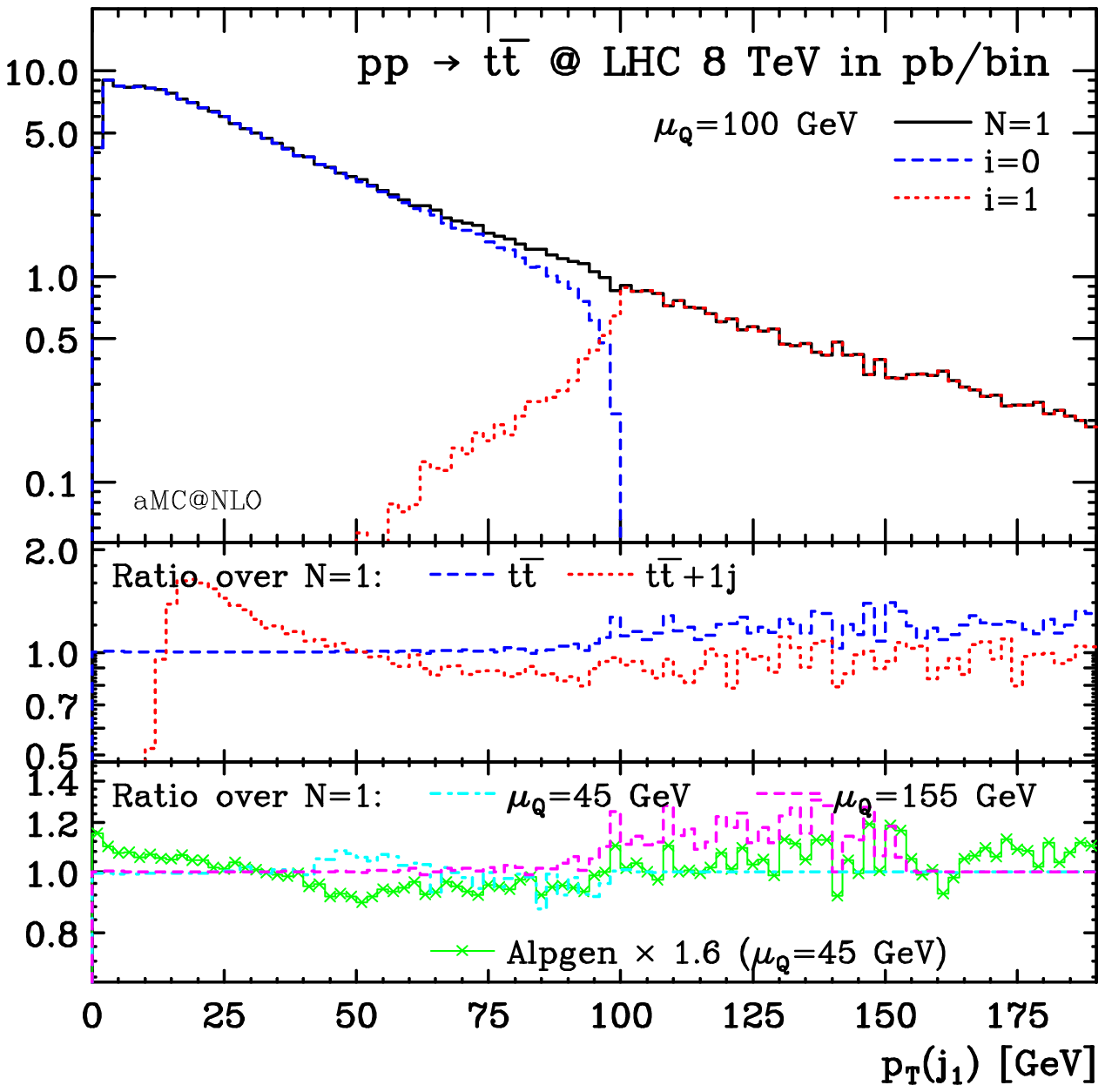, width=0.475\textwidth}
  \end{center}
  \begin{center}
        \epsfig{file=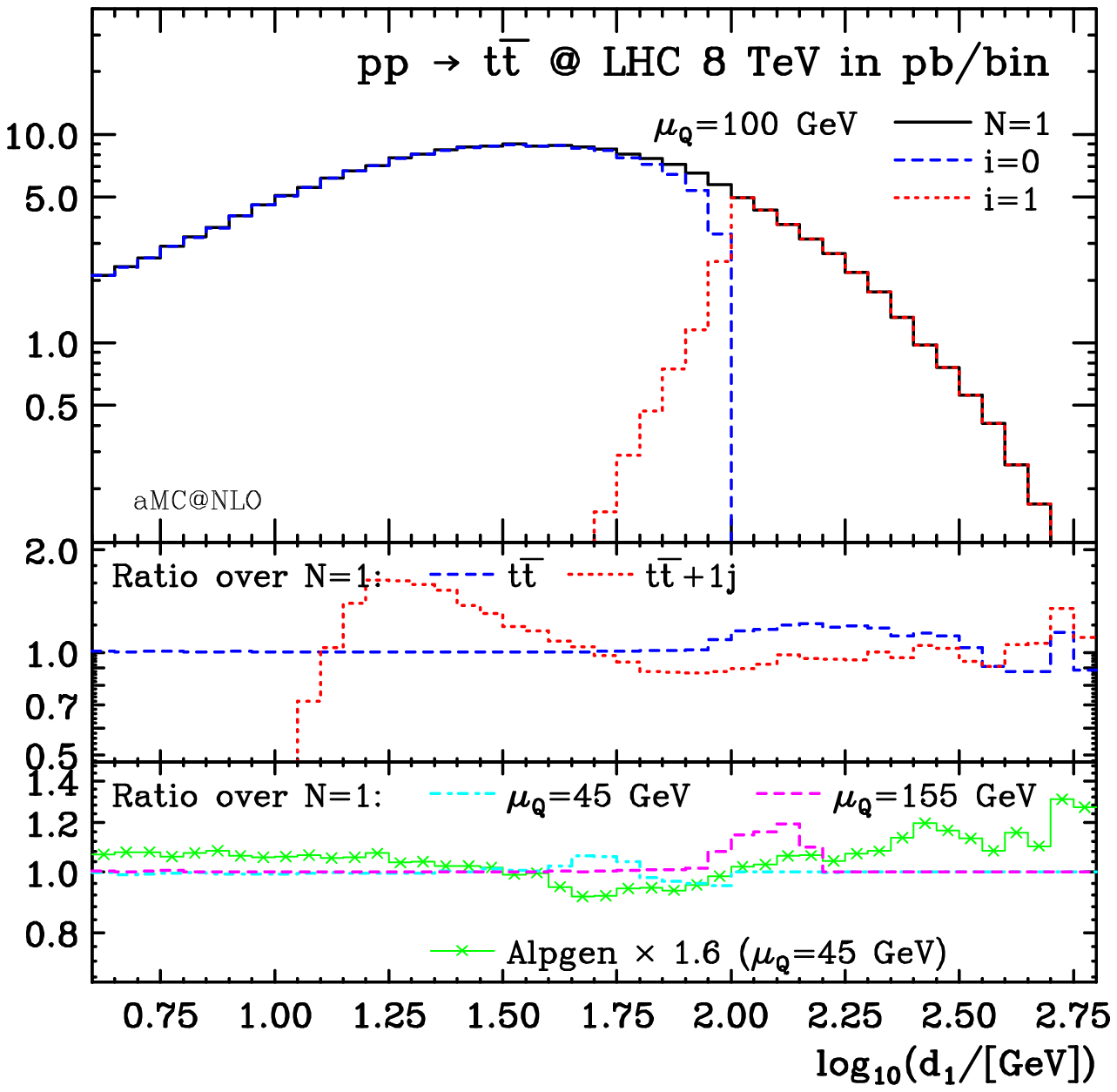, width=0.475\textwidth}
        \epsfig{file=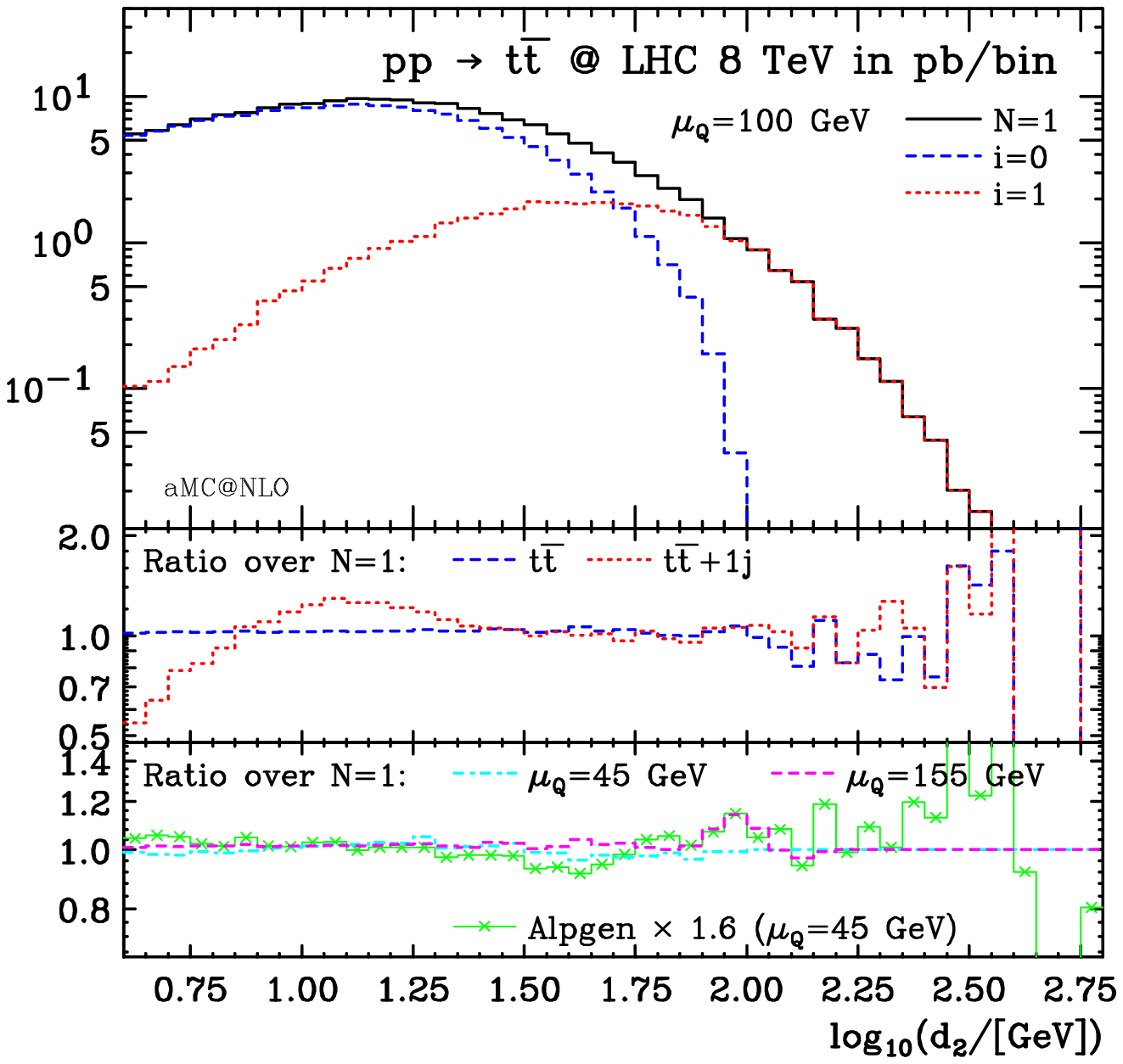, width=0.49\textwidth}
  \end{center}
  \vspace{-20pt}
  \caption{$t\tb$ production, with $N=1$, sharp $D$ function, and 
    Sudakov reweighting. Pair $\pt$ (upper left), hardest-jet $\pt$ 
    (upper right), $d_1$ (lower left), and $d_2$
    (lower right) are shown.}
\label{fig:12}
\end{figure}
\begin{figure}[htb!]
  \begin{center}
        \epsfig{file=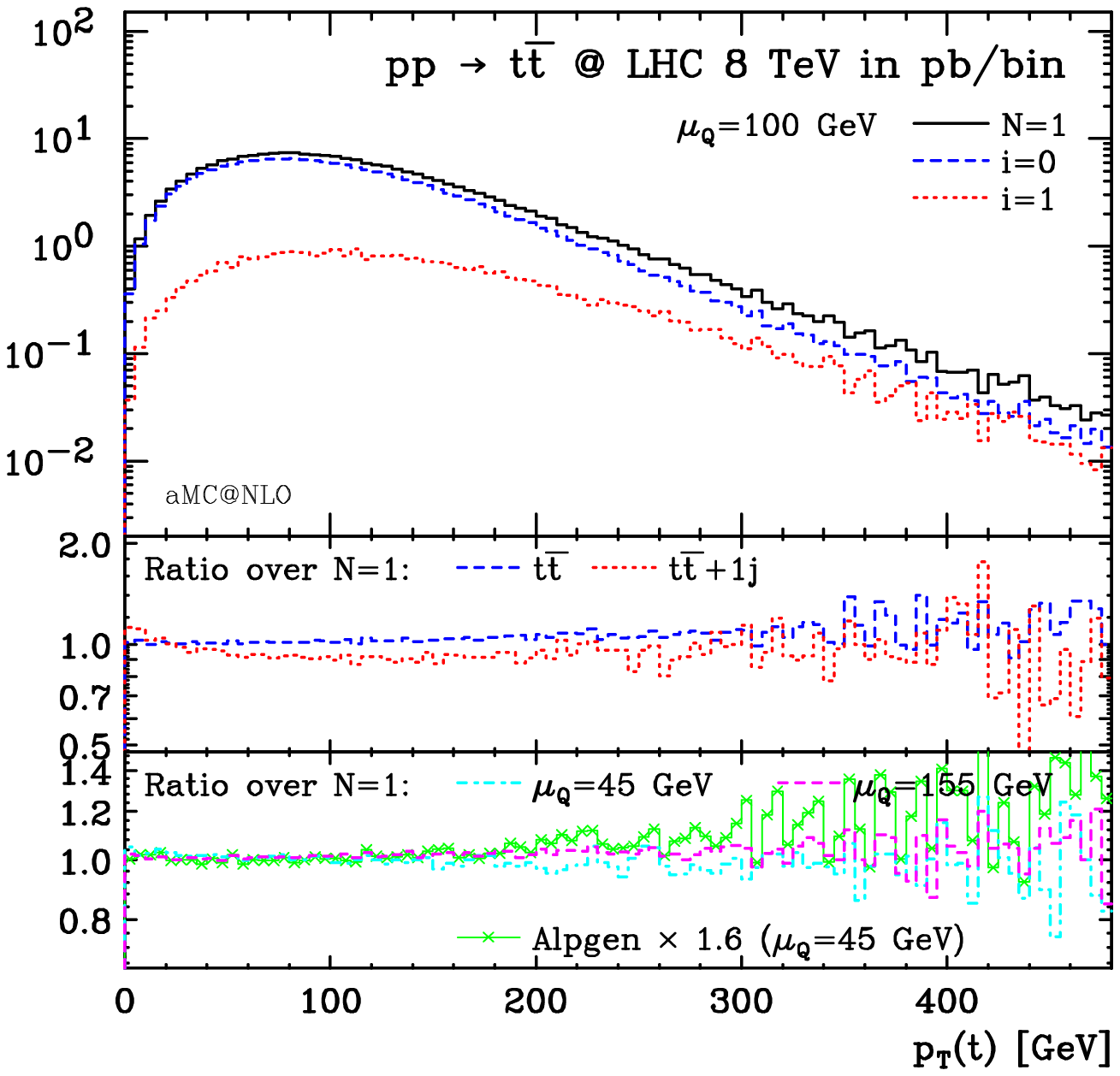, width=0.49\textwidth}
        \epsfig{file=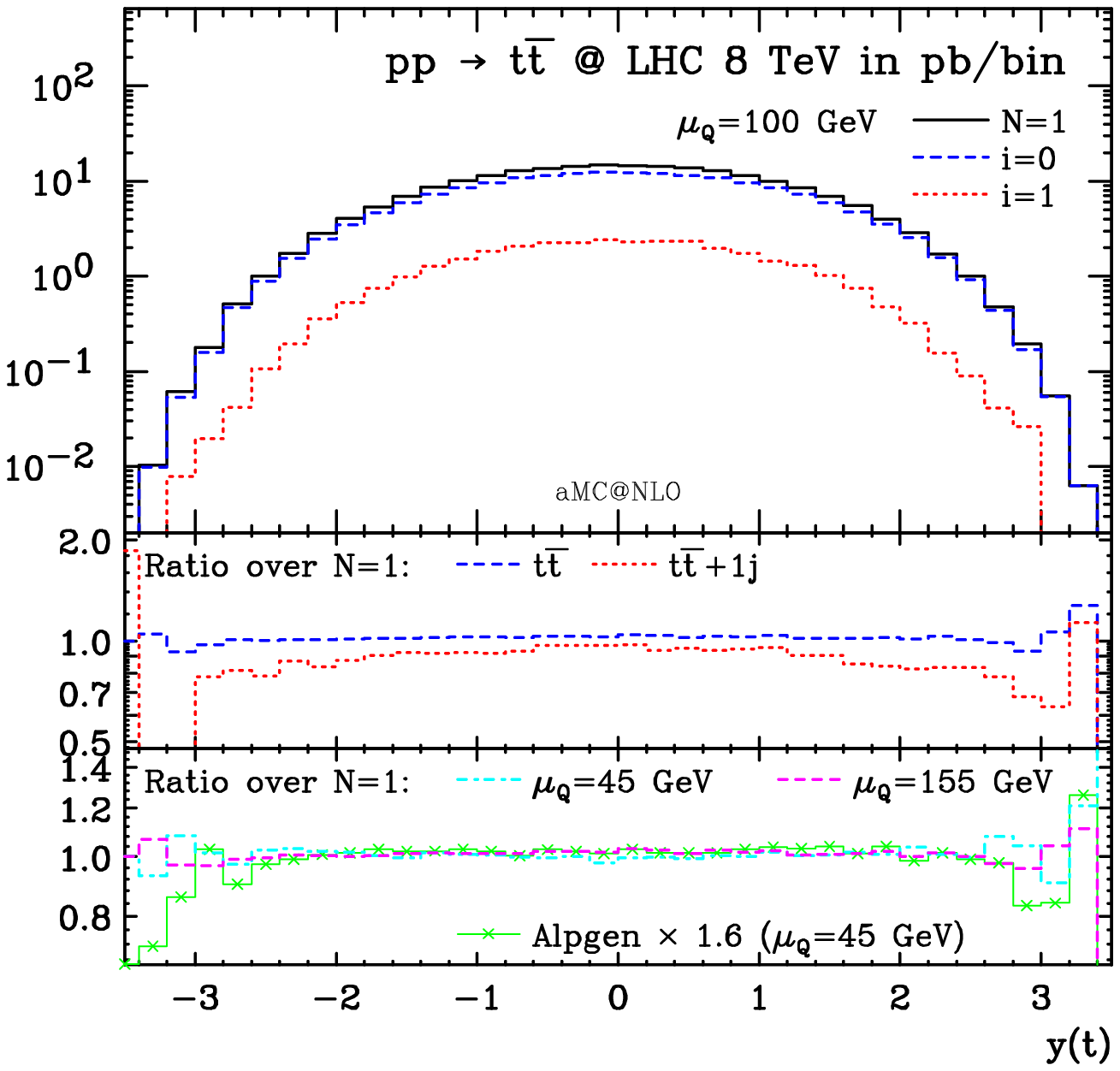, width=0.49\textwidth}
  \end{center}
  \begin{center}
        \epsfig{file=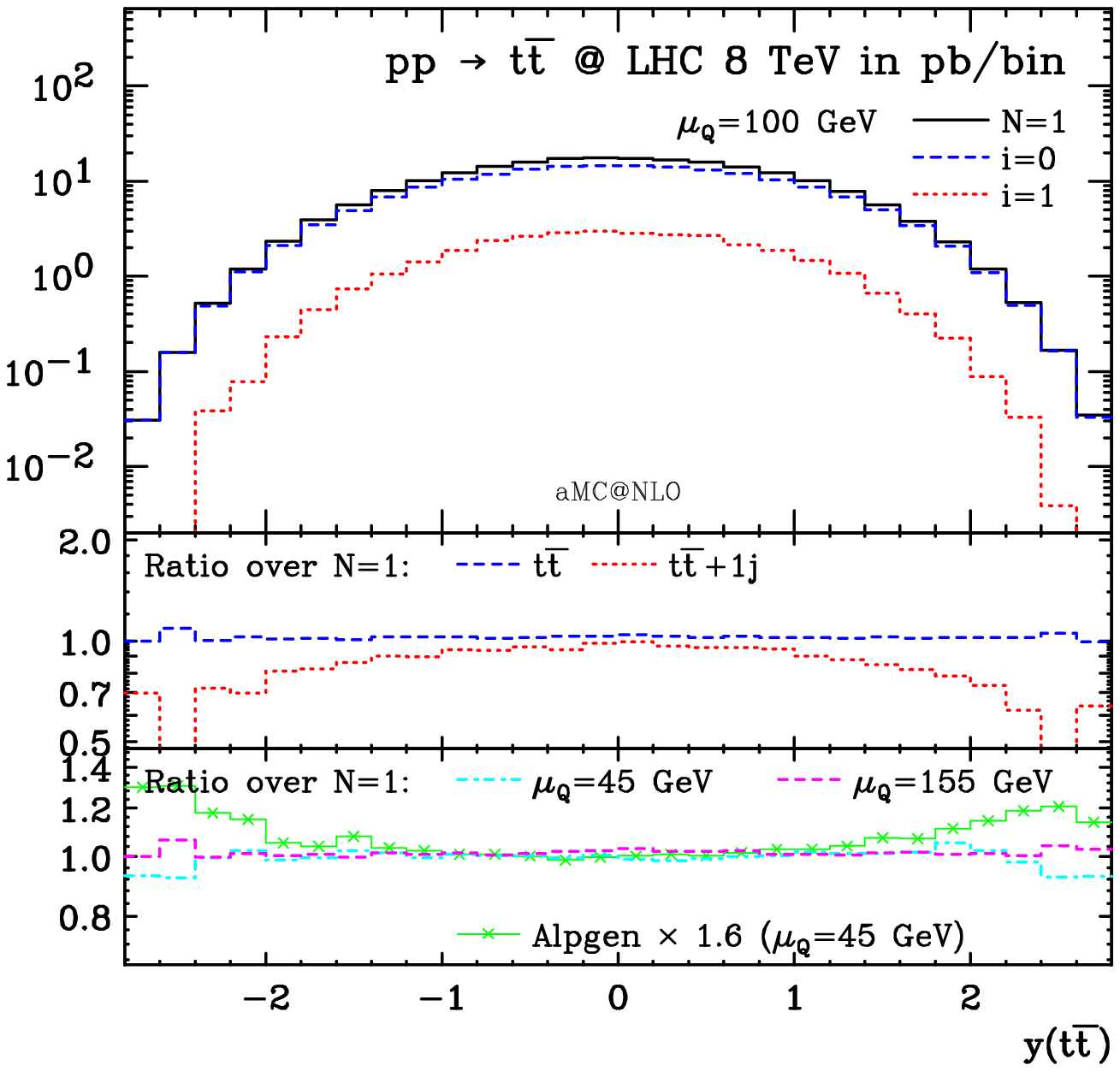, width=0.49\textwidth}
        \epsfig{file=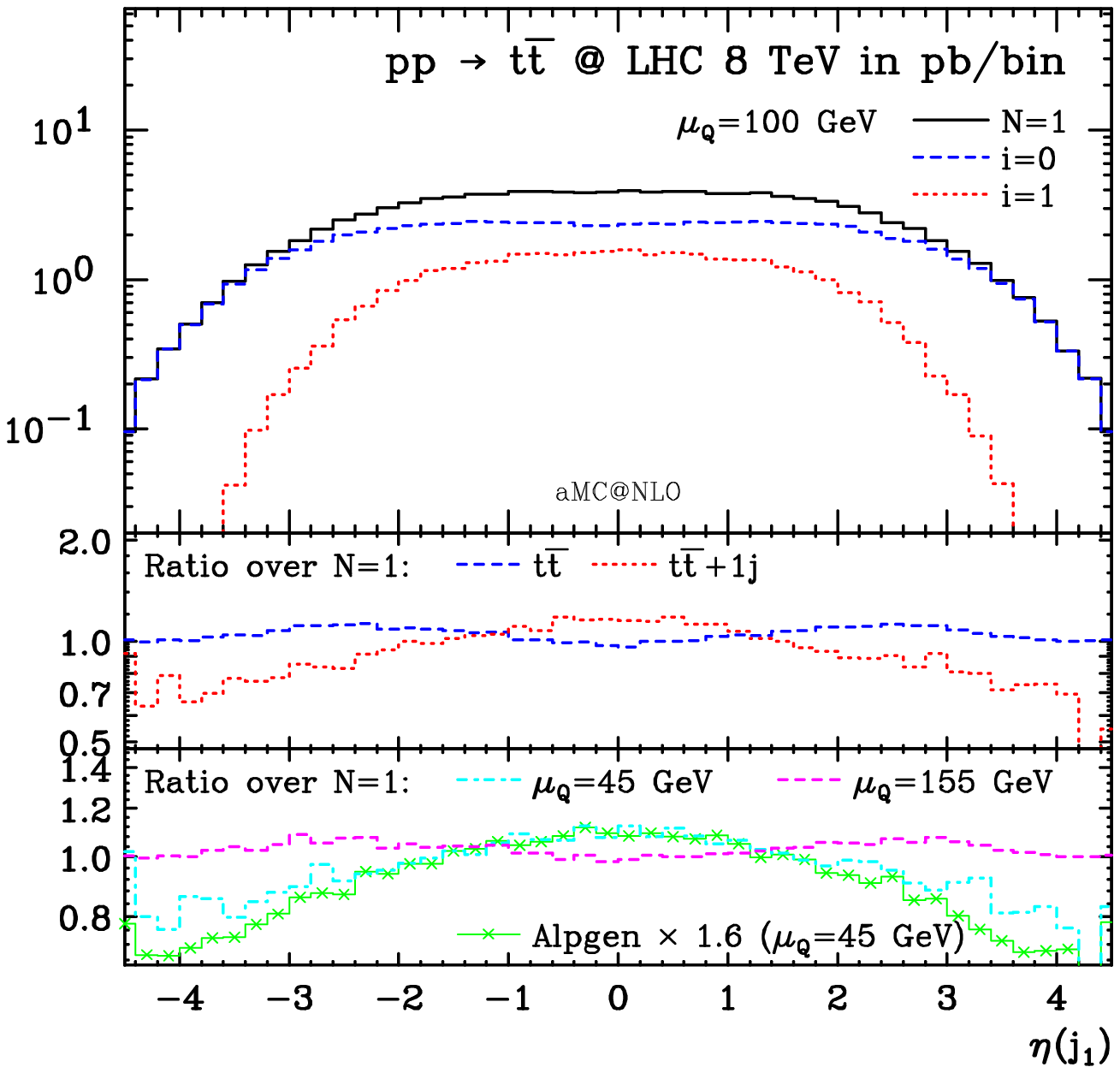, width=0.49\textwidth}
  \end{center}
  \vspace{-20pt}
  \caption{As in fig.~\protect\ref{fig:12}, for the top-quark $\pt$
    (upper left) and rapidity (upper right), the pair rapidity
    (lower left), and the hardest-jet pseudorapidity (lower right).
    The latter observable is obtained with a $\pt(j_1)>30$~GeV cut.}
\label{fig:13}
\end{figure}
\begin{figure}[htb!]
  \begin{center}
        \epsfig{file=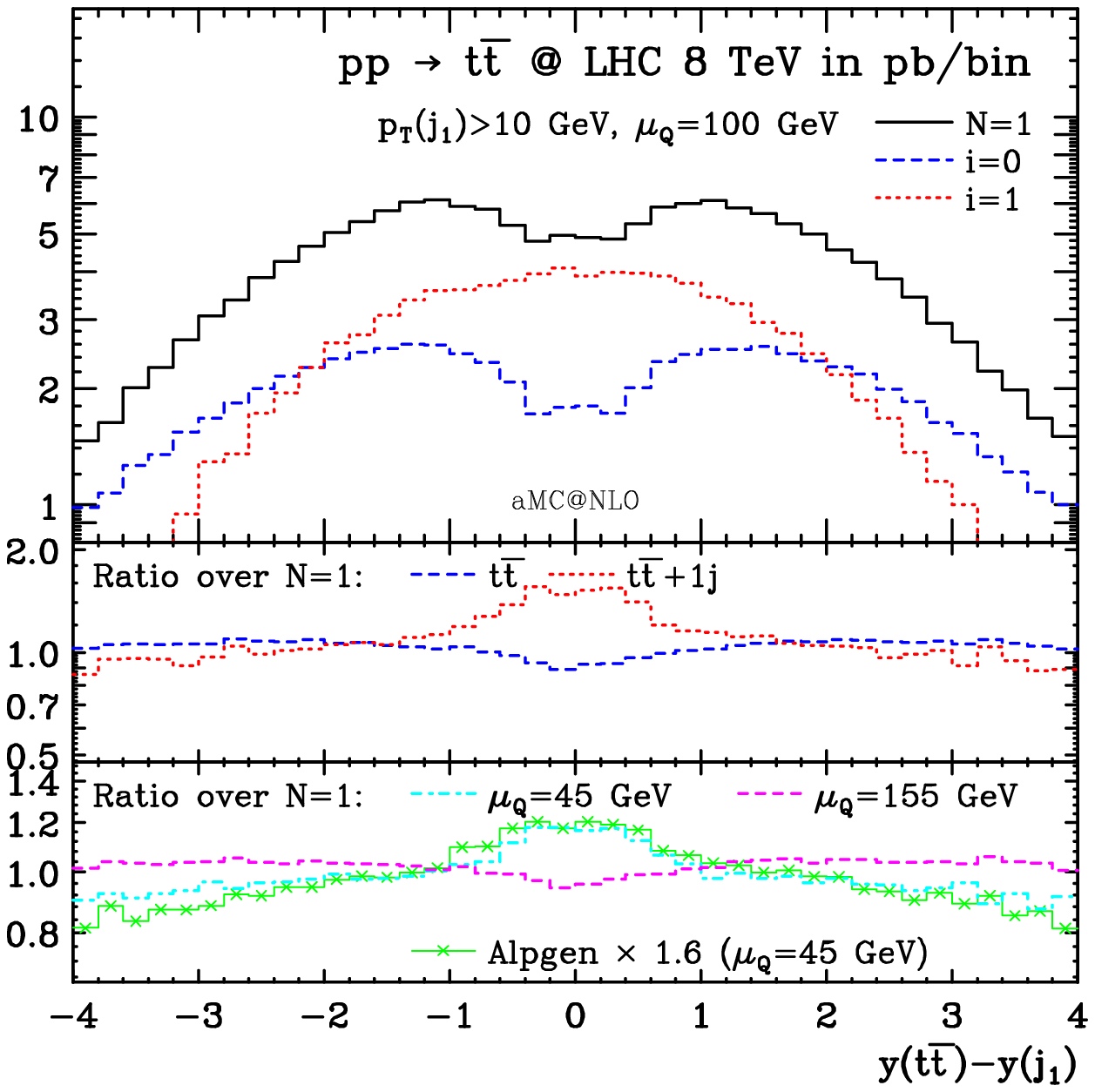, width=0.48\textwidth}
        \epsfig{file=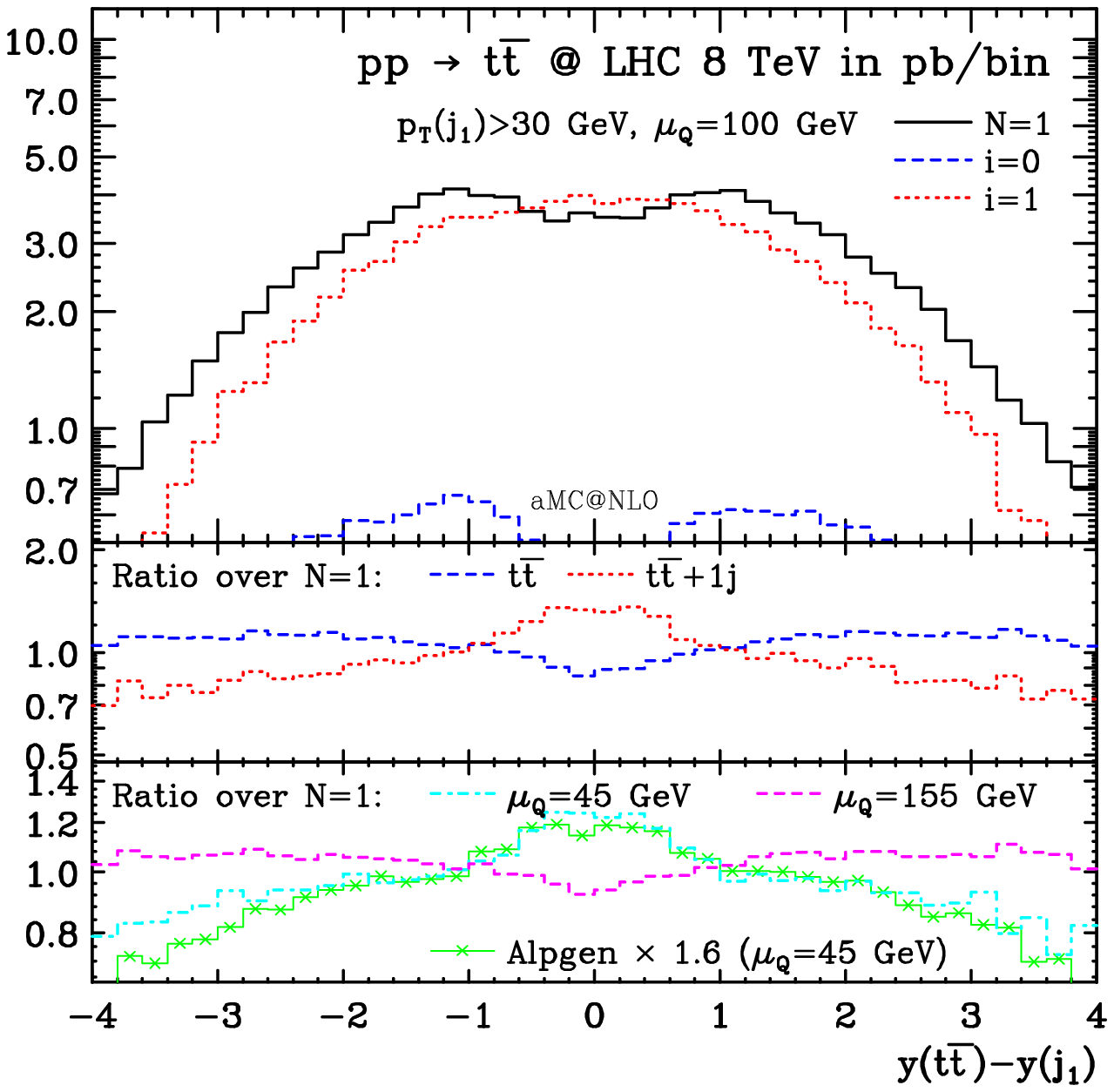, width=0.49\textwidth}
  \end{center}
  \begin{center}
        \epsfig{file=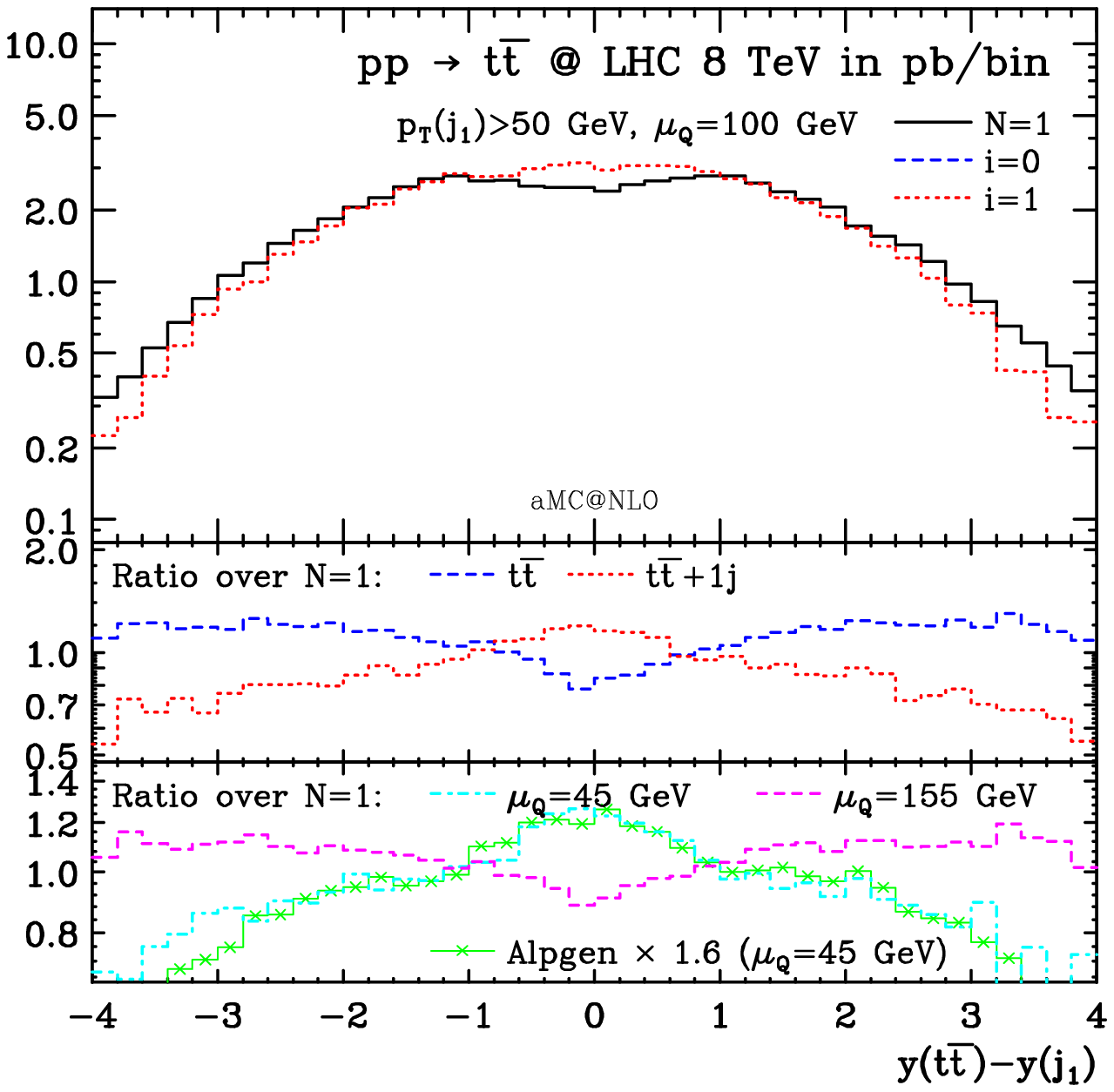, width=0.49\textwidth}
        \epsfig{file=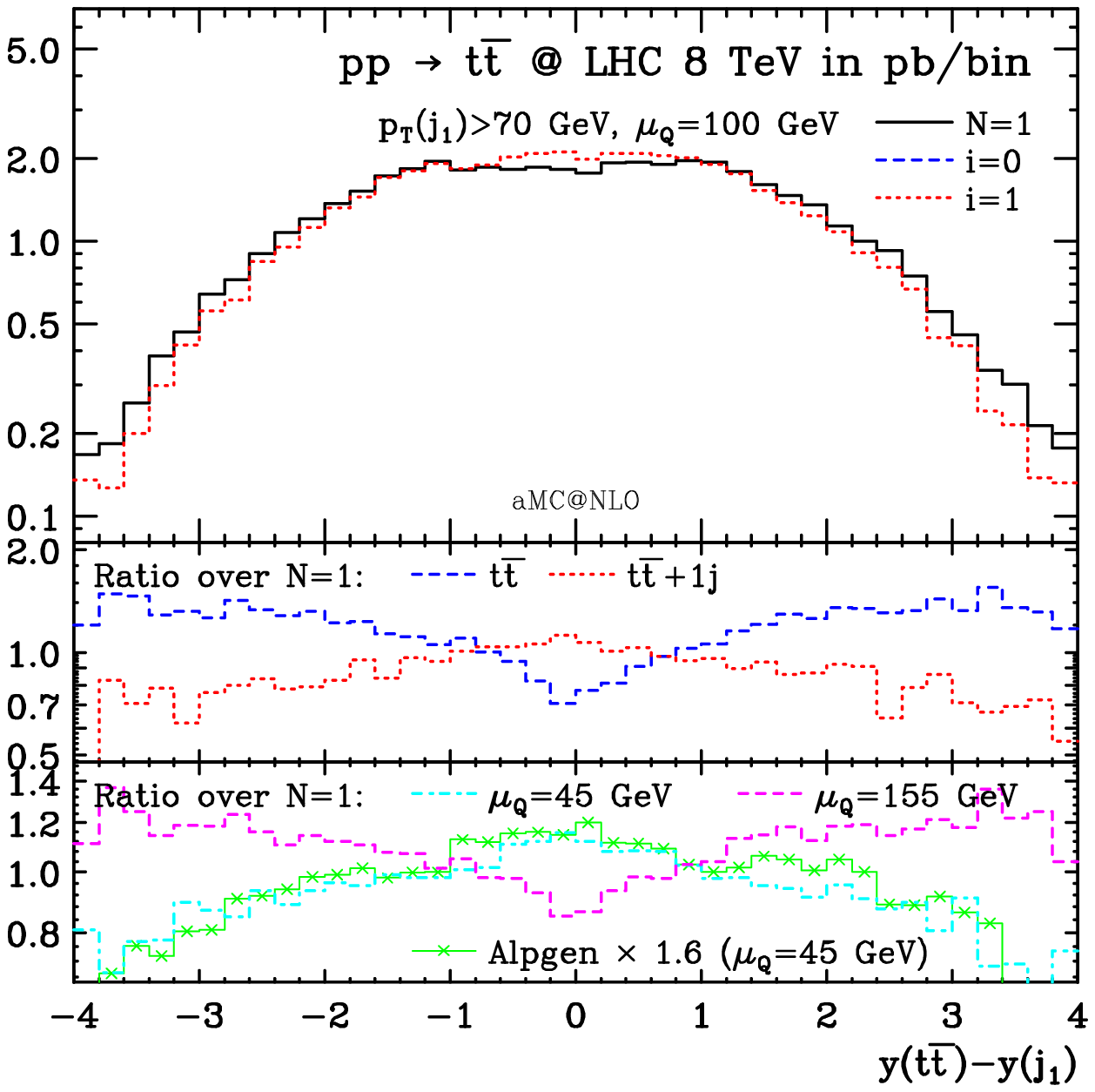, width=0.48\textwidth}
  \end{center}
  \vspace{-20pt}
  \caption{As in fig.~\protect\ref{fig:12}, for the difference in rapidity
    between the $t\tb$ pair and the hardest jet, for four different $\pt$
    cuts on the latter.}
\label{fig:14}
\end{figure}
We point out that the top quarks are not included in 
the clustering algorithm that determines the $d_i$'s of 
sect.~\ref{sec:tech}, which enter in the merging procedure.
This implies that gluon radiation off $t$ and $\tb$ is not
constrained, which is an acceptable (and approximate) solution
only because such radiation is quite unlikely to be hard in
the case of the very massive actual top. 
The one-loop matrix elements which contribute to the $0$- and
$1$-parton samples are those originally computed in 
refs.~\cite{Nason:1987xz} and~\cite{Dittmaier:2007wz} respectively. 
We have considered three values for the matching scale, 
$\mu_Q=100$ (our default), $45$, and $155$~GeV. We have also obtained
results with \alpgen, using $45$~GeV as matching scale in the
MLM procedure, and generating the $t\tb+0$, $t\tb+1$ and
$t\tb+2$ parton samples for consistency with the $N=1$ case.

For $t\tb$ production, one could repeat most of what has been
said in sects.~\ref{sec:resH} and~\ref{sec:resW}. However, there 
are a few specific features that are worth stressing. Firstly,
the merging systematics is greater than previously observed.
In part, this is due to the very large range of matching scales
adopted here, but it is also related to the dynamics of the present process.
Namely, up to quite large $\pt$ values (one can use the pair transverse
momentum to be definite) the standalone MC@NLO $t\tb+0j$ result is 
larger in absolute normalization than the $t\tb+1j$ one; this is the 
combined effect of the fact that the shower easily produces hard radiation 
(as a consequence of the top mass driving the setting of the
shower scale to relatively large values), and of the large $K$ factor 
in the $t\tb+0j$ NLO computation. This feature is easily seen e.g.~in
the upper inset of the upper-left panel of fig.~\ref{fig:12} -- the
relative difference between the dashed blue and dotted red histograms
is about 30\% for $\pt(t\tb)\ge 100$~GeV. Secondly, there is a good 
agreement between \alpgen, and the merged-NLO results obtained
with the same matching scale $\mu_Q=45$~GeV (see the comparison between
the dot-dashed blue and the solid-plus-crosses green histograms in the
lower insets). It will be interesting to see whether this agreement
holds also for much larger matching scales, such as those adopted
here at the NLO. However, one starts to see deviations between 
the two computations when approaching large $\pt$'s, 
which can be in part due to the feature of the NLO
cross sections outlined above, and in part to the different scale and PDF
choices made in \alpgen\ and MC@NLO. Thirdly, the agreement between the 
NLO-merged  and \alpgen\ results for $\eta(j_1)$ (see the lower right panel 
in fig.~\ref{fig:13}) shows how the inclusion of the $1$-parton sample
addresses the issue raised for standalone MC@NLO at the end of sect.~4
of ref.~\cite{Mangano:2006rw} for such a 
variable\footnote{Ref.~\cite{Mangano:2006rw} actually discussed the
case of rapidity (and with different jet-finding parameters), 
which is obviously fully analogous to what is done here.}. However, 
from that figure one can also see that the systematics affecting the
shape is quite large. Finally, for the $y(t\tb)-y(j_1)$
difference, shown in fig.~\ref{fig:14}, a dip or depletion is present
for all the $\pt(j_1)$ cuts if $\mu_Q=100$~GeV is used, while it
disappears in all cases except in the $\pt(j_1)>10$~GeV one when 
$\mu_Q=45$~GeV, for {\em both} the NLO-merged {\em and} \alpgen\ results.

\section{Conclusions\label{sec:conc}}
We have presented a procedure for the merging of 
MC@NLO simulations characterized by underlying processes with
different parton multiplicities. It entails only minor changes
to the standard MC@NLO short-distance cross sections, and therefore 
is not difficult to incorporate into existing implementations.
We have done so in the automated \amcatnlo\ framework, which 
guarantees maximum flexibility and independence of the process.

We have provided no proof, and made no claims, on the formal accuracy
of the merging. Instead, we have thoroughly compared the merged
results with those of standalone MC@NLO, for the three sample cases 
of Standard Model Higgs, $e^+\nu_e$, and $t\tb$ hadroproduction,
and found agreement in shape and normalization where relevant.
We have also studied the theoretical uncertainties that affect
the merging, and they have turned out to be quite small.
Although the agreement with LO-merged results computed with
\alpgen\ and the MLM approach is generally good, there are 
a few differences which will deserve further investigations.

There are obviously several open questions, both theoretical
and phenomenological, and this work should be seen only as a first
step towards further improvements. To name just a few: the application 
to processes that feature $b$ quarks or light jets at the Born level of 
the lowest multiplicity; the use of alternative definitions of 
the $d_i$'s; the use of different scales in the short-distance
cross sections and in the showers; the possibility of including other 
features of the CKKW scheme, such as PDF reweighting; the potential 
role of vetoed-truncated showers;  the use of tree-level matrix elements 
only (as opposed to NLO cross sections) for large multiplicities;
the issue of logarithms of higher orders in showers.
In the near future, we plan to apply the method proposed here to 
the \HWpp\ and \PYe\ Monte Carlos (which should not require 
any modifications), to extensively compare our results with other 
LO and NLO merging techniques, and to validate our approach using LHC data.

\section*{Acknowledgments}
We wish to thank F.~Maltoni, M.~Mangano, P.~Torrielli, B.~Webber
and M.~Zaro for the discussions we have had with them during the 
course of this work.

\end{document}